\documentclass[twoside,12pt,a4paper]{article}
\usepackage{a4wide}
\usepackage{graphicx,natbib}
\usepackage[T1]{fontenc}
\usepackage{lmodern}
\usepackage[latin1]{inputenc}
\usepackage{latexsym,enumitem}
\usepackage{amsmath,amsthm,amsopn,amstext,amscd,amsfonts,amssymb}
\usepackage{xcolor,colortbl,caption,wrapfig,wasysym}
\usepackage{booktabs,mm}
\usepackage{pstricks-add}
\usepackage{algorithm}
\usepackage[noend]{algpseudocode}
\algnewcommand{\LeftComment}[1]{\Statex \(\triangleright\) #1}
\algnewcommand\algorithmicto{\textbf{to}}
\algrenewtext{For}[3]%
{$\algorithmicfor\ #1 \gets #2\ \algorithmicto\ #3\ \algorithmicdo$}
\usepackage{enumitem}

\title{Approximate computation of projection depths}
\author{Rainer Dyckerhoff$^1$, Pavlo Mozharovskyi$^2$, Stanislav Nagy$^3$ \\
\indent\\
\small{$^1$Institute of Econometrics and Statistics, University of Cologne} \\ 
\small{$^2$LTCI, Telecom Paris, Institut Polytechnique de Paris} \\
\small{$^3$Faculty of Mathematics and Physics, Charles University} \\
}

\allowdisplaybreaks[2]
\newcommand{\D}{\mathrm{D}}
\newcommand{\ip}[2]{\left\langle#1,#2\right\rangle}

\newtheorem{definition}{Definition}
\newtheorem{proposition}{Proposition}

\renewcommand{\phi}{\varphi}

\newcommand{\dist}{\mathop{\text{dist}}\nolimits}

\newcommand{\linspan}{\mathop{\mathrm{span}}}

\DeclareMathOperator*{\argmin}{arg\,min}
\def\mc#1#2#3{\multicolumn{#1}{#2}{#3}}

\definecolor{mark}{RGB}{219,166,25}
\newcommand{\md}{\cellcolor{mark!75}}

\newcommand{\bestPerc}{\Circle}
\newcommand{\bestRank}{\CIRCLE}        
\newcommand{\bestRankPerc}{\CIRCLE}
\begin{document}
\date{}
\thispagestyle{empty}
\maketitle
\thispagestyle{empty}


\begin{abstract}
Data depth is a concept in multivariate statistics that measures the centrality 
of a point in a given data cloud in $\IR^d$.
If the depth of a point can be represented as the minimum of the
depths with respect to all one-dimensional projections of the data, then 
the depth satisfies the so-called projection property. 
Such depths form an important class that includes many of the depths
that have been proposed in literature.
For depths that satisfy the projection property an approximate algorithm 
can easily be constructed since taking the minimum of the depths with 
respect to only a finite number of one-dimensional projections
yields an upper bound for the depth with respect to the multivariate data. 
Such an algorithm is particularly useful if no exact algorithm exists or if 
the exact algorithm has a high computational complexity, as is the case
with the halfspace depth or the projection depth. To compute these depths
in high dimensions, the use of an approximate algorithm with better
complexity is surely preferable.
Instead of focusing on a single method we provide a comprehensive and fair 
comparison of several methods, both already described in the literature  
and original.
\end{abstract}



\section{Introduction}\label{sec:Intro}


A statistical depth function is a generalization of the concept of quantiles to multivariate data. Given a probability measure or a data sample in $\mathbb{R}^d$, the depth assigns to any point of the space $\mathbb{R}^d$ a real number, usually scaled to $[0,1]$, which characterizes its degree of centrality w.r.t. this distribution or data set. By providing a non-parametric, affine-invariant, and (often) robust multivariate ordering, data depth finds numerous applications, e.g., in descriptive statistics, statistical inference, or risk measurement to name only a few \citep{LiuPS99,Cascos09}. For more information on applications of data depth we refer to surveys \cite{ZuoS00} and \cite{Mosler13}.

Application of the depth-based methods requires efficient algorithms for the computation of depths. Since many depth notions are fully data-driven and affine-invariant, their computation may constitute a challenge. For example, computation of the halfspace depth~\citep{Tukey75} --- one of the most important depth notions in the literature --- is an NP-hard problem~\citep{JohnsonP78}, and the only exact existing algorithm for computation of the projection depth~\citep{LiuZ14} is still very slow. For this reason, theoretical developments on the data depth are accompanied by a substantial body of literature on its computation, which still contains a number of open problems. While exact computation of certain depth notions can come at a very high computational cost (see, e.g., \citealp{DyckerhoffM16} for the halfspace depth and~\citealp{LiuZ14} for projection depth), approximations have been proposed.

\cite{Dyckerhoff04} described a class of depths which satisfy the weak projection property. Out of the existing variety, these can be calculated in a universal way by minimizing depth in univariate projections. For this, in each direction, only the univariate depth of $n$ observations should be computed, which often has computational time complexity only $O(n)$. This class of depths includes in particular Mahalanobis~\citep{Mahalanobis36}, zonoid~\citep{KoshevoyM97}, halfspace~\citep{Tukey75}, projection~\citep{ZuoS00} and asymmetric projection~\citep{Dyckerhoff04} depth, which constitute the focus of the current work. 

Several authors have already applied approximation techniques to the (sample) depth computation, notably for the halfspace depth and projection depth. Purely random methods seem most intuitive and have been used, e.g., by \cite{CuestaAlbertosNR08} for the halfspace depth and \cite{LiuZ14} for the projection depth. More sophisticated procedures were proposed as well. \cite{RousseeuwS98} minimize univariate halfspace depth after projecting data onto directions based on a random combination of sample points. \cite{ChenMW13} first project data on subspaces orthogonal to linear spans of $0 < k \le d$ points from the sample, and then employ a brute-force approximation of the halfspace depth in these projections. \cite{MozharovskyiML15} accelerate the problem of approximation of the halfspace depth of many points (and of the sample itself) w.r.t. a sample by preliminary sorting the data in all projections. \cite{DuttaG12} use the Nelder-Mead algorithm \citep[][run in $\mathbb{R}^d$]{NelderM65} for approximation of the projection depth.

In this work, a systematic experimental approach is used to study the behavior of the approximation of the sample depth by minimizing it on univariate directions. For minimization of the univariate depth, we contrast eight approximation algorithms in an extensive simulation study. As algorithms we considered: (i) random search, (ii) grid search, (iii) refined random search, (iv) refined grid search, (v) random simplices, (vi) simulated annealing, (vii) coordinate descent, (viii) Nelder-Mead. Since the performance of the algorithms depends on chosen parameters, we start by fine-tuning of the algorithms. After this, the algorithms are compared in different settings to provide a trustworthy conclusion.

Throughout this paper we use the following notation. 
Vectors are notated with bold letters, e.g., $\bmx,\bmy,\bmz$. 
$\IR^+_0$ denotes the set of non-negative real numbers. The $d-1$ dimensional
unit sphere in $\IR^d$ is denoted by $\IS^{d-1}$.
The transpose of a vector $\bmx$ is denoted by $\bmx\T$. 
The $d\times d$ identity matrix is denoted by $\bmI_{d}$, or shortly $\bmI$, and $\boldsymbol{0}$ or $\boldsymbol{0}_d$ denotes the origin in $\mathbb{R}^d$.
For a set $A$ we denote by $\mathcal{U}(A)$ the uniform distribution on the set $A$.
$\text{conv}\{A\}$ denotes the convex hull of $A$, i.e., the smallest convex set containing $A$.
The elements of a vector $\bmx \in \mathbb{R}^d$ are denoted by $x_1,\dots,x_d$.
By $\overset{\mathrm{d}}{=}$ we mean equality in distribution.

\section{Preliminary material}
In Section~\ref{ssec:datadepth} we give a short introduction into the
notion of \emph{depth} and discuss the \emph{projection property} and its
implications for the computation of depth. We will see that for depths 
satisfying the projection property, the computation of the depth is equivalent
to minimizing an objective function over the unit sphere. Therefore, in
Section~\ref{ssec:GeomSphere} we will have a closer look on the unit sphere
and describe its geometry.    

\subsection{Data depth}\label{ssec:datadepth}
Depth is a concept that measures the centrality of a given point $\bmz\in\IR^d$
w.r.t. a probability distribution $P$ on $(\IR^d,\mathcal{B}^d)$ where
$\mathcal{B}^d$ denotes the Borel $\sigma$-algebra on $\IR^d$. A generic depth is denoted by $\D(\bmz|P)$. In applications, the probability measure $P$ is often
the empirical measure on a set of data points $\bmX=(\bmx_1,\dots,\bmx_n)$.  
In that case we write $\D(\bmz|\bmX)$.
Every reasonable notion of depth should satisfy the following set of axioms.
\begin{description}
\item[D1: Affine invariance.] For every non-singular $d\times d$-matrix $\bmA$ and 
$\bmb\in\IR^d$ it holds true that $\D(\bmz\,|\,\bmX)=\D(\bmA\bmz+\bmb\,|\,\bmA\bmX+\bmb)$, where $\bmA\bmX+\bmb = (\bmA\bmx_1 + \bmb, \dots, \bmA\bmx_n + \bmb)$.
\item[D2: Vanishing at infinity.] $\lim_{\|\bmz\|\to\infty}\D(\bmz\,|\,\bmX)=0$.
\item[D3: Upper semicontinuity.] For each $\alpha>0$ the set 
$\{\bmz\in\IR^d\,|\,\D(\bmz|\bmX)\ge\alpha\}$ is closed.
\item[D4: Monotone decreasing on rays.] For each point $\bmx_0$ of maximal depth
and each $\bmr\in \IR^d$, $\bmr\ne\bmNull$, the function 
$\lambda\mapsto \D(\bmx_0+\lambda\bmr\,|\,\bmX)$, $\lambda\ge0$, is monotone
non-increasing.
\end{description}
Properties D1, D2, and D4 have been introduced by \cite{Liu90}. A further
set of axioms for a depth has been given by \cite{ZuoS00}.
The main difference between their axioms and ours is that they do not require a
depth to be upper semicontinuous. In addition, they require that for
distributions having a properly defined unique center of symmetry, the depth attains its
maximum value at this center. However, for centrally symmetric distributions,
this follows already from our axioms. 
For further discussion on these axioms, see e.g., \cite{Dyckerhoff04}. In the rest
of this paper we assume that a depth satisfies the four axioms D1, D2, D3 and D4.

We consider five depth notions; 
they all satisfy the four above mentioned properties.

For a point $\bmz\in\mathbb{R}^d$, its Mahalanobis depth (MD) \citep{Mahalanobis36} w.r.t. a data set $\bmX=(\bmx_1,...,\bmx_n)$ is defined as follows:
\begin{equation*}
	\D_M(\bmz|\bmX) = \bigl(1 + (\bmz - \ol{\bmx})\T\bmS_{\bmX}^{-1}(\bmz - \ol{\bmx})\bigr)^{-1}\,.
\end{equation*}
Here, $\ol{\bmx}$ and $\bmS_{\bmX}$ denote the mean and the empirical covariance matrix of $\bmX$, respectively.

The halfspace depth (HD) \citep{Tukey75,DonohoG92} is defined by
\begin{equation*}
	\D_{H}(\bmz|\bmX)=\min_{\bmp\in\mathbb{S}^{d - 1}} \frac{1}{n}\# \{i\,:\,\ip{\bmx_i}{\bmp} \ge \ip\bmz\bmp,\,i=1,...,n\}
\end{equation*}
where $\# A$ denotes the number of elements of a set $A$.

The zonoid depth (ZD) \citep{KoshevoyM97} 
is given by
\begin{equation*}
    \D_{Z}(\bmz|\bmX)=\sup\{\alpha\in(0,1]:\,\boldsymbol{z}\in Z_{\alpha}(\bmX)\}\,,
\end{equation*}
where $Z_{\alpha}(\bmX)$ is the zonoid $\alpha$-trimmed region, defined 
by 
\[
Z_{\alpha}(\bmX)=\left\{\sum_{i=1}^n\lambda_i\bmx_i:0\le\lambda_i\le\frac{1}{n\alpha},
\sum_{i=1}^n\lambda_i=1\right\}
\]
and the convention $\sup\emptyset=0$ is used.

The projection depth (PD) \citep{ZuoS00} is given by
\begin{equation*}
	    \D_{P}(\bmz|\bmX)=\min_{\bmp\in\mathbb{S}^{d - 1}}\Bigl(1 + \frac{|\ip\bmz\bmp - \text{med}(\ip\bmX\bmp)|}{\text{MAD}(\ip\bmX\bmp)}\Bigr)^{-1}\,,
\end{equation*}
where $\ip\bmX\bmp$ is understood to be the univariate data set obtained by projecting each point of $\bmX$ on $\bmp$, med is the univariate median, and MAD is the median absolute deviation from the median.

Since PD is always symmetric around its deepest point, the asymmetric projection depth (APD) \citep{Dyckerhoff04} has been defined as
\begin{equation*}
	    \D_{AP}(\bmz|\bmX)=\min_{\bmp\in\mathbb{S}^{d - 1}}\Bigl(1 + \frac{(\ip\bmz\bmp - \text{med}(\ip\bmX\bmp))_+}{\text{MAD}^+(\ip\bmX\bmp)}\Bigr)^{-1}\,,
\end{equation*}
with $(a)_+=\max\{a,0\}$ being the positive part of $a$ and MAD$^+$ being the median of the positive deviations from the median.

All of the above depths satisfy the \emph{(weak) projection property}, defined
as follows.
\begin{definition}
A depth $\D$ satisfies the (weak) \emph{projection property}, if for each point 
$\bmy\in\IR^d$ and each sample $\bmX$ it holds:
\[
\D(\bmy|\bmX)=\inf\{\D(\ip{\bmp}{\bmy}| \ip{\bmp}{\bmX})|\bmp\in \IS^{d-1}\}\,.
\]
\end{definition}

If a depth satisfies the projection property, its computation is
equivalent to minimization of the (possibly non-differentiable) objective
function 
\[
\phi_{\bmz,\bmX}:\IS^{d-1}\to\IR^+_0,\,\bmp\mapsto \D(\ip{\bmp}{\bmz}|\ip{\bmp}{\bmX})\,.
\]
Therefore, classical optimization methods can be used to compute the depth.
Particular attention should be paid here to the domain of the function $\phi_{\bmz,\bmX}$
which is the the unit sphere $\IS^{d-1}$. Of course, the function $\phi_{\bmz,\bmX}$ could 
be easily extended to a function $\tilde\phi_{\bmz,\bmX}$ defined on 
$\IR^d\setminus\{\bmNull\}$ by setting $\tilde\phi_{\bmz,\bmX}(\bmp)=
\D(\ip{\bmp}{\bmz}|\ip{\bmp}{\bmX})$. However, because of the affine invariance of
the depth, $\tilde\phi_{\bmz,\bmX}$ is constant on lines through the origin. Therefore,
we claim that it should be advantageous to use optimization methods which are
adapted to the particular domain $\IS^{d-1}$. This claim will be
confirmed in the simulation studies in Section~\ref{sec:simulation}.      

\begin{figure}[ht]
\centerline{%
\begin{tabular}{@{}ccc@{}}
\toprule
Zonoid depth & Halfspace depth & Projection depth\\
\midrule
\mc{3}{c}{Trivariate normal distribution}\\
\includegraphics[width=0.29\textwidth]{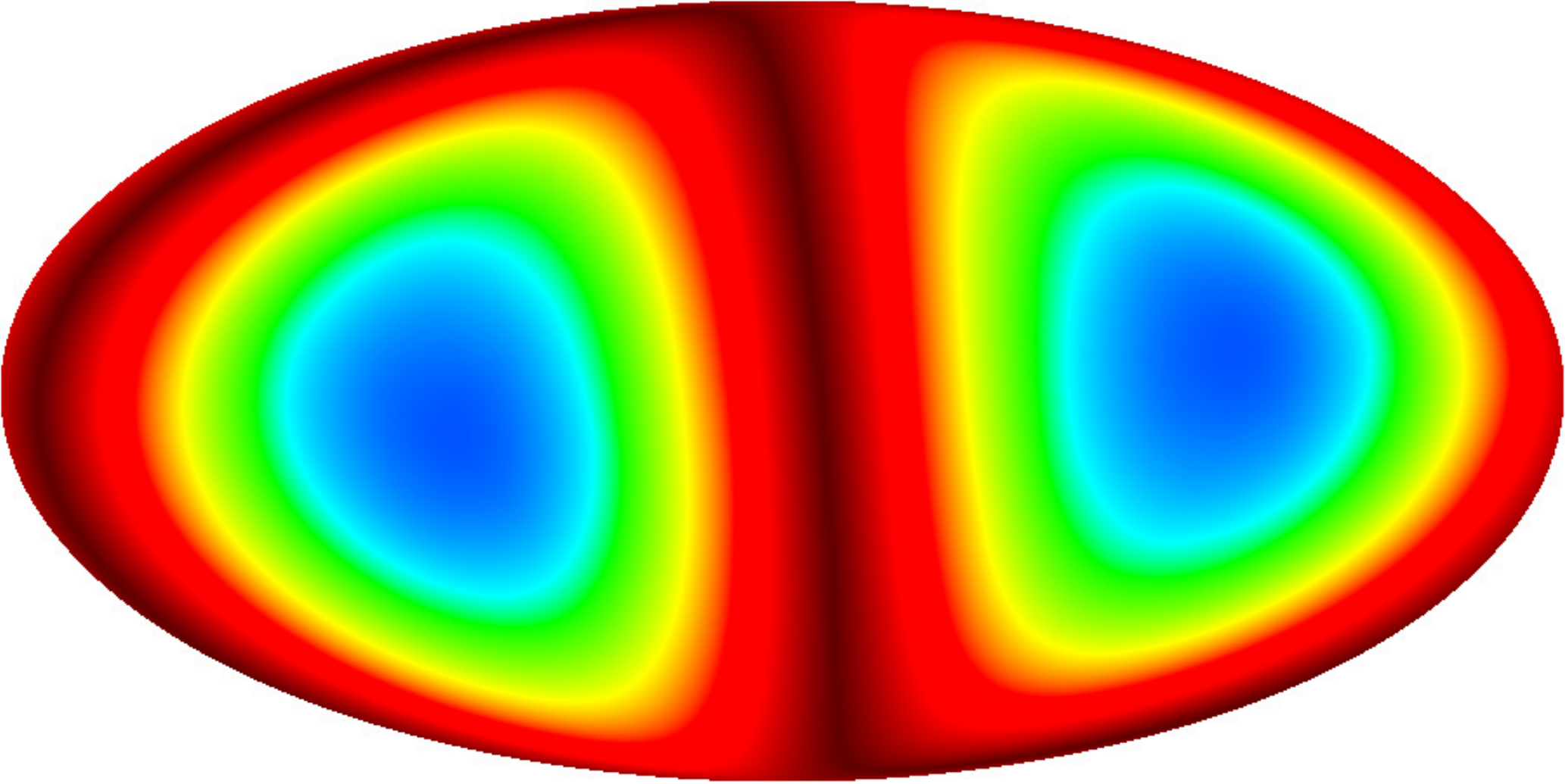} &
\includegraphics[width=0.29\textwidth]{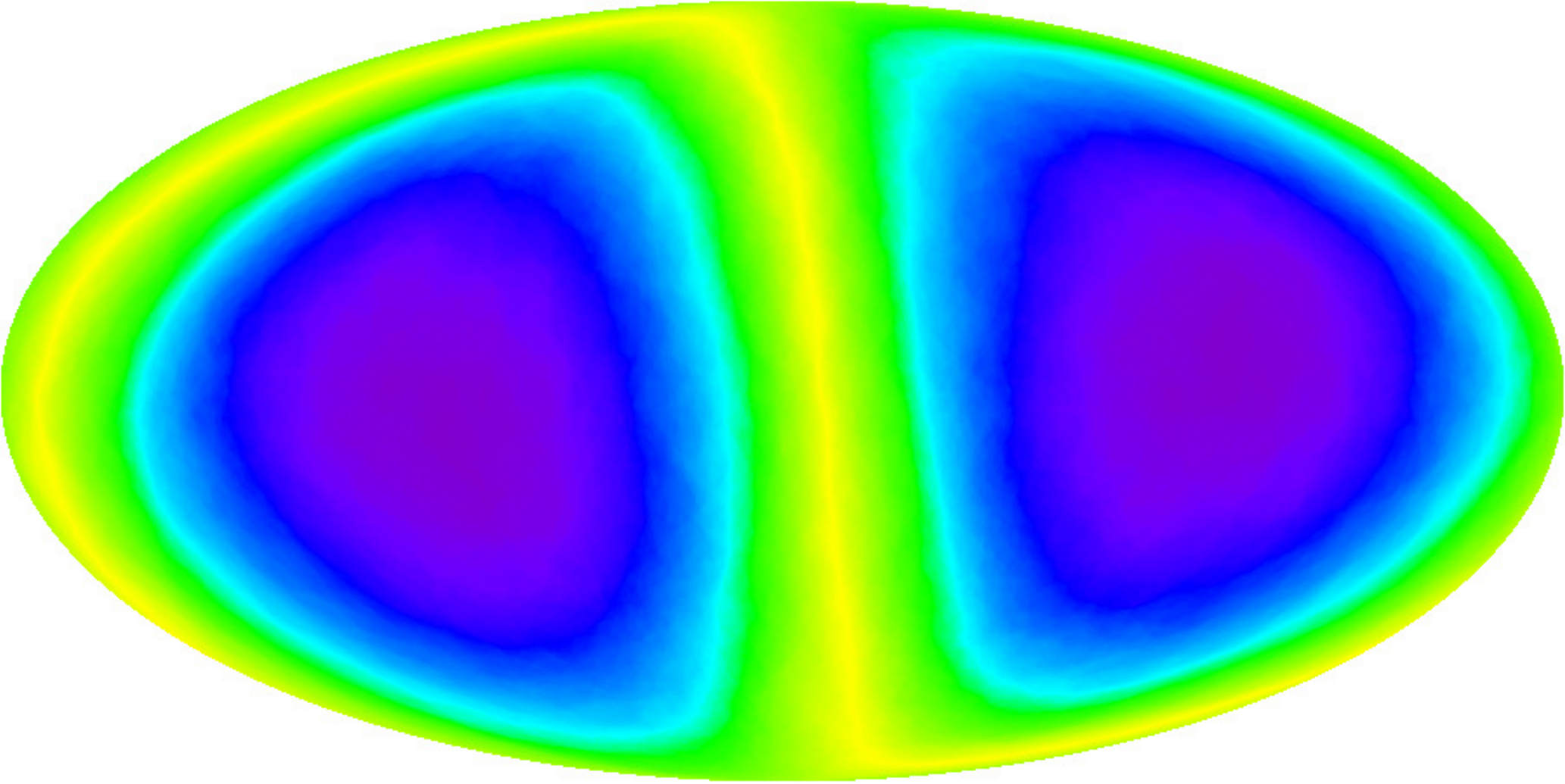} &
\includegraphics[width=0.29\textwidth]{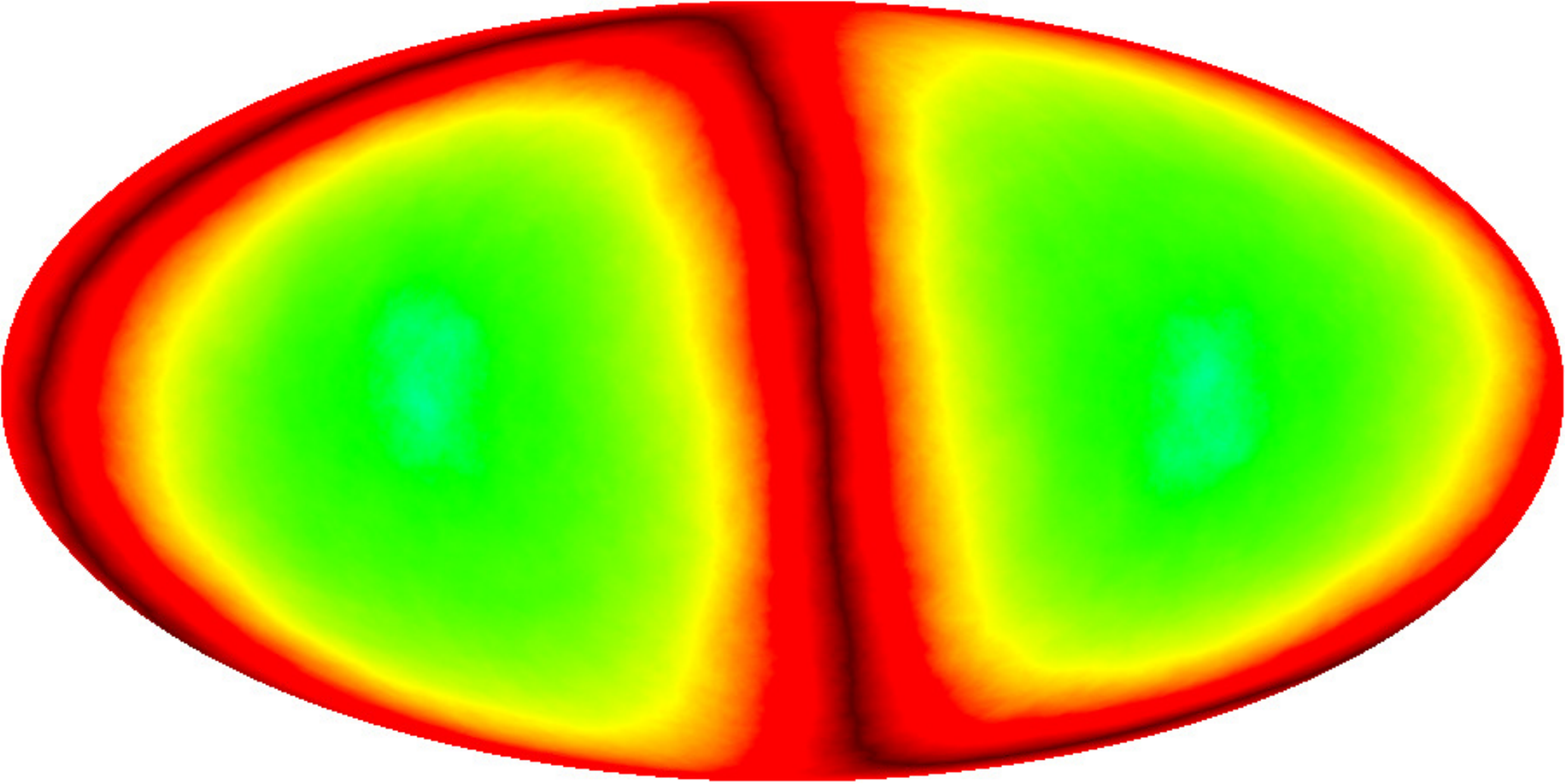}\\
\midrule
\mc{3}{c}{Trivariate Cauchy distribution}\\
\includegraphics[width=0.29\textwidth]{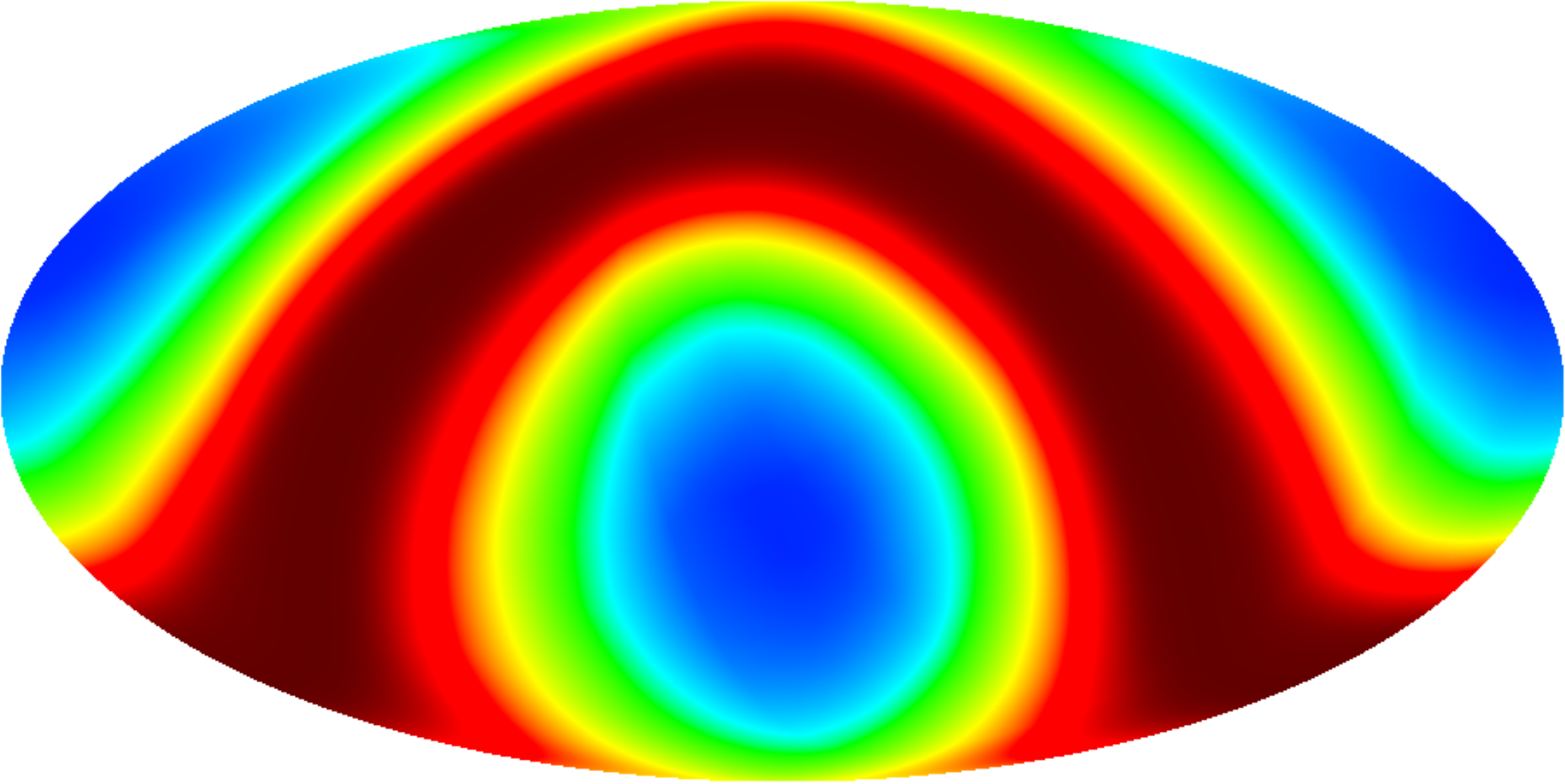} &
\includegraphics[width=0.29\textwidth]{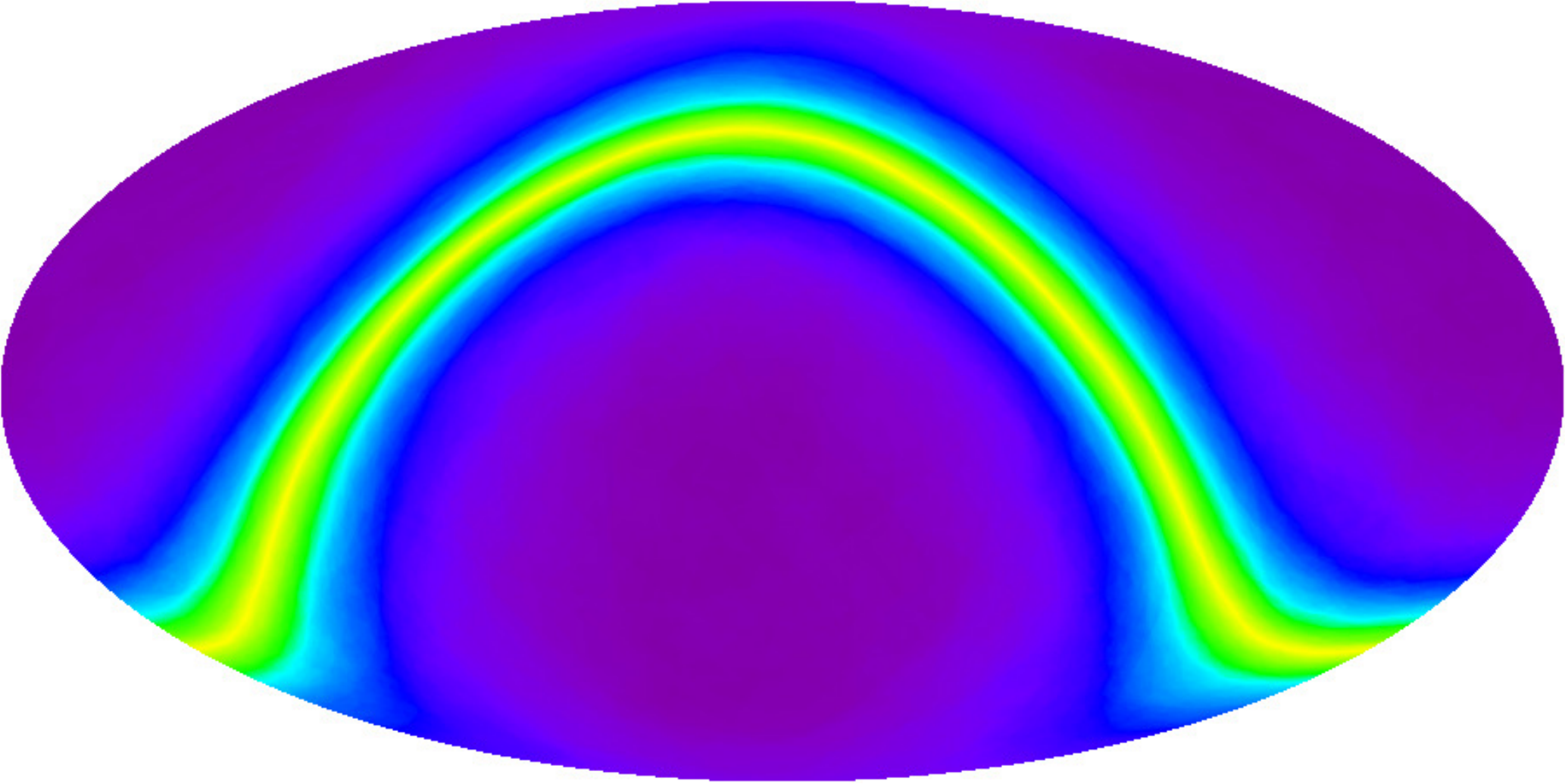} &
\includegraphics[width=0.29\textwidth]{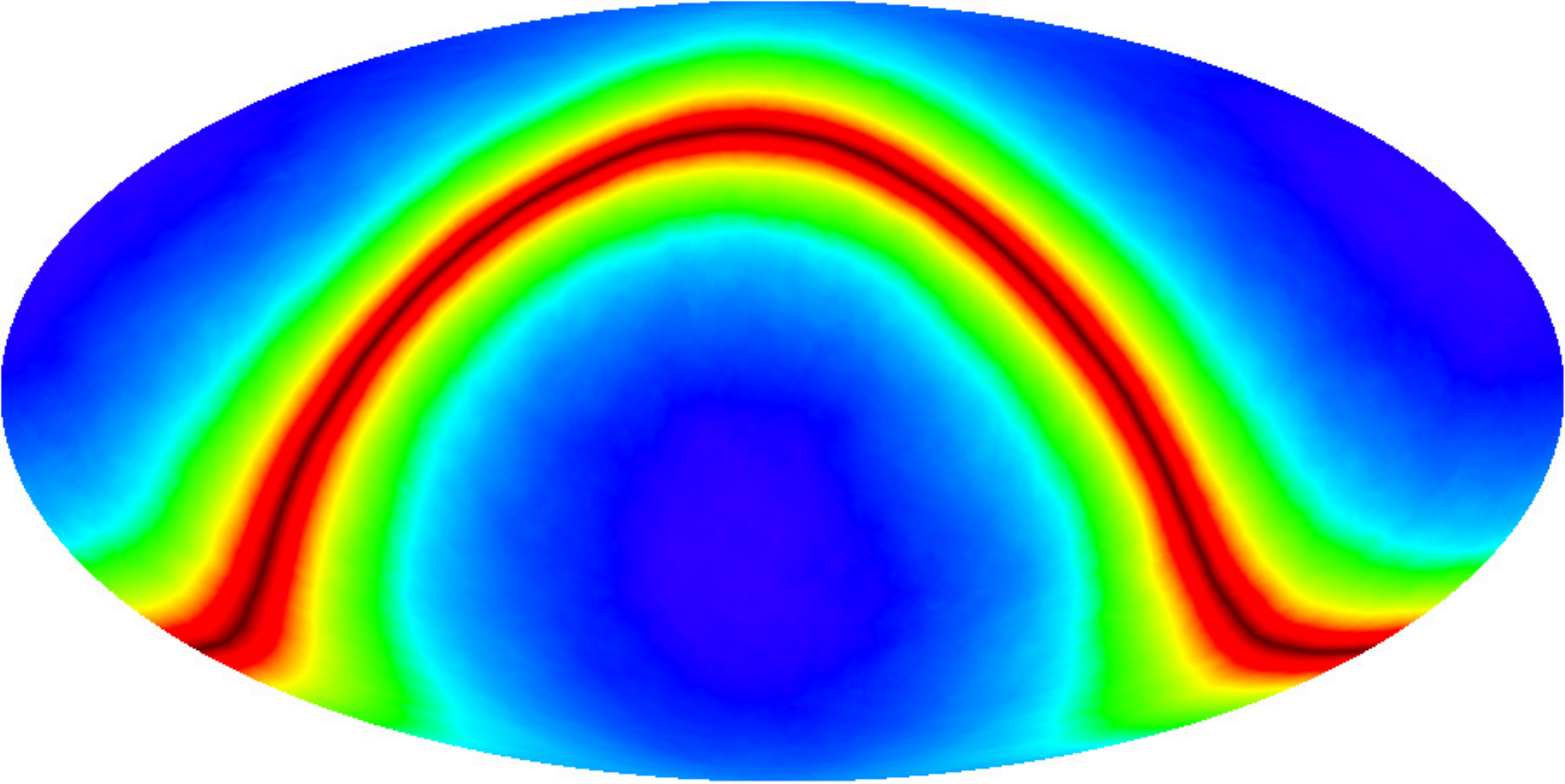}\\
\midrule
\mc{3}{c}{Trivariate uniform distribution}\\
\includegraphics[width=0.29\textwidth]{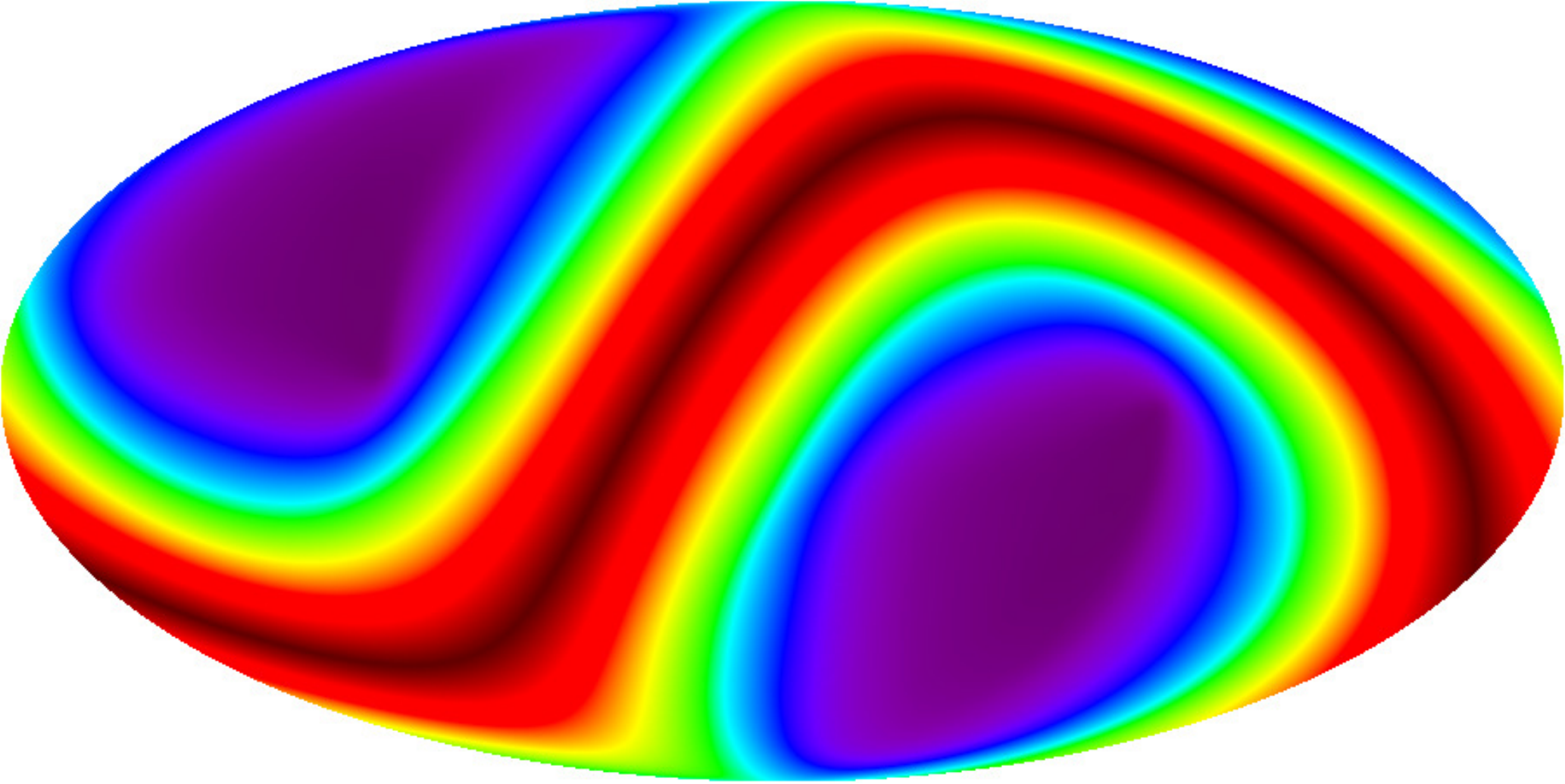} &
\includegraphics[width=0.29\textwidth]{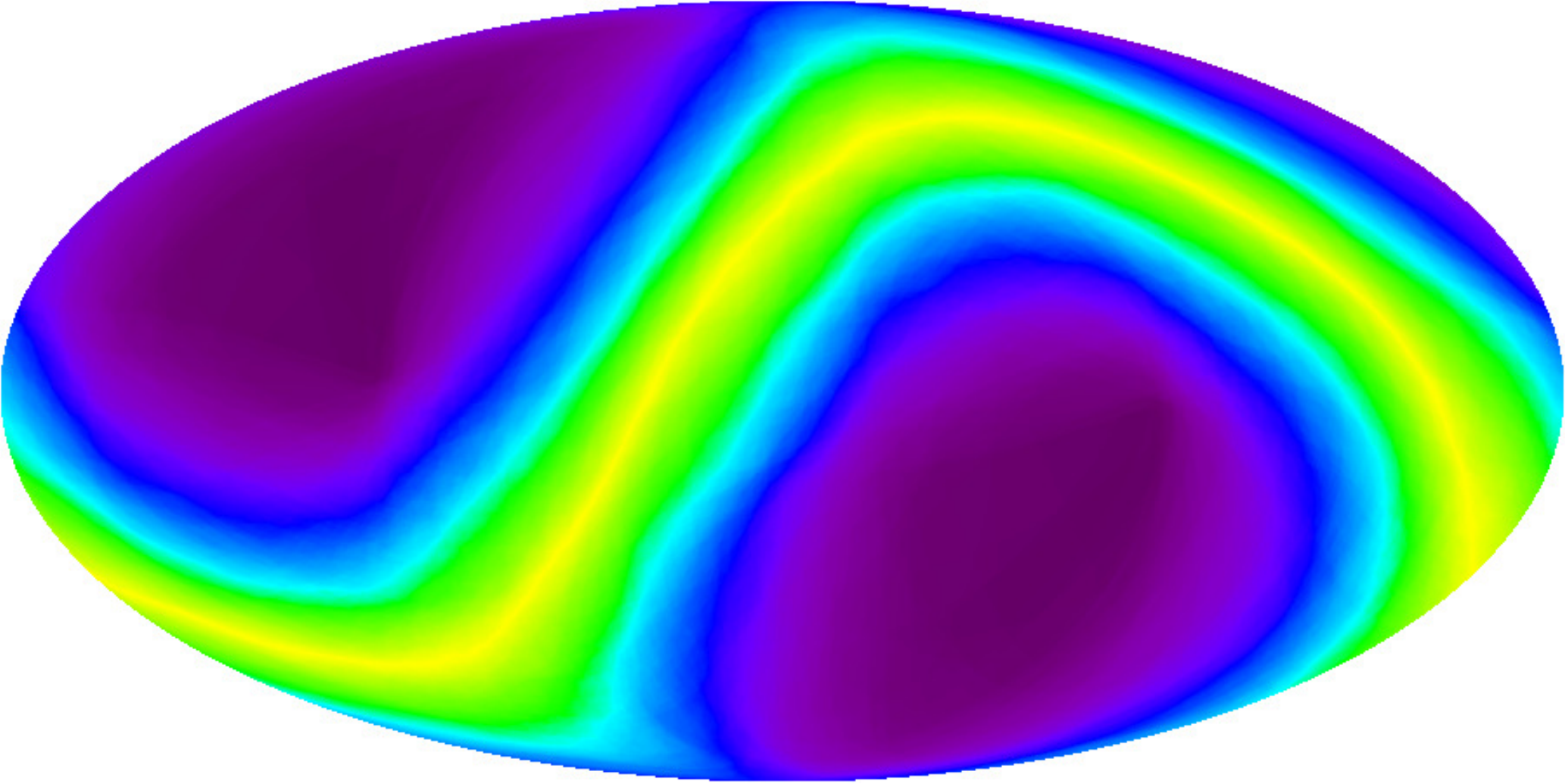} &
\includegraphics[width=0.29\textwidth]{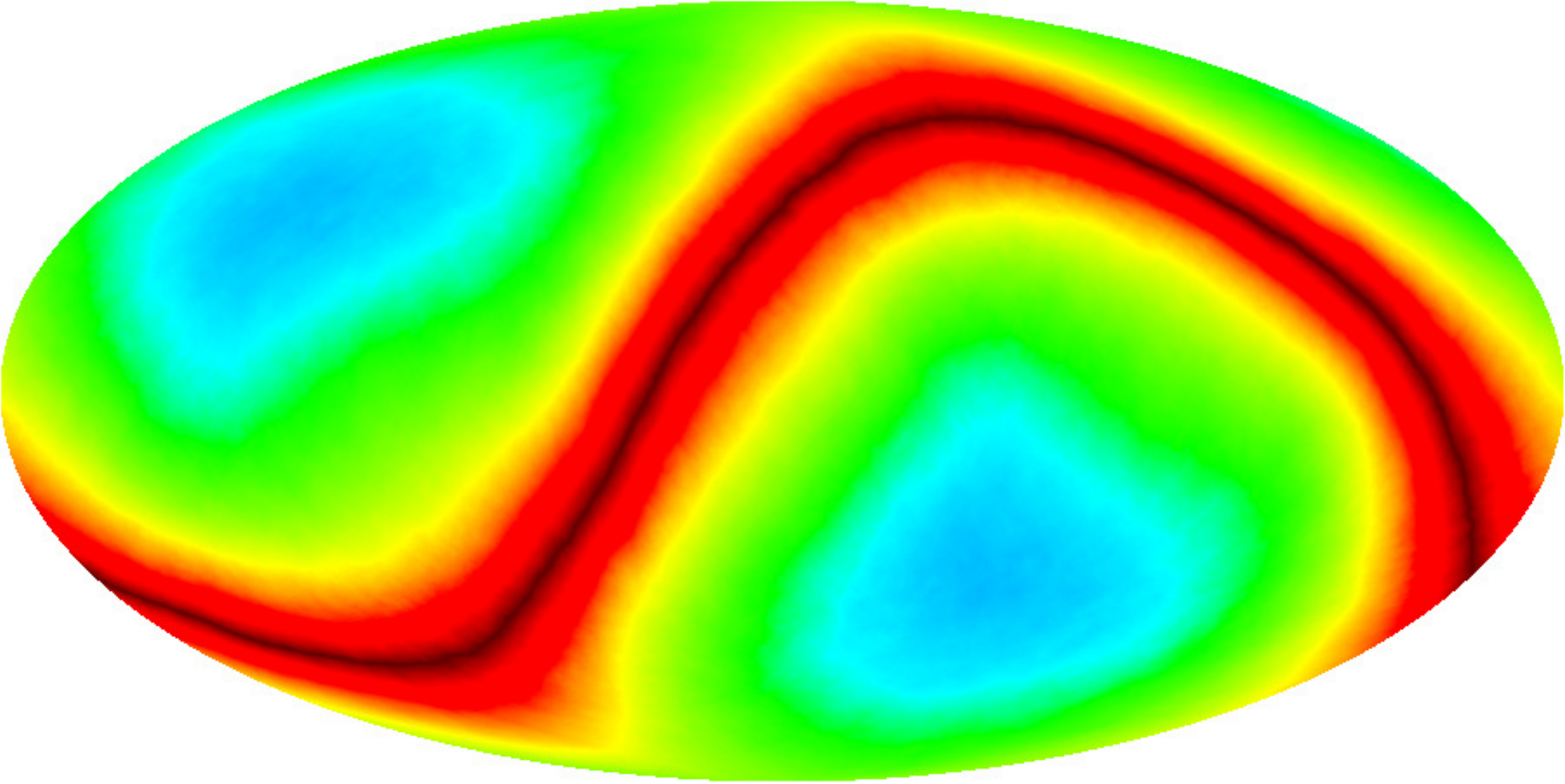}\\
\bottomrule
\end{tabular}}
\caption{The map $\phi_{\bmz,\bmX}$ for trivariate data. A total of $n=1000$ data points were 
simulated from a trivariate distribution.
The univariate depth (of a single randomly chosen point $\bmz$) in direction $\bmp$ 
is shown on a color scale from violet (low depth) to dark red (high
depth). The sphere $\IS^2$ is mapped on the plane using the so-called
Mollweide projection, see \cite{Snyder87}. }\label{fig:Mollweide}
\end{figure}

To get some insights in the behavior of the objective function $\phi_{\bmz,\bmX}$, e.g., 
whether there are local minima or not, we present several plots of $\phi_{\bmz,\bmX}$
in the case $d=3$ for different depths and data sets
in Figure~\ref{fig:Mollweide}.
The figures suggest that (at least
for common distributions and in the case $d=3$) local minima seem not to be a
major problem. 

A further important observation is the following. All of the above depths are
bounded above by unity. Therefore, the range of $\phi_{\bmz,\bmX}$ depends
on the depth of $\bmz$, i.e., if $\D(\bmz\,|\,\bmX)=c_0$, then the range of
$\phi_{\bmz,\bmX}$ is a subset of $[c_0,1]$. The larger the depth of $\bmz$, 
the smaller is the variation of $\phi_{\bmz,\bmX}$. Hence, it should be easier
to compute the depth of a point with a high depth.


\subsection{The geometry of the unit sphere}\label{ssec:GeomSphere}
The set
\[
S(\bma,r)=\{\bmx\in\IR^d|\|\bmx-\bma\|=r\}
\]
is called the \emph{sphere} with center $\bma\in\mathbb{R}^d$ and radius $r\ge 0$. 
The sphere with radius one and center $\bmNull$ is called the 
\emph{unit sphere} in $\IR^d$, it is denoted by $\IS^{d-1}$.

The intersection of the unit sphere with an affine subspace is called a
\emph{small sphere}. If the affine subspace contains the origin (i.e.,
if it is a linear subspace), then the intersection is called a \emph{great sphere}.
In the special case that the affine subspace is a plane, i.e., has dimension $2$,
then the intersection is called \emph{small circle} (if the plane does not
pass through the origin) or \emph{great circle} (if the plane passes through the
origin).   

The intersection of the unit sphere with a closed (open) halfspace is called
closed (open) \emph{spherical cap}. If the bounding hyperplane of the halfspace
passes through the origin, the spherical cap is called a \emph{hemisphere}.  

The distance between two points $\bmu,\bmv\in\IS^{d-1}$ on the sphere
as \emph{measured in the ambient space} $\IR^d$ is given by the Euclidean
distance,
\[
d_e(\bmu,\bmv)=\|\bmu-\bmv\|=\sqrt{\sum_{i=1}^d(u_i-v_i)^2}\,.
\]
However, when we measure the distance between two points $\bmu,\bmv\in\IS^{d-1}$
\emph{in the sphere itself}, then the distance is given by the
\emph{great-circle distance}, that is the
length of the shorter arc of a great circle passing through $\bmu$ and $\bmv$,
\[
d_g(\bmu,\bmv)=\arccos(\ip{\bmu}{\bmv})\,.
\]
It holds $0\le d_g(\bmu,\bmv)\le\pi$, i.e., the great-circle distance between
two points on the unit sphere is at most $\pi$.
 
The Euclidean distance and the great-circle distance are related as follows:
\[
d_e(\bmu,\bmv) = 2\sin\left(\frac{d_g(\bmu,\bmv)}2\right)\,.
\]  
The transformation $[0,\pi]\to[0,2]$, $\phi\mapsto 2\sin(\phi/2)$, is
continuous and strictly increasing. Therefore, both metrics generate the
same topology of open sets on $\IS^{d-1}$. The $\epsilon$-neighborhood of a point 
$\bma\in\IS^{d-1}$ is given by
\[
B(\bma,\epsilon)
=\{\bmp\in\IS^{d-1}|d_g(\bmp,\bma)<\epsilon\}
=\{\bmp\in\IS^{d-1}|\ip\bma\bmp>\cos(\epsilon)\}\,.
\]
It consists of all points in the sphere that have a great-circle
distance less than $\epsilon$ to $\bma$.
Since $B(\bma,\epsilon)$ is the intersection of the unit sphere $\IS^{d-1}$
with the halfspace
\[
\{\bmx\in\IR^d|\ip\bma\bmx>\cos(\epsilon)\}
\]
the set $B(\bma,\epsilon)$ is a spherical cap and its topological
boundary is the small sphere
\[
\{\bmx\in\IS^{d-1}|\ip\bma\bmx=\cos(\epsilon)\}\,.
\]
The shortest path between two points $\bmx$ and $\bmy$ on the unit sphere
$\IS^{d-1}$ is given by the shorter arc of a great circle which passes through 
$\bmx$ and $\bmy$. Therefore, the great circles on a sphere are the
\emph{geodesics}, i.e., generalizations of straight lines from the usual Euclidean
space.
A great circle through $\bmx$ and $\bmy$ is unique as long as $\bmy\ne \pm\bmx$.
Let $\tilde\bmy=\bmy-\ip{\bmx}{\bmy}\bmx$. Then, $\bmx$ and $\tilde\bmy$ are
orthogonal. Therefore, the (unique) great circle through $\bmx$ and $\bmy$ is
given by 
\begin{equation}\label{eq.greatcircle}
\left\{\bmz(\phi)=\cos(\phi)\bmx+\sin(\phi)\frac{\tilde\bmy}
{\|\tilde\bmy\|}\,\middle|\,\phi\in(-\pi,\pi]\right\}\,.
\end{equation}
Simple algebra shows that the great-circle distance between $\bmx$ and
$\bmz(\phi)$ is $|\phi|$. Further, $\bmz(0)=\bmx$ and $\bmz(d_g(\bmx,\bmy))=\bmy$.
 
\section{Algorithms for approximating projection depths}\label{sec:Algorithms}
In this section we present several algorithms to compute the depth of a given
point $\bmz$ w.r.t. a data cloud $\bmX=(\bmx_1,\dots,\bmx_n)$. We assume that
the depth satisfies the projection property so that the depth can be computed
as the minimum of the projected univariate depths over the unit sphere
$\IS^{d-1}$. Most of the algorithms are presented in pseudocode.
  
\subsection{Simple random search (RS)}\label{ssec:RS}
If a depth satisfies the projection property, then for each $\bmp\in \IS^{d-1}$
the value $\D(\ip{\bmp}{\bmz}|\ip{\bmp}{\bmX})$ is an upper bound for $\D(\bmz|\bmX)$.
Therefore, it seems reasonable to compute the univariate depths $\D(\ip{\bmp}{\bmz}\,
|\,\ip{\bmp}{\bmX})$ for several values of $\bmp$. The minima of these values
constitute a decreasing sequence of upper bounds for the true value of the
depth. One can show that under weak conditions this sequence of upper bounds
converges to the true value. The following proposition can be found in
\cite{Dyckerhoff04}.

\begin{proposition}\label{pr.ProjAlg}
Let $\bmz\in\IR^d$ and $\D$ a depth that satisfies the projection property.
Further, let the mapping $\IS^{d-1}\to[0,\infty)$, $\bmp\mapsto
\D(\ip{\bmp}{\bmz}|\ip{\bmp}{\bmX})$, be upper semicontinuous. If $\bmp_1,\bmp_2,\dots$ is a
sequence of independent random vectors distributed uniformly on $\IS^{d-1}$,
then, almost surely,
\[
\lim_{N\to\infty}\min_{1\le i\le N}\D(\ip{\bmp_i}{\bmz}|\ip{\bmp_i}{\bmX})
=
\D(\bmz|\bmX)\,.
\]
\end{proposition} 

Because of the affine invariance of the depth it holds true that 
$\D(\ip{\bmp}{\bmz}|\ip{\bmp}{\bmX})=\D(\ip{-\bmp}{\bmz}|\ip{-\bmp}{\bmX})$.
Therefore, one can restrict $\bmp$ to a hemisphere.

Proposition~\ref{pr.ProjAlg} motivates the following algorithm which we will call 
\emph{simple random search}.
Generate a ``large number'' $N$ of random directions $\bmp_1,\dots,\bmp_N\in
\IS^{d-1}$, independently drawn from the uniform distribution on $\IS^{d-1}$.
For each of these directions compute the univariate depth. The
(multivariate) depth is then approximated by
\[
\min_{1\le i\le N}\D(\ip{\bmp_i}{\bmz}|\ip{\bmp_i}{\bmX})\,.
\]
It is well known that random vectors from the uniform distribution on the
sphere $\IS^{d-1}$ can easily be simulated by generating $d$ independent
random numbers from a standard normal distribution and normalizing the
resulting vector to have norm one. For the sake of completeness we present this
algorithm in pseudocode as Algorithm~\ref{alg.rndSphere}. 
\begin{algorithm}[ht]
\caption{Random vectors from uniform distribution on the sphere}\label{alg.rndSphere}
\begin{algorithmic}[1]
\Function{rndSphere}{$d$}
\State $s \gets 0$
\For{i}{1}{d} 
\State $x_i \gets \textsc{rndNormal()}$
\Comment{random number from a $\mathcal{N}(0,1)$-distribution}
\State $s \gets s + x_i^2$
\EndFor
\For{i}{1}{d} $x_i \gets x_i / \sqrt{s}$
\EndFor
\State \Return $(x_1,\dots,x_d)$
\EndFunction
\end{algorithmic}
\end{algorithm}

The simple random search is presented in pseudocode as 
Algorithm~\ref{alg.SimpleRandomSearch}.
\begin{algorithm}[ht]
\caption{Simple random search}\label{alg.SimpleRandomSearch}
\begin{algorithmic}[1]
\Function{randomSearch}{$\bmz,\bmX$}
\State $d_{min} \gets \infty$
\For{j}{1}{N}
\State $\bmp\gets\textsc{rndSphere}(d)$
\State $d_{min} \gets \min(\D(\ip\bmp\bmz|\ip\bmp\bmX),d_{min})$
\EndFor
\State\Return $d_{min}$
\EndFunction
\end{algorithmic}
\end{algorithm}

In implementing the above algorithm, one could think of two possibilities for
choosing the number of iterations $N$. First, the number $N$ could be chosen in
advance, depending on $n$, $d$ and the desired accuracy. Second,
random directions would be generated until a
certain stopping criterion is satisfied.  
In the first case, it seems reasonable to assume that $N$ (for a given
accuracy) should depend on $d$ but not on $n$. It further seems reasonable
to require $N(d)\propto N(2)^{d-1}$. This would result in an overall
complexity of the algorithm of order $O(N(2)^{d-1}n)$ provided the complexity of
computing the univariate depth is of order $O(n)$, which holds for all the considered depths.
However, this is not desirable since the complexity grows exponentially with
the dimension $d$. 


The convergence of $\min_{1\le i\le N}\D(\ip{\bmp_i}{\bmz}|\ip{\bmp_i}{\bmX})$
for the halfspace depth and the projection depth has been extensively studied in 
\cite{NagyDM19}. Given a precision $\epsilon$, these results can be used to 
find $N(\epsilon)$ such that the error is approximately $\epsilon$.

Of course, an algorithm that uses a larger number of univariate depth
evaluations should give a better result than with a smaller number. Therefore,
for a fair comparison of different algorithms one should limit the number of
depth evaluations to a given number. Following this logic, we choose not to use
a stopping criterion in the random search but rather use a fixed number $N$
of iterations. This number should be the same for all compared algorithms. 
In Sections~\ref{ssec:tuning} and \ref{ssec:comparison} where we compared
several algorithms and parameter combinations, $N\in\{100,1000,10000\}$ was
used.
 
\subsection{Simple grid search (GS)}\label{ssec:GS}
Since in the random search the directions are distributed randomly on the
sphere, it might be tempting to use a deterministic grid of directions instead.
Thus, the second algorithm that we discuss is a grid search on the sphere. 
%
A parametrization of the sphere $\IS^{d-1}$ is given by 
\emph{generalized spherical coordinates}:
\begin{align*}
x_1&= \cos(\phi_1),\\
x_2&= \sin(\phi_1)\cos(\phi_2),\\
\vdots&\qquad\qquad\vdots\\
x_{d-1} &= \sin(\phi_1)\dots\sin(\phi_{d-2})\cos(\phi_{d-1}),\\
x_{d}   &= \sin(\phi_1)\dots\sin(\phi_{d-2})\sin(\phi_{d-1}),
\end{align*}
where $\phi_1,\dots,\phi_{d-2}\in[0,\pi]$, $\phi_{d-1}\in[0,2\pi)$.
Here, $\phi_1$ is called \emph{polar angle}. 
For $\phi_1=0$ we get the \emph{north pole} $(1,0,\dots,0)\T=\bme_1$, 
for $\phi_1=\pi$ we get the \emph{south pole} $(-1,0,\dots,0)\T=-\bme_1$.
Since
\[
d_g(\bmx,\bme_1)
=\arccos(\ip{\bmx}{\bme_1}) 
=\arccos(x_1)
=\phi_1\,,
\]
the polar angle $\phi_1$ of a point $\bmx\in \IS^{d-1}$ is the great-circle
distance of $\bmx$ from the north pole.

As was mentioned earlier, we can always restrict the direction $\bmp$ to
a hemisphere. 
In the implementation we used a grid where the first angle $\phi_1$ was
restricted to the interval $[0,\pi/2]$ which corresponds to a grid on the
northern hemisphere. 

However, a severe drawback of using generalized spherical coordinates is the
fact that the resulting grid is not uniform, the so-called \emph{pole problem}.
This is best illustrated in the case of the usual $2$-sphere in $\IR^3$.
The meridians (the lines for which $\phi_2$ is constant) converge at the two
poles. Thus, near the poles the distance between grid points is smaller than at
the equator which leads to an (unwanted) oversampling in the neighborhood of the
poles. Further, the representation of the poles is obviously not unique since all
coordinate vectors $(0,\phi_2,\dots,\phi_{d-1})$, $\phi_2,\dots,\phi_{d-2}
\in[0,\pi]$, $\phi_{d-1}\in[0,2\pi)$, represent the north pole. 
More generally, if an angle $\phi_i$ is equal to $0$ or $\pi$, then for all
values  of $\phi_{i+1},\dots,\phi_{d-1}$ we get the same point on the sphere.  


Contrary to the trivial case of the $1$-sphere, it is for dimension $d>2$ not
possible to perfectly uniformly distribute an arbitrary number of points on
$\IS^{d-1}$. This is closely connected to the existence of convex regular
polytopes. As a consequence, a perfectly uniform grid on $\IS^{d-1}$ is for
$d>2$ only possible for finitely many values of $N$. In particular, a regular
grid on $\IS^{d-1}$ cannot be made arbitrarily fine.

However, there are grids on the sphere which are nearly uniform.
Such quasi-uniform grids do not suffer from the pole problem. They
are used in geophysics, climate modeling, or astronomy. Examples of quasi-uniform 
grids include the Kurihara grid \citep{Kurihara95}, 
the icosahedral grid \citep{Williamson68},
the cubed sphere grid \citep{RonchiIP96}, 
or the Yin-Yang grid \citep{KageyamaS04}. 
The disadvantage of these grids is that they are more complicated to implement
or that they only work for the case of the $2$-sphere.
Since it will become obvious that there are methods that are clearly superior 
to the simple random search or the simple grid search for the computation of
depth, we decided not to spend much time on implementing these more complicated
quasi-uniform grids. In our simulations we use a simple uniform grid in the
hyper-cubic domain that is based on generalized spherical coordinates.

A major drawback of the grid search is that it suffers from the 
\emph{curse of dimensionality}. For example assume that we only use four
subdivisions per angle.  Then, for a data set of dimension $d$, which means
that the grid based on spherical coordinates is of dimension $d-1$, the grid
consists of $4^{d-1}$ grid points. For $d=11$ we already have more than
a million grid points, for $d=20$ we already have $2.7\cdot10^{11}$ grid points,
a number far exceeding the capacity of most computers.
Therefore, we choose not to use the grid search for dimensions larger than
$d=10$. For the same reason we do not provide a description of the grid search
in pseudocode.

In the simulations in Sections~\ref{ssec:tuning} and~\ref{ssec:comparison}
we choose the mesh size of the grid such that the number of depth evaluations
was approximately equal to a given number $N$.

\subsection{Refined random search (RRS)}\label{ssec:RRS}
In the simple random search the whole  surface of the sphere is searched with
the same intensity. Therefore, a lot of time is wasted searching areas which
are far away from the minimum of the objective function. The idea of the
\emph{refined random search} is to concentrate the search in the neighborhood of
directions with low depth.

At the start of the algorithm the neighborhood of a point on the sphere is
defined to be the hemisphere with pole at that point.
We choose $N_1$  directions at random in the
neighborhood of the current best point. Every time a new direction with minimum
depth is found, this point is chosen as the new center of the neighborhood.
As the search continues we choose the neighborhood smaller and smaller.  

We first discuss how to sample from a neighborhood of the north pole $\bme_1$.
The $\epsilon$-neighborhood $B(\bme_1,\epsilon)$ of the north pole 
consists of all points whose great-circle distance
to the north pole is less than $\epsilon$. Since the distance to the north pole
is the polar angle, the \emph{$\epsilon$-polar cap} is the set of all points
whose polar angle is less than $\epsilon$.
Therefore, in a first step we choose the polar angle $\phi_1$ from a uniform
distribution on $[0,\epsilon]$ and get the first coordinate of the sampled 
point, $x_1=\cos(\phi_1)$. In the second step we choose a point from a
uniform distribution on the small sphere $\{\bmx\in\IS^{d-1}|x_1=\cos(\phi_1)\}$
which has radius $\sqrt{1-\cos^2(\phi_1)}$. This is described in pseudocode
in Algorithm~\ref{alg.rndPolarCap}.

Sampling from a uniform distribution in the first step means that the distance of
the points from $\bme_1$ is uniformly distributed. Note that this does not yield
a uniform distribution on the $\epsilon$-polar cap but a distribution where
points near the pole have a higher density than points farther away.
However, this is desirable in the refined random search since it means that
points near the current minimum have a higher probability to get drawn. 

On the contrary, assume that we would choose new points from a uniform
distribution on the $\epsilon$-polar cap. In high dimensions, most of the
surface area of the spherical cap is concentrated near the base of the cap.
Therefore, with a very high probability we would get points that have a large
distance from the north pole $\bme_1$. For example, if $d=10$ and $\epsilon=0.1$
there is a probability of $0.998$ to draw a point which has a distance
between $0.05$ and $0.1$ from $\bme_1$, whereas the probability to draw a
point that has a distance between $0$ and $0.05$ is only $0.002$. This is even
worse in higher dimensions. Therefore, we decided to draw the polar angle from
a uniform distribution. 

Note further that the density of $\bmx$ depends only on the polar angle
$\phi_1$ which measures the distance from the pole.

To sample from the neighborhood of an arbitrary point $\bmp$, we look for
a transformation $\bmQ$ of the sphere that
transforms the north pole $\bme_1$ to $\bmp$ and
does not change distances.
Such a transformation is given by a \emph{Householder matrix} \citep{GolubL89}: 
\begin{equation}\label{eq.Householder}
\bmQ=\bmI-\frac{2}{\bmv^T\bmv}\bmv\bmv^T\,,
\end{equation}
where $\bmv$ is some non-zero vector and $\bmI$ is the identity matrix.
A Householder matrix is symmetric,
orthogonal and thus involutory. Geometrically, $\bmQ$ is a reflection at the
hyperplane through the origin whose normal is $\bmv$. It can easily be seen that
choosing $\bmv=\bmp-\bme_1$ 
does the trick, i.e., $\bmQ\bme_1=\bmp$. Therefore, if $\bmx$ is sampled
from the $\epsilon$-polar cap $B(\bme_1,\epsilon)$, then $\bmQ\bmx$ is
sampled from the spherical cap $B(\bmp,\epsilon)$. To compute $\bmQ\bmx$
we first note that $\bmv\T\bmv=(\bmp-\bme_1)\T(\bmp-\bme_1)=\|\bmp\|^2-2p_1+1
=2(1-p_1)$ since $\|\bmp\|^2=1$. Therefore,
\[
\bmQ\bmx
=\bmx-\frac{2}{2(1-p_1)}(\bmp-\bme_1)(\bmp-\bme_1)\T\bmx
=\bmx-\frac{\bmp\T\bmx-x_1}{1-p_1}(\bmp-\bme_1)
=\bmx-\lambda (\bmp-\bme_1)
\]
where $\lambda=(\ip\bmp\bmx-x_1)/(1-p_1)$.
For the first component we get $(\bmQ\bmx)_1=x_1-\lambda p_1+\lambda$
whereas for the remaining components we have $(\bmQ \bmx)_i=x_i-\lambda p_i$. 
Algorithm~\ref{alg.Householder} describes the Householder transformation in
pseudocode. 
\begin{algorithm}[ht]
\caption{Householder transformation}\label{alg.Householder}
\begin{algorithmic}[1]
\LeftComment{The Householder transformation that maps $\bme_1$ to $\bmp$ is applied 
to $\bmx$}
\Function{Householder}{$\bmx,\bmp$}
\If{$p_1=1$} \Return $\bmx$\EndIf
\State $\lambda\gets (\ip\bmp\bmx - x_1) / (1 - p_1)$
\State $x_1\gets x_1+\lambda$
\For{i}{1}{d} $x_i \gets x_i -\lambda\cdot p_i$
\EndFor
\State \Return $\bmx$
\EndFunction
\end{algorithmic}
\end{algorithm}
The generation of a random point from the spherical cap is 
described in Algorithm~\ref{alg.rndPolarCap}.
\begin{algorithm}[ht]
\caption{Random vectors from a spherical cap}\label{alg.rndPolarCap}
\begin{algorithmic}[1]
\LeftComment{Generate a random number from a spherical cap with size $\epsilon$ around $\bmp$}
\Function{rndSphericalCap}{$\bmp,\epsilon$}
\State $x_1 \gets \textsc{rndUnif}()$
\Comment{random number from a $\mathcal U([0,1])$-distribution}
\State $x_1 \gets \cos(\epsilon\cdot x_1)$
\State $(x_2,\dots,x_d)\gets \sqrt{1-x_1^2}\cdot \textsc{rndSphere}(d-1)$ 
\State $\bmx\gets (x_1,\dots,x_d)$ 
\State \Return $\textsc{Householder}(\bmx,\bmp)$
\EndFunction
\end{algorithmic}
\end{algorithm}
The refined random search is presented in Algorithm~\ref{alg.refRandomSearch}.
\begin{algorithm}[ht]
\caption{Refined random search}\label{alg.refRandomSearch}
\begin{algorithmic}[1]
\Function{refinedRandomSearch}{$\bmz,\bmX$}
\State $\bmp_{min} \leftarrow \bme_1$ \Comment{start with the North Pole}
\State $d_{min} \leftarrow \D(\ip{\bmp_{min}}{\bmz}|\ip{\bmp_{min}}{\bmX})$
\State $\epsilon \leftarrow \pi / 2$ 
\Comment{initial neighborhood is a hemisphere}    
\For{i}{1}{N_{ref}}
\For{j}{1}{N_{it}}
\State $\bmp_{cur}\gets\textsc{rndSphericalCap}(\bmp_{min},\epsilon)$
\State $d_{cur} \leftarrow \D(\ip{\bmp_{cur}}{\bmz}|\ip{\bmp_{cur}}{\bmX})$
\If{$d_{cur} < d_{min}$}  
\State $d_{min} \leftarrow d_{cur}$
\State $\bmp_{min} \leftarrow \bmp_{cur}$
\EndIf
\EndFor
\State $\epsilon \leftarrow \epsilon\cdot\alpha$ \Comment{shrinking the neighborhood}
\EndFor
\State\Return $d_{min}$
\EndFunction
\end{algorithmic}
\end{algorithm}
There, 
a geometric shrinking of the
neighborhood is applied, i.e., in each refinement step the size of the
neighborhood is multiplied by a factor $\alpha<1$. Of course, other shrinking
schemes are possible. For example, one could apply an arithmetic shrinking
scheme. For our simulations we choose a geometric shrinking scheme. 

The refined random search depends on the parameters $N_{ref}$, $N_{it}$ and
$\alpha$. For $N_{ref}$ and $\alpha$ we tried several parameter combinations and
compared them in Section~\ref{ssec:tuning}. The parameter $N_{it}$ was always
chosen such that the total number of depth evaluations was equal to a given 
number $N$, i.e.,
$N_{it}=N/N_{ref}$ was used. The fine-tuning of these parameters is described in
Section~\ref{ssec:tuning}, see also the figures in the Supplementary Materials.

A drawback of the refined random search is that the algorithm might be trapped
in a local minimum. However, the plots in Figure~\ref{fig:Mollweide} suggest
that -- at least in simple situations -- this should not be a major problem. 

\subsection{Refined grid search (RGS)}\label{ssec:RGS}
Since the grid search suffers from the curse of dimensionality, we may use the
same ideas as in the refined random search, i.e., we start with a relatively
coarse grid and apply successive refinements of the grid in the neighborhood
of the current minimum.
As in the refined random search the \emph{spherical cap neighborhood} may be used.
The oversampling near the pole may be sensible in this case, since it means 
oversampling near the current minimum.

As the refined random search, the refined grid search depends on the parameters
$N_{ref}$, the number of refinements, and $\alpha$, the shrinking factor of the
spherical cap. The mesh size of the grids was chosen such that the total number
of depth evaluations was approximately equal to the given value of $N$.
Again, the fine-tuning of these parameters is described in
Section~\ref{ssec:tuning}. For $N_{ref}$ and $\alpha$ the same parameter sets as
for the refined random search were used. For the same reasons as in the case of
the grid
search, the refined grid search was used only for $d\le 10$ in the simulations
in Section~\ref{sec:simulation}. Because of this limitation and for the sake 
of brevity we also do not provide a description of the algorithm in pseudocode. 

\subsection{Random simplices (RaSi)}\label{ssec:RaSi}
In this algorithm random directions are used that are derived from the data
points $\bmx_1,\dots,\bmx_n$ themselves. Similar strategies have already been used
by \cite{RousseeuwS98} for the halfspace depth and by \cite{ChristmannFJ02} for
the regression depth and classification.

To be more specific, a random sample $\bmx_{i_0},\bmx_{i_1},\dots,\bmx_{i_d}$
of size $d+1$ is drawn without replacement from the $n$ data points. 
If these points are in general position, they form a simplex. In the next
step, on the facet opposite to $\bmx_{i_0}$ a point $\bmx_o=\sum_{j=1}^dw_j\bmx_{i_j}$
is chosen, where the weights $(w_1,\dots,w_d)\T$ are drawn from a symmetric
Dirichlet distribution with parameter $\alpha$. Finally, the direction 
$\bmp=\bmx_{i_0}-\bmx_o$ is used to project the data and compute the
univariate depth. This process is repeated $N$ times and the minimum attained
depth is returned. The algorithm is described in pseudocode as Algorithm~\ref{alg.RaSi}.
\begin{algorithm}[ht]
\caption{Random simplices}\label{alg.RaSi}
\begin{algorithmic}[1]
\Function{randomSimplices}{$\bmz,\bmX$}
\State $d_{min} \gets \infty$
\For{i}{1}{N}
\State $(i_0,i_1,\ldots,i_d)\gets\textsc{rndSubset}(d+1,\{1,\dots,n\})$
\State $(w_1,\dots,w_d)\gets\textsc{rndDirichlet}(d,\alpha)$
\State $\bmp\gets\sum_{k=1}^d\bmx_{i_k}w_k-\bmx_{i_0}$
\State $\bmp\gets\bmp/\|\bmp\|$
\State $d_{min} \gets \min(\D(\ip\bmp\bmz|\ip\bmp\bmX),d_{min})$
\EndFor
\State\Return $d_{min}$
\EndFunction
\end{algorithmic}
\end{algorithm}

Besides $N$, the number of iterations, the only parameter is the parameter
$\alpha$ of the Dirichlet distribution. For $\alpha=1$ we get a uniform
distribution on the facet defined by $\bmx_{i_1},\dots,\bmx_{i_d}$. The higher
$\alpha$, the more the distribution is concentrated around the center of
that facet. Fine-tuning of the parameter $\alpha$ is described in Section~\ref{ssec:tuning}.

\subsection{Simulated annealing (SA)}\label{ssec:SA}
In the refined random search we might get 
trapped in a local minimum.
\emph{Simulated annealing} avoids this by accepting also worse
solutions with a certain probability.   
Let $d_{cur}$ denote the current depth and  $d_{new}$ the depth of a new 
trial solution. Then, $d_{new}$ is chosen as the new solution with 
probability 
\[
p = \min\left\{\exp\left(-\frac{d_{new}-d_{cur}}{T}\right),1\right\}\,.
\]
If $d_{new}<d_{cur}$, the new solution is always accepted. However,
if $d_{new}>d_{cur}$, the new solution is still accepted with positive
probability $p$.

The parameter $T$ is called \emph{temperature}. 
If $T$ is high, the probability of accepting a worse solution is high. 
Conversely, if $T$ is low, worse solutions are accepted only with small
probability. In the course of the algorithm the temperature (and thus the
probability of accepting worse solutions) is slowly decreased. 
The way the parameter $T$ is decreased is called
\emph{cooling schedule}. 
Often a linear
cooling schedule, $T(t)=T(0)-\eta\, t$ where $\eta>0$ is a chosen parameter, or
a geometric cooling schedule, $T(t)=T(0)\alpha^t$ where $\alpha\in(0,1)$, 
is used. Note, that the choice of the cooling schedule
can have a significant impact on the performance of simulated annealing.
If the temperature is decreased too fast, then the algorithm may be trapped in
a local minimum. If the temperature is decreased too slowly, the convergence of the
algorithm is also very slow. A description of simulated annealing where an
exponential cooling schedule is used is shown in Algorithm~\ref{alg.SA}.
 
Simulated annealing has already been applied to the computation of a special
projection depth by \cite{ShaoZ12}. 

\begin{algorithm}[ht]
\caption{Simulated annealing}\label{alg.SA}
\begin{algorithmic}[1]
\Function{simulatedAnnealing}{$\bmz,\bmX$}
\If{\texttt{Start = Mn}}\ $\bmu_{cur}\gets\bmz-\ol{\bmx}$\EndIf
\If{\texttt{Start = Rn}}\ $\bmu_{cur}\gets\textsc{rndSphere}(d)$\EndIf
\State $\epsilon\gets(\pi/2)/\beta$ \Comment{size of the spherical cap}
\State $d_{cur} \leftarrow \D(\ip{\bmu_{cur}}{\bmz}|\ip{\bmu_{cur}}{\bmX})$
\State $d_{min} \gets d_{cur}$
\State $T \leftarrow T_0$             \Comment{starting temperature, $T_0=1$}    
\Repeat
\For{j}{1}{N_{it}}
\State $\bmu_{new}\gets\textsc{rndSphericalCap}(\bmu_{cur},\epsilon)$
\State $d_{new} \leftarrow \D(\ip{\bmu_{new}}{\bmz}|\ip{\bmu_{new}}{\bmX})$
\State $p \leftarrow \min\left(\exp\left(-\frac{d_{new}-d_{cur}}{T}\right),1\right)$ 
\State \textbf{with probability} $p$ \textbf{do}
\State \hspace*{2em}$\bmu_{cur} \leftarrow \bmu_{new}$
\State \hspace*{2em}$d_{cur} \leftarrow d_{new}$
\State \hspace*{2em}$d_{min} \gets \min(d_{cur},d_{min})$
\EndFor
\State $T \leftarrow T\cdot \alpha$ \Comment{cooling schedule} 
\Until $T < T_{min}$                \Comment{$T_{min}=0.001$}
\State\Return $d_{min}$
\EndFunction
\end{algorithmic}
\end{algorithm}
Since the sequence of values $d_{cur}$ is in general not decreasing, the last
value $d_{cur}$ needs not necessarily be the minimum depth found during the
course of the algorithm. Therefore, in Algorithm~\ref{alg.SA} we keep track of
the minimum depth found so far and return this value.
 
Simulated annealing depends on a lot of parameters. The performance of
the algorithm depends strongly on the fine-tuning of these parameters. 
In particular, to apply the algorithm (i) the
starting solution, (ii) the used neighborhood, (iii) the probability distribution for
generating a new candidate, (iv) the number of iterations, (v) the starting temperature,
(vi) the temperature at which the algorithm is stopped, and (vii) the cooling schedule have
to be specified. 

In our simulations the temperature $T(0)$ at the start of the algorithm was
always chosen as $1$, the temperature $T_{min}$ at which the algorithm stops
was always chosen as $0.001$. For the starting solution a parameter \texttt{Start}
having two possible values \texttt{Mn} and \texttt{Rn} was passed. 
For \texttt{Start=Rn} the algorithm was started with a direction randomly drawn
from $\mathcal U(\IS^{d-1})$. For \texttt{Start=Mn} the direction $\bmu=\bmz-\ol{\bmx}$
joining the mean of the data points and $\bmz$ was used. Especially for the
case of spherical distributions this should give a good starting solution. 
For the neighborhood we used the spherical cap of size $\epsilon$. The size
$\epsilon$ of the spherical cap was controlled by a parameter $\beta$ via
$\epsilon=(\pi/2)/\beta$. A new candidate solution was drawn from the spherical
cap using the function \textsc{rndSphericalCap} described in Section~\ref{ssec:RRS}
and in Algorithm~\ref{alg.rndPolarCap}. In the simulations we always used a
geometric cooling schedule. The speed of cooling was controlled by a
parameter $\alpha$. The number $N_{it}$ of iterations was
chosen such that the total number of univariate depth evaluations was equal to 
a given value $N$. The fine-tuning of the parameters $\alpha$, $\beta$ and 
\texttt{Start} is described in Section~\ref{ssec:tuning}.

\subsection{Coordinate descent (CD)}\label{ssec:CD}
In the coordinate descent algorithm for finding the minimum of a function
$f:\IR^d\to\IR$ in the Euclidean space $\IR^d$  one proceeds
as follows. Start with an initial value $\bmx^0=(x_1^0,\dots,x_d^0)\T$. In the
$k$-th iteration we solve $d$ minimization problems,
\[
x_j^{k+1}=\argmin_{y\in\IR}f(x_1^{k+1},\dots,x_{j-1}^{k+1},y,x_{j+1}^k,\dots,x_d^k)\,,
\quad j=1,\dots,d.
\]
This yields a sequence $\bmx^0,\bmx^1,\bmx^2,\dots$ for which
$f(\bmx^0)\ge f(\bmx^1)\ge f(\bmx^2) \ge\dots$.

The function $\phi_{\bmz,\bmX}:\IS^{d-1}\to\IR^+_0$, $\bmp\mapsto\D(\ip{\bmp}{\bmz}|\ip{\bmp}{\bmX})$
can be extended to a function $\tilde\phi_{\bmz,\bmX}$ on the domain $\IR^d\setminus\{\bmNull\}$.
To the function $\tilde\phi_{\bmz,\bmX}$ the coordinate descent could be applied without major modifications.
However, because of the affine invariance of the depth, $\tilde\phi_{\bmz,\bmX}$ is constant on
lines, i.e., for $\lambda\ne0$ holds
$\tilde\phi_{\bmz,\bmX}(\lambda\bmp)=\tilde\phi_{\bmz,\bmX}(\bmp)$.  We believe that the performance of the coordinate descent
should profit from taking into account the special geometry of the domain of $\phi_{\bmz,\bmX}$.
Therefore, we adapted the coordinate descent for the special case that
the domain of the objective is the unit sphere $\IS^{d-1}$. Hence, we cannot use
the coordinate directions since they are not contained in the unit sphere. 
Instead, we replace the straight lines by \emph{great circles} which are the
geodesics on the unit sphere. As noted earlier, we can always restrict
ourselves to minimize the univariate depth over a hemisphere. So we do not
have to minimize over a whole great circle, but only over a great semi-circle. 

From \eqref{eq.greatcircle} it follows that for two orthogonal directions 
$\bmu,\bmv\in \IS^{d-1}$ the great semi-circle between $\bmv$ and $-\bmv$
passing through $\bmu$ is given by
\[
S(\bmu,\bmv)=\left\{\cos(\alpha)\bmu+\sin(\alpha)\bmv\middle|\alpha\in\left(-\frac\pi2,\frac\pi2\right]\right\}\,.
\]
In the $k$-th iteration we solve $d-1$ (univariate) minimization problems.
Denote by $\bmu^{(k,0)}$ the current point at the beginning of the $k$-th 
iteration. We choose $d-1$ directions $\bmp_1,\dots,\bmp_{d-1}$ such
that the vectors $\bmu^{(k,0)},\bmp_1,\dots,\bmp_{d-1}$ form an orthonormal
system of vectors in $\IR^d$. Thus, $\bmu^{(k,0)}+\linspan\{\bmp_1,\dots,\bmp_{d-1}\}$ is a
hyperplane which is tangent to the unit sphere at $\bmu^{(k,0)}$.  
Together with the current point the directions $\bmp_j$ determine the great
circles for the univariate optimization problems. 

Denote by $\bmu^{(k,j)}$, $j=1,\dots,d-1$, the solution of the $j$-th
minimization problem in the $k$-th iteration,
\[
\bmu^{(k,j)}=\argmin_{\bmv\in S(\bmu^{(k,j-1)},\bmp_j)}\D(\ip\bmv\bmz|\ip\bmv\bmX).
\]
Note that, when we move from $\bmu^{(k,j-1)}$ to $\bmu^{(k,j)}$, also
the tangent hyperplane has to be rotated accordingly. However, this rotation
takes place in the plane spanned by $\bmu^{(k,j-1)}$ and $\bmp_j$ and affects 
only these two vectors, leaving the other vectors of the orthonormal system
intact. Denote the image of $\bmp_j$ under this rotation by 
$\tilde\bmp_j$. Then, after $j$ minimizations, the system of vectors is mapped
to
\[
\bmu^{(k,j)},\tilde\bmp_1,\dots,\tilde\bmp_j,\bmp_{j+1},\dots,\bmp_{d-1}\,.
\]
It is easy to see that this is still an orthonormal system of vectors.
Note also, that there is no need to compute the vectors $\tilde\bmp_j$ since
for the $(j+1)$-th minimization we only need $\bmu^{(k,j)}$ and $\bmp_{j+1}$.

We still need to find the vectors $\bmp_1,\dots,\bmp_{d-1}$. If $\bmH$ is a
Householder matrix from equation~\eqref{eq.Householder} that maps $\bmu^{(k,0)}$
to the unit vector $\bme_d$, then $\bmH\bmu^{(k,0)}=\bme_d$ which is
equivalent ($\bmH$ is orthogonal and symmetric) to $\bmu^{(k,0)}=\bmH\bme_d$.
Thus, the last column of $\bmH$ is equal to $\bmu^{(k,0)}$. Since the
columns of $\bmH$ form an orthonormal system of vectors, 
$\bmp_1,\dots,\bmp_{d-1}$ can be chosen as the first $d-1$ columns of $\bmH$.
An easy calculation shows that
\[
(\bmp_j)_i=\begin{cases}
-\frac{u_ju_i}{1-u_d}\,&\text{if $i\ne j,d$,}\\[1ex]
1-\frac{u_j^2}{1-u_d}\,&\text{if $i=j$,}\\
u_j\,&\text{if $i=d$,}
\end{cases}
\]
where $u_i$ denotes the $i$-th component of $\bmu^{(k,0)}$ and $(\bmp_j)_i$ is
the $i$-th component of $\bmp_j$. A description of the coordinate descent in
pseudocode is given in Algorithm~\ref{alg.CD}.
The stopping criterion that we used in our simulations guaranteed that a 
specified number $N$ of evaluations of the univariate depth was not exceeded.
In Algorithm~\ref{alg.CD} a procedure \textsc{LineSearch} is used.
$\textsc{LineSearch}(\bmu,\bmp_j)$ tries to find the
minimum  of $\D(\ip\bma\bmz|\ip\bma\bmX)$ for $\bma$ in the great semi-circle
$S(\bmu,\bmp_j)$. For this line search several strategies are possible, e.g., 
a uniform search (see Algorithm~\ref{alg.LSUnif}) over the semi great-circle
or a golden section search (see Algorithm~\ref{alg.LSGS}).

\begin{algorithm}[ht]
\caption{Line search (uniform spacing)}\label{alg.LSUnif}
\begin{algorithmic}[1]
\Function{lineSearch}{$\bmu,\bmp$}
\Comment search along the great circle defined by $\bmu$ and $\bmp$
\State $f_{min}\gets\infty$
\For{i}{0}{n_{LS}}
\State $\lambda\gets -\pi/2 + i\cdot \pi / n_{LS}$
\State $\bmw\gets \cos(\lambda)\bmu + \sin(\lambda)\bmp$
\State $f\gets \D(\ip\bmw\bmz|\ip\bmw\bmX)$ 
\If{$f < f_{min}$}
\State $f_{min}\gets f$
\State $\bmu_{min}\gets \bmw$
\EndIf
\EndFor
\State\Return $(\bmu_{min},f_{min})$
\EndFunction
\end{algorithmic}
\end{algorithm}

\begin{algorithm}[ht]
\caption{Line search (golden section)}\label{alg.LSGS}
\begin{algorithmic}[1]
\Function{lineSearch}{$\bmu,\bmp$}
\Comment search along the great circle defined by $\bmu$ and $\bmp$
\State $\alpha\gets (\sqrt5-1)/2$
\State $(a,b)\gets (-\pi/2,\pi/2)$
\State $\lambda\gets\alpha a + (1-\alpha)b$
\State $\mu\gets(1-\alpha) a + \alpha b$

\State $\bmw\gets \cos(\lambda)\bmu + \sin(\lambda)\bmp$
\State $f_1\gets \D(\ip\bmw\bmz|\ip\bmw\bmX)$ 

\State $\bmw\gets \cos(\mu)\bmu + \sin(\mu)\bmp$
\State $f_2\gets \D(\ip\bmw\bmz|\ip\bmw\bmX)$ 

\State $f_{min}\gets\min(f_1,f_2)$

\While{$b-a > \epsilon$}  
\If{$f_1 > f_2$}
\State $(a,\lambda,f_1) \gets (\lambda,\mu,f_2)$
\State $\mu \gets (1-\alpha)a+\alpha b$
\State $\bmw\gets \cos(\mu)\bmu + \sin(\mu)\bmp$
\State $f_2\gets \D(\ip\bmw\bmz|\ip\bmw\bmX)$ 
\State $f_{min}\gets\min(f_2,f_{min})$
\Else
\State $(b,\mu,f_2) \gets (\mu,\lambda,f_1)$
\State $\lambda \gets \alpha a+(1-\alpha)b$
\State $\bmw\gets \cos(\lambda)\bmu + \sin(\lambda)\bmp$
\State $f_1\gets \D(\ip\bmw\bmz|\ip\bmw\bmX)$ 
\State $f_{min}\gets\min(f_1,f_{min})$
\EndIf
\EndWhile
\State\Return $(\bmw,f_{min})$
\EndFunction
\end{algorithmic}
\end{algorithm}

In Section~\ref{ssec:tuning} we compare several strategies for the line search.  
\begin{algorithm}[ht]
\caption{Coordinate descent}\label{alg.CD}
\begin{algorithmic}[1]
\Function{coordinateDescent}{$\bmz,\bmX$}
\State $\bmu\gets\textsc{rndSphere}(d)$ \Comment{start with a random point}
\Repeat
  \State $\bmv\gets \bmu$
  \Comment{$\bmv$ is $\bmu^{(k,0)}$} 
  \For{j}{1}{d-1}
    \For{i}{1}{d-1}
      \State $p_i\gets -v_i\cdot v_j / (1 - v_d)$
    \EndFor
    \State $p_j\gets 1 + p_j$
    \State $p_d\gets v_j$ \Comment{$\bmp$ is $\bmp_j$}
    \State $(\bmu,d_{cur})\gets\textsc{LineSearch}(\bmu,\bmp)$
  \EndFor
\Until{stopping criterion is satisfied}
\State\Return $d_{cur}$
\EndFunction
\end{algorithmic}
\end{algorithm}
To test our hypothesis that the coordinate descent should profit from taking
into account the spherical geometry of the domain, we implemented the 
coordinate descent such that by passing a parameter \texttt{Space}, having two
possible values \texttt{Ec} (Euclidean space) or \texttt{Sp} (Sphere), we could
choose between the na\"ive application of the coordinate descent and the version
specifically adapted to the sphere. Further, a parameter \texttt{LS} can be
passed that controls the line search algorithm used. For \texttt{LS} equal to
\texttt{Eq} the line search is done on an equally spaced grid, for \texttt{LS}
equal to \texttt{GS} the golden section method is used. 

\subsection{Nelder-Mead method (NM)}\label{ssec:NM}
\begin{figure}[ht]
\centerline{%
\includegraphics[height=5.4cm]{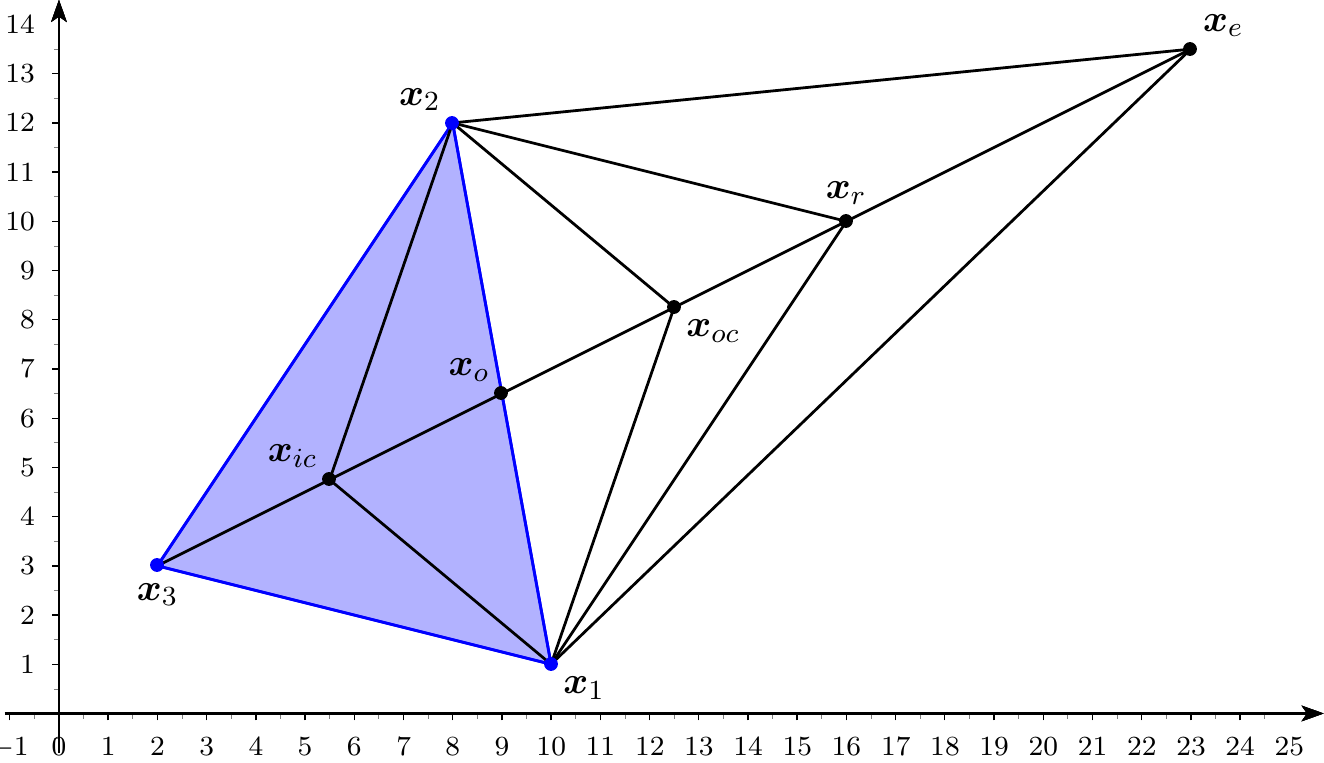}\quad
\includegraphics[height=5.4cm]{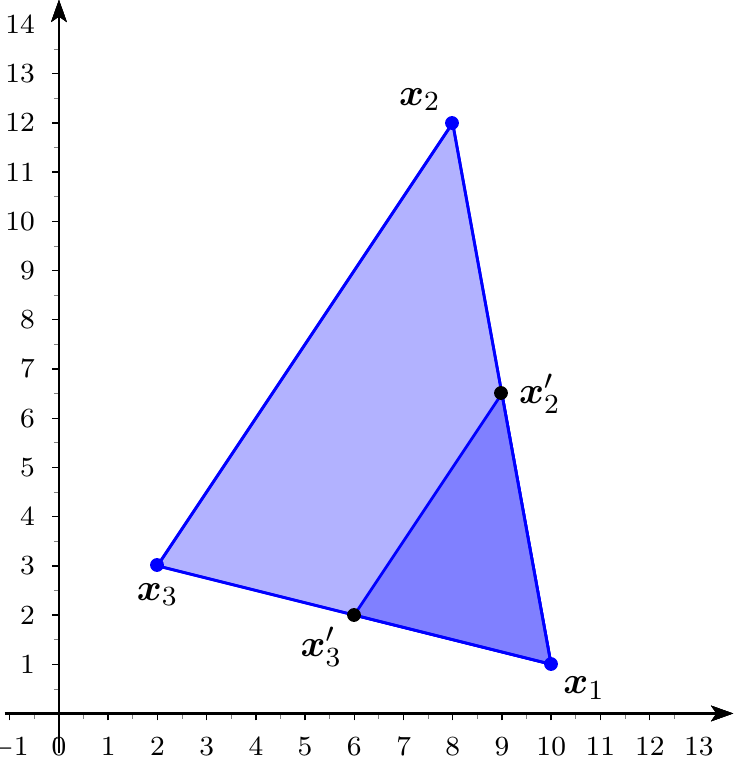}}
\caption{Illustration of the Nelder-Mead algorithm. The original simplex is
given in light blue. In the left panel the points $\bmx_r,\bmx_e,\bmx_{ic}$
and $\bmx_{oc}$ are shown together with the newly formed simplices. In the right
panel the shrunk simplex is shown in dark blue.}\label{fig.NM}
\end{figure}

The \emph{Nelder-Mead method} (also known as \emph{downhill simplex method}), 
originally proposed by \cite{NelderM65}, is a well-known optimization method
that does not rely on derivatives. It is based on a simplex i.e., a polytope that
is defined by $d+1$ vertices $\bmx_1,\dots,\bmx_{d+1}\in\IR^d$. Assume that
the vertices are ordered in such a way that the corresponding function values 
of an objective function $f$ are increasing, $f(\bmx_1)\le\dots\le 
f(\bmx_{d+1})$. In each step of the
algorithm the vertex $\bmx_{d+1}$ with the worst function value is replaced by 
a new vertex in a specified manner. The new vertex is chosen in such a way that
the simplex typically approaches a minimum of the function $f$. 
Denote by $\bmx_o=\frac{1}{d}\sum_{i=1}^{d}\bmx_i$ the centroid of all but the
worst vertices. The updating of the simplex is based on the following
operations:
\begin{itemize}
\item \textbf{Reflection:} $\bmx_r =\bmx_o+\alpha(\bmx_o-\bmx_{d+1})$, $\alpha>0$;
\item \textbf{Expansion:} $\bmx_e =\bmx_o+\gamma(\bmx_r-\bmx_o)$, $\gamma>1$;
\item \textbf{Inside Contraction:} $\bmx_{ic} =\bmx_o+\rho(\bmx_{d+1}-\bmx_o)$, $0<\rho<1$;
\item \textbf{Outside Contraction:} $\bmx_{oc} =\bmx_o+\rho(\bmx_r-\bmx_o)$, $0<\rho<1$;
\item \textbf{Shrinking:} $\bmx_i' =\bmx_1+\sigma(\bmx_i-\bmx_1)$, $i=2,\dots,d+1$, $0<\sigma<1$. 
\end{itemize}  
A common choice for the parameters is $\alpha=1$, $\gamma=2$, $\rho=\sigma=0.5$,
see Figure~\ref{fig.NM}.
In the first four operations the new point lies on the straight line through
$\bmx_o$ and $\bmx_{d+1}$, in the last operation new points are computed
on the straight lines through $\bmx_1$ and $\bmx_i$, $i=2,\dots,d+1$. The new
simplex is then formed by the points $\bmx_1$ and $\bmx_2',\dots,\bmx_{d+1}'$.

As in the case of the coordinate descent, it is possible to apply the 
Nelder-Mead algorithm without any modifications to the function $\tilde \phi_{\bmz,\bmX}$
that extends $\phi_{\bmz,\bmX}$ on the domain $\IR^d\setminus\{\bmNull\}$. Again, we 
believe that it is better to take care of the special geometry of the
domain $\IS^{d-1}$ and adapt the Nelder-Mead method in a proper way.
To do this the straight lines along which new points are computed have to be
replaced by proper curves. The natural choice are again the geodesics on the
sphere, i.e., the great circles. 

Following \eqref{eq.greatcircle} the great circle defined by $\bmx$ and $\bmy$
is given for $\tilde\bmy=\bmy-\ip{\bmx}{\bmy}\bmx$ by
\[
\left\{\bmz(\phi)=\cos(\phi)\bmx+\sin(\phi)\frac{\tilde\bmy}
{\|\tilde\bmy\|}\,\middle|\,\phi\in(-\pi,\pi]\right\}\,.
\]
Thus, the analogue of the point $\bmx+t(\bmy-\bmx)$ is given by
$\bmz(t\alpha)$, where $\alpha=\arccos(\ip{\bmx}{\bmy})$ is the great-circle
distance between $\bmx$ and $\bmy$.
We define the function $\bmgamma_{\bmx,\bmy}:\IR\to\IS^{d-1}$ by
\[
\bmgamma_{\bmx,\bmy}(t)
=\cos(t\alpha)\bmx+\sin(t\alpha)\frac{\tilde\bmy}{\|\tilde\bmy\|}\,.
\]
Clearly, $\bmgamma_{\bmx,\bmy}(t)$ is a point that lies on the great circle
defined by $\bmx$ and $\bmy$, and the great-circle distance between 
$\bmgamma_{\bmx,\bmy}(t)$ and $\bmx$ is $t$ times the great-circle distance between 
$\bmy$ and $\bmx$ (provided that $|t\alpha|<\pi$). For $t\in[0,1]$, the
point $\bmgamma_{\bmx,\bmy}(t)$ can be seen as a \emph{spherical convex
combination} of $\bmx$ and $\bmy$. 

For computing $\bmgamma_{\bmx,\bmy}(t)$, we first have to compute $\alpha$.
Although it is tempting to simply compute $\alpha$ using the inverse cosine,
there is a problem when $\bmx$ and $\bmy$ are nearly parallel since the inverse
cosine is numerically instable for values of the argument near unity. 
Therefore, we recommend to use the following formula to compute $\alpha$:
\begin{equation}\label{eq.arcsin}
\alpha=\begin{cases}
2\arcsin\left(\frac12\|\bmx-\bmy\|\right)&\text{if $\ip{\bmx}{\bmy}\ge0$,}\\
\pi-2\arcsin\left(\frac12\|\bmx+\bmy\|\right)&\text{if $\ip{\bmx}{\bmy}<0$.}
\end{cases}
\end{equation}
For computing $\bmgamma_{\bmx,\bmy}(t)$ we proceed as follows.
First note that
\[
\|\tilde\bmy\|^2
=\|\bmy\|^2-2\ip{\bmx}{\bmy}^2+\ip{\bmx}{\bmy}^2\cdot\|\bmx\|^2\\
=1-\ip{\bmx}{\bmy}^2
=1-\cos^2(\alpha)
=\sin^2(\alpha)
\]
and therefore $\|\tilde\bmy\|=\sin(\alpha)$. Hence, $\bmgamma_{\bmx,\bmy}(t)$
can be computed as follows
\begin{align*}
\bmgamma_{\bmx,\bmy}(t)
&=\cos(t\alpha)\bmx+\sin(t\alpha)\frac{\tilde\bmy}{\|\tilde\bmy\|}\\
&=\cos(t\alpha)\bmx+\frac{\sin(t\alpha)}{\sin(\alpha)}(\bmy-\ip{\bmx}{\bmy}\bmx)\\
&=\left[\cos(t\alpha)-\frac{\sin(t\alpha)}{\sin(\alpha)}\ip{\bmx}{\bmy}\right]\bmx
+\frac{\sin(t\alpha)}{\sin(\alpha)}\bmy\\
&=\left[\frac{\cos(t\alpha)\sin(\alpha)-\sin(t\alpha)\cos(\alpha)}{\sin(\alpha)}\right]\bmx
+\frac{\sin(t\alpha)}{\sin(\alpha)}\bmy\\
&=\frac{\sin(\alpha-t\alpha)}{\sin(\alpha)}\bmx
+\frac{\sin(t\alpha)}{\sin(\alpha)}\bmy\\
&=\frac{\sin((1-t)\alpha)}{\sin(\alpha)}\bmx
+\frac{\sin(t\alpha)}{\sin(\alpha)}\bmy\,.
\end{align*}
Note that $\sin(\alpha)$ can be computed as a byproduct when we compute
$\alpha$ by \eqref{eq.arcsin}, 
\[
\sin(\alpha)=\begin{cases}
\|\bmx-\bmy\|\sqrt{\frac{1+\ip{\bmx}{\bmy}}2}\,&\text{if $\ip{\bmx}{\bmy}\ge0$,}\\
\|\bmx+\bmy\|\sqrt{\frac{1-\ip{\bmx}{\bmy}}2}\,&\text{if $\ip{\bmx}{\bmy}<0$.}
\end{cases}
\]
In the case $\ip{\bmx}{\bmy}\ge0$, this follows from
\[
\sin(\alpha)
=2\sin\left(\frac\alpha2\right)\cos\left(\frac\alpha2\right)
=2\frac{\|\bmx-\bmy\|}2
\sqrt{\frac{1+\cos(\alpha)}2}
=\|\bmx-\bmy\|\sqrt{\frac{1+\ip{\bmx}{\bmy}}2}
\]
and analogously when $\ip{\bmx}{\bmy}<0$. Therefore, to compute
$\bmgamma_{\bmx,\bmy}(t)$ we use the function \textsc{greatCircle}, given in
pseudocode in Algorithm~\ref{alg.greatCircle}.
\begin{algorithm}[ht]
\caption{Point on a great circle}\label{alg.greatCircle}
\begin{algorithmic}[1]
\Function{greatCircle}{$\bmx,\bmy,t$}
\State $sp \gets \ip{\bmx}{\bmy}$
\If{$sp\ge 0$}
\State $sum\gets\|\bmx-\bmy\|^2$
\State $\alpha\gets 2 \cdot \arcsin(0.5 \cdot \mathrm{sqrt}(sum))$
\State $sina\gets \mathrm{sqrt}(sum \cdot (1 + sp) / 2)$ 
\Else
\State $sum\gets\|\bmx+\bmy\|^2$
\State $\alpha\gets \pi - 2 \cdot \arcsin(0.5 \cdot \mathrm{sqrt}(sum))$
\State $sina\gets \mathrm{sqrt}(sum \cdot (1 - sp) / 2)$ 
\EndIf
\State $gx\gets (1-t)\cdot\alpha$
\State $gy\gets t\cdot\alpha$
\If{(\texttt{Bound = y}) and ($\mathrm{abs}(gy)>\pi/2$)}
\If{$gy > 0$} $gy\gets\pi/2$ \textbf{else} $gy\gets-\pi/2$\EndIf
\State $gx\gets \alpha-gy$
\EndIf
\State $cx \gets \sin(gx) / sina$
\State $cy \gets \sin(gy) / sina$
\State $\bmz \gets cx \cdot \bmx + cy \cdot \bmy$
\State \Return $\bmz$
\EndFunction
\end{algorithmic}
\end{algorithm}

In the Nelder-Mead method, the procedure \textsc{greatCircle} will be used
for the operations Reflection, Expansion, Contraction and Shrinking.
For $t$ such that $t\alpha\notin[-\pi/2,\pi/2]$ the new point lies outside
the closed hemisphere with pole $\bmx$. To avoid such a behavior the value of
$t\alpha$ could be limited to the interval $[-\pi/2,\pi/2]$ in the routine
\textsc{greatCircle}. This is described in lines 13 to 15 of
Algorithm~\ref{alg.greatCircle}. 

Besides generalizing the basic operations of the Nelder-Mead algorithm,
we also have to generalize the notion of a centroid to the case of a sphere.
Given points $\bmp_1,\dots,\bmp_n\in\IS^{d-1}$, we have to define
a \emph{spherical centroid} $c(\bmp_1,\dots,\bmp_n)\in\IS^{d-1}$. 
It is not quite clear how to do this in a sensible way.
A natural generalization seems to be the so-called \emph{Fréchet mean}
(also called \emph{Riemannian center of mass}  or \emph{Karcher mean}, 
see \citealp{Afsari11,GroveK73,GroveKR74a,GroveKR74b}) 
of points $\bmp_1,\dots,\bmp_n$ on a Riemannian manifold $\mathcal{M}$ with 
distance function $\dist(\cdot,\cdot)$, which is defined to be the minimizer
of the sum of squared distances, 
\[
c_F(\bmp_1,\dots,\bmp_n)=\argmin_{\bmx\in\mathcal{M}}\sum_{i=1}^n\dist^2(\bmx,\bmp_i)\,.
\]
One problem with the Fréchet mean is that it needs not be unique. To see this,
consider the case of two points on the 2-sphere $\IS^2$. If the two points
are the north and the south pole, respectively, then the whole equator  minimizes the sum
of squared distances which shows that the Fréchet mean is not unique in this
case. However, if the points are contained in an open hemisphere,
then the Fréchet mean on the sphere is unique, see \cite{BussF01}.
Nevertheless, there is no closed
form expression for the Fréchet mean, instead it has to be computed by solving
an optimization problem.

A much easier possibility would be to compute the centroid in the ambient
space $\IR^d$ and then project the point back on the unit sphere $\IS^{d-1}$,
i.e.,
\[
c_N(\bmp_1,\dots,\bmp_n)=\frac{\ol{\bmp}}{\|\ol{\bmp}\|}\,,\quad\text{where}\
\ol{\bmp}=\frac1n\sum_{i=1}^n\bmp_i\,.
\]
Although this mean is also not unique when $\ol{\bmp}=\bmNull$, it is
widely used in directional statistics, see, e.g., \cite{LeyV17}.
We will call this point the \emph{na\"ive mean}.


Comparing the na\"ive mean and the Fréchet mean, we decided to choose the
na\"ive mean in our simulations. The reason is that in the long run the simplex
formed by $\bmp_1,\dots,\bmp_d$ should be small. If we only
consider a small region of the sphere then its geometry is ``nearly flat''.
Therefore, if the simplex is small, the difference between the Fréchet mean
and the na\"ive mean should also be small. Thus, we expect no great 
difference in performance between using the na\"ive mean and the 
Fréchet mean in the Nelder-Mead method. 

The Nelder-Mead algorithm, adopted to
the case of $\IS^{d-1}$ as the domain, is given as Algorithm~\ref{alg.spNelderMead}.
Since in the operations Reflection, Expansion and Contraction only the worst
point $\bmp_d$ is changed, the sequence $(\bmp_1,f_1),\dots,(\bmp_{d-1},f_{d-1})$
is still ordered so that $f_1\le\dots\le f_d$. Therefore in line~28 of
Algorithm~\ref{alg.spNelderMead} the routine \textsc{inPlaceMerge} can be used.
This routine expects two sorted sequences as arguments and merges them into one
sorted sequence. This is more efficient than sorting the whole sequence.
Only in the case when Shrinking is applied the subsequence 
$(\bmp_1,f_1),\dots,(\bmp_{d-1},f_{d-1})$ has to be sorted which is done in line~27.     

\begin{algorithm}[ht]
\caption{Spherical Nelder-Mead method}\label{alg.spNelderMead}
\begin{algorithmic}[1]
\Function{sphericalNelderMead}{$\bmz,\bmX$}
\If{\texttt{Start = Mn}}\ $\bmu\gets\bmz-\ol{\bmx}$\EndIf
\If{\texttt{Start = Rn}}\ $\bmu\gets\textsc{rndSphere}(d)$\EndIf
\State $\epsilon\gets(\pi/2)/\beta$ \Comment{size of the spherical cap}
\For{i}{1}{d}
\Comment finding the starting simplex
\State $\bmp_i\gets\textsc{rndSphericalCap}(\bmu,\epsilon)$
\State $f_i\gets \D(\ip{\bmp_i}\bmz|\ip{\bmp_i}\bmX)$
\EndFor
\State $\textsc{sort}([(\bmp_1,f_1),\dots,(\bmp_d,f_d)])$
\Comment sort pairs $(\bmp_i,f_i)$ such that $f_1\le\dots\le f_d$
\Repeat
\State $\bmx_o\gets c_N(\bmp_1,\dots,\bmp_{d-1})$
\State $\bmx_r\gets\textsc{greatCircle}(\bmx_o,\bmp_d,-\alpha)$
\Comment{reflected point}
\State $f_r\gets \D(\ip{\bmx_r}\bmz|\ip{\bmx_r}\bmX)$ 
\If{$f_1\le f_r <f_{d-1}$} $(\bmp_d,f_d)\gets(\bmx_r,f_r)$
\ElsIf{$f_r<f_1$}
\State $\bmx_e\gets\textsc{greatCircle}(\bmx_o,\bmx_r,\gamma)$
\Comment{expanded point}
\State $f_e\gets \D(\ip{\bmx_e}\bmz|\ip{\bmx_e}\bmX)$ 
\If{$f_e<f_r$} $(\bmp_d,f_d)\gets(\bmx_e,f_e)$ \textbf{else} $(\bmp_d,f_d)\gets(\bmx_r,f_r)$\EndIf
\Else \Comment{$f_{d-1}\le f_r$}
\If{$f_r<f_d$} $\bmx_h\gets\bmx_r$ \textbf{else} $\bmx_h\gets\bmp_d$\EndIf
\State $\bmx_c\gets\textsc{greatCircle}(\bmx_o,\bmx_h,\rho)$
\Comment{contracted point}
\State $f_c\gets \D(\ip{\bmx_c}\bmz|\ip{\bmx_c}\bmX)$
\If{$f_c<f_d$} $(\bmp_d,f_d)\gets(\bmx_c,f_c)$
\Else \Comment{reduction}
\For{i}{2}{d}
\State $\bmp_i\gets\textsc{greatCircle}(\bmp_1,\bmp_i,\sigma)$
\State $f_i\gets \D(\ip{\bmp_i}\bmz|\ip{\bmp_i}\bmX)$
\EndFor
\State $\textsc{sort}([(\bmp_1,f_1),\dots,(\bmp_{d-1},f_{d-1})])$
\EndIf
\EndIf
\State $\textsc{inPlaceMerge}([(\bmp_1,f_1),\dots,(\bmp_{d-1},f_{d-1})],[(\bmp_d,f_d)])$
\Comment put $(\bmp_d,f_d)$ in the

\Comment correct position 
\Until stopping criterion is satisfied
\EndFunction
\end{algorithmic}
\end{algorithm}
In our implementation we used a common choice for the parameters, namely
$\alpha=1$, $\gamma=2$, $\rho=\sigma=0.5$.
The stopping criterion was again chosen in order to guarantee that a 
specified number $N$ of evaluations of the univariate depth was not exceeded.
The starting simplex was chosen as follows.
From an $\epsilon$-spherical cap $d$ points were chosen randomly using the
procedure \textsc{rndSpericalCap}. The pole of this cap was determined by a 
parameter \texttt{Start} with possible values \texttt{Mn} and \texttt{Rn}.  
For \texttt{Start=Rn} the pole of the cap was randomly drawn from
$\mathcal U(\IS^{d-1})$ whereas for \texttt{Start=Mn} the pole is given by
$\bmu=\bmz-\ol{\bmx}$.
The size $\epsilon$ of the cap was controlled by a parameter $\beta$ via 
$\epsilon=(\pi/2)/\beta$. A further parameter \texttt{Bound} controlled whether
the movement along the great circle in the routine
\textsc{greatCircle}($\bmx,\bmy,t$) was limited to a maximum distance
of $\pi/2$ between $\bmx$ and the new point (\texttt{Bound=y}) or
not (\texttt{Bound=n}).
Further, we compared the adapted version of the Nelder-Mead
algorithm with applying the ordinary Nelder-Mead algorithm to the function
$\tilde\phi_{\bmz,\bmX}$. This was controlled by a parameter \texttt{Space} having
possible values \texttt{Ec} (Euclidean space) and \texttt{Sp} (Sphere).
The tuning of the parameters \texttt{Start}, $\beta$, \texttt{Bound} and
\texttt{Space} is again described in Section~\ref{ssec:tuning}. 

\section{Simulation comparison}\label{sec:simulation}

This section comprises the description and the results of our simulation comparison. In Section~\ref{ssec:dists} we define the benchmark distributions and describe the experimental study. Section~\ref{ssec:tuning} indicates the sets over which parameters are tuned, and concludes on their final choice. Section~\ref{ssec:comparison} analyses the results of the simulation study.

\subsection{Distributional and simulation settings}\label{ssec:dists}

The simulation study is based on the depth computation of a point $\bmz$ w.r.t. a sample $\bmX$ of $n$ i.i.d. $d$-variate points. The point $\bmz$ is taken to be the average of $10$ arbitrary points of the sample. This guarantees that it belongs to the convex hull of the data so that the depth is always strictly positive, but also does not place it too deep in the data set to preserve the random nature of the choice of $\bmz$ (since only $10$ points out of $1000$ are averaged). The following six distributions are used for the comparison (due to the affine-invariance of the considered depths we do not introduce any correlation structure):
\begin{itemize}
	\item the standard normal distribution $\mathcal{N}(\boldsymbol{0}_d,\boldsymbol{I}_d)$;
	\item the spherically-symmetric Student $t_5$ distribution;
	\item the spherically-symmetric Student $t_1$ (Cauchy) distribution;
	\item the skewed normal distribution generated in the following way \citep{Azzalini2013}: let $U_i\sim\mathcal{U}([0,1])$, $\bmZ_i\sim\mathcal{N}(\boldsymbol{0}_d,\boldsymbol{I}_d)$, $U_i$ and $\bmZ_i$ stochastically independent, and $\boldsymbol{\delta}\in\mathbb{R}^d$ be a skewness parameter. Then the skewed normal random vector equals
	\begin{equation*}
		\bmX_i \overset{\mathrm{d}}{=} \begin{cases}\hphantom{-}\bmZ_i & \text{ if } U_i \le \Phi(\boldsymbol{\delta}\T \bmZ_i)\,, \\ -\bmZ_i & \text{ if } U_i > \Phi(\boldsymbol{\delta}\T \bmZ_i)\,, \end{cases}
	\end{equation*}
	where $\Phi(\cdot)$ is the c.d.f. of the standard normal distribution. We set $\boldsymbol{\delta} = (5,0,...,0)^T$;
	\item the uniform distribution on $[0,1]^d$;
	\item a product of $d$ independent exponential distributions (with parameter $\lambda=1$).
\end{itemize}

In the simulation study, we consider the five depth notions described in Section~\ref{ssec:datadepth}. An explicit formula makes it unnecessary to approximate the Mahalanobis depth, which is in addition a quadratic (and thus everywhere smooth) function. Since rather good results for optimization techniques are expected, we include it for a qualitative comparison with random algorithms. While the zonoid depth can be computed efficiently even in higher dimensions using the algorithm of~\cite{DyckerhoffKM96}, it is also included as a benchmark.

When fine-tuning the algorithms, for each depth, for $n=1000$ points, in dimensions $d=5,10,15,20$, we use $N=1000$ random directions for each of the algorithms and each combination of parameters. For each of the distributions, we repeat the computation of the depth $1000$ times and summarize the results using two statistics based on the following idea. Since all the considered algorithms report an upper bound on the actual depth, for the same $\bmz$ and a data set one can compare parameters according to the obtained depth values, because lower obtained depth is always closer to the exact value. We do it by reporting:
\begin{itemize}
	\item the average rank of the obtained depth approximation (among all considered parameter combinations) over $1000$ runs (the lower the better, with $1$ being the best), which we shortly denote as \textsc{AveRank};
	\item the percentage when the considered set of parameters achieved the smallest depth value (among all parameter combinations) over $1000$ runs (the higher the better, with $100\%$ being the best), which we shortly denote as \textsc{PercBest}.
\end{itemize}
We consider \textsc{AveRank} as the more important criterion. The reason is that \textsc{AveRank} not only focuses on how many times a method is the best method (as does \textsc{PercBest}), but also takes into account the rank of a method when it is not the best method. Often the best methods according to \textsc{AveRank} and \textsc{PercBest} coincide. Therefore, we restrict to \textsc{AveRank} in the presentation, and present both statistics in the Supplementary Materials. 


\subsection{Fine-tuning of the algorithms}\label{ssec:tuning}

Before running the simulation study in Section~\ref{ssec:comparison}, the variety of possible settings of the algorithms should be reduced to a proper choice of (nearly optimal) parameters. We do this by means of a comparative benchmark. The chosen parameters are then fixed for each algorithm throughout the subsequent sections. Since the task of parameter tuning is to choose parameters for a single algorithm, we run (and analyze) each algorithm separately.

For each algorithm, out of a preliminary chosen range, we select a suitable set of parameter values based on visualization and additional analysis in detail if necessary. Note, that even with such an extensive simulation the validity of the chosen parameters is still limited. We restrict the tuning process to rather small ranges of parameters since only several distributions are considered. Also, such a tuning provides a certain degree of robustness, which is especially desirable because the single chosen set of parameters will be used for all further experiments. Finally, this simplifies visual presentation and manual analysis. Table~\ref{tab:parsChosen} summarizes the parameters' ranges and their choices.

\begin{table}[h!]
	\begin{center}
	\begin{tabular}{llll}
	\toprule
		Method & Parameters & Values & Description\\
        \midrule
		RRS  & $N_{\text{ref}}$ & $5$, $\boldsymbol{10}$, $15$    & Number of refinement steps\\
		     & $\alpha$ & $0.2$, $\boldsymbol{0.5}$, $0.8$        & Shrinking factor of the spherical cap\\ 
        \midrule                                            
		RGS  & $N_{\text{ref}}$ & $5$, $\boldsymbol{10}$, $15$    & Number of refinement steps\\                
		     & $\alpha$ & $0.2$, $\boldsymbol{0.5}$, $0.8$        & Shrinking factor of the spherical cap\\    
        \midrule                                            
		RaSi & $\alpha$ & $\boldsymbol{1.25}$, $1.5$, $1.8$       & Parameter of the Dirichlet distribution\\
        \midrule
		SA   & $\alpha$ & $0.5$, $0.8$, $\boldsymbol{0.95}$       & Cooling factor\\
		     & $\beta$ & $5$, $\boldsymbol{10}$, $15$             & Size of the spherical cap\\
		     & $\texttt{Start}$ & \texttt{\bf Mn}, $\texttt{Rn}$  & Starting value (mean, random)\\
        \midrule
		CD   & $\texttt{Space}$ & $\texttt{Ec}$, \texttt{\bf Sp}  & Euclidean space or sphere\\
		     & $\texttt{LS}$ & $\texttt{Eq}$, \texttt{\bf GS}     & Line search: equally spaced or golden section\\
        \midrule
		NM   & $\texttt{Space}$ & $\texttt{Ec}$, \texttt{\bf Sp}  & Euclidean space or sphere\\ 
		     & $\texttt{Start}$ & \texttt{\bf Mn}, $\texttt{Rn}$  & Starting value (mean, random)\\
		     & $\beta$ & $\boldsymbol{1}$, $2$, $4$               & Size of the spherical cap\\
		     & $\texttt{Bound}$ & \texttt{\bf y}, $\texttt{n}$ & Bound movement on great circles (yes/no)\\
        \bottomrule
	\end{tabular}
	\end{center}
	\caption{Considered and selected (in bold) parameters, obtained by the fine-tuning.}
	\label{tab:parsChosen}
\end{table}
 
\begin{figure}[h!]
	\includegraphics[width=\textwidth]{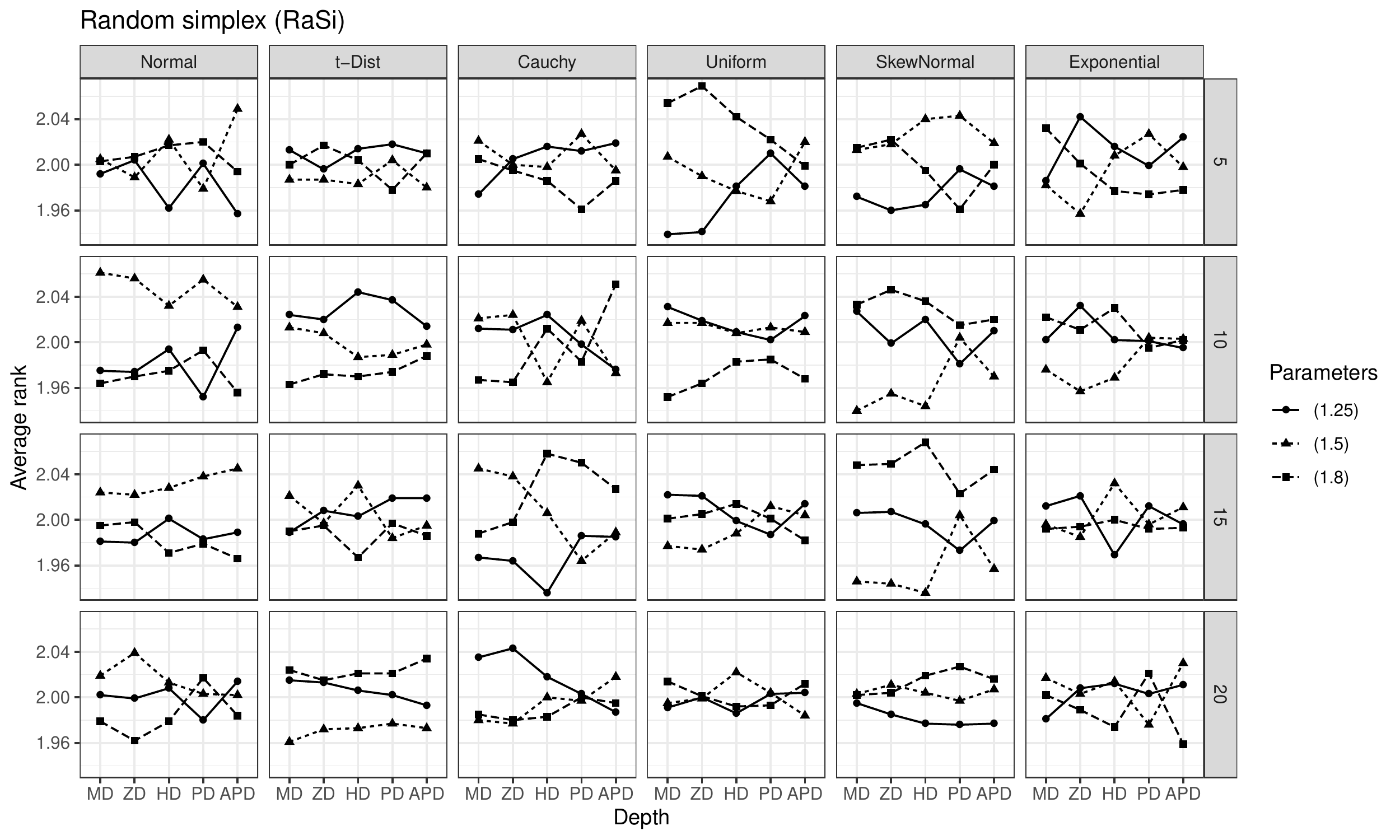}
	\caption{\textsc{AveRank} for the random simplices (RaSi) algorithm. The only parameter is $\alpha$, taking values in $\{1.25, 1.5, 1.8\}$.}\label{fig:sim1RaSiAveRank}
\end{figure}

\begin{figure}[hb!]
	\includegraphics[width=\textwidth]{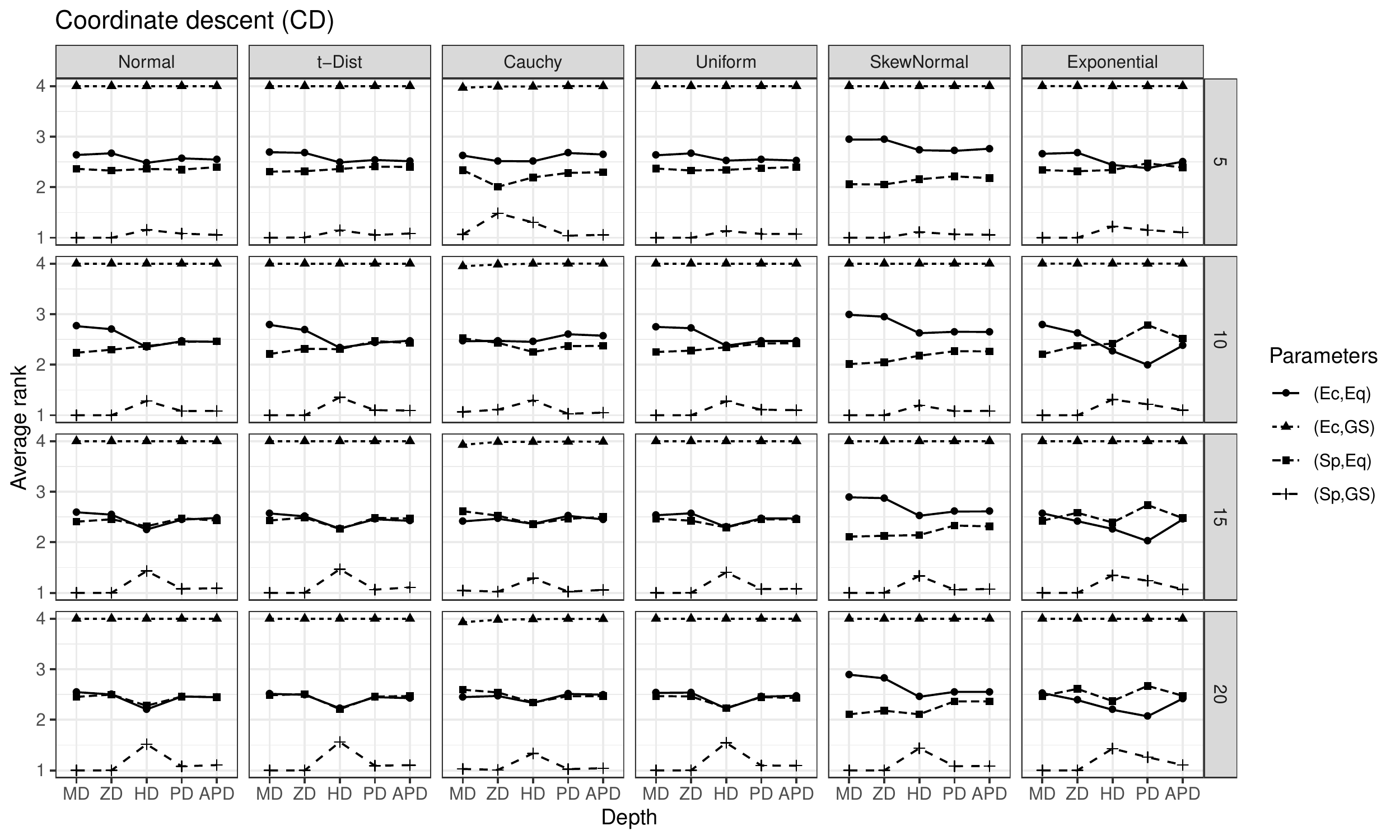}
	\caption{\textsc{AveRank} for the coordinate descent (CD) algorithm. The first parameter (\texttt{Space}) determines the space to be used: Euclidean space (\texttt{Ec}) or sphere (\texttt{Sp}). The second parameter (\texttt{LS}) defines the algorithm to be used for the line search: evaluation over an equidistant grid (\texttt{Eq}) or the golden section algorithm (\texttt{GS}).}\label{fig:sim1CDAveRank}
\end{figure}

Algorithms \emph{random search} (RS) and \emph{grid search} (GS) do not require hyper-parameters and thus need not to be tuned. For the six remaining algorithms, we detail the parameter-tuning process below. All figures indicating computed statistics can be found in the Supplementary Materials. Here we only place two illustrative ones: an example of a mixed result (random simplices, Figure~\ref{fig:sim1RaSiAveRank}) and of a clear winner (coordinate descent, Figure~\ref{fig:sim1CDAveRank}).
\begin{itemize}
	\item \emph{Refined random search} (RRS): Visual inspection and summary statistics (average over all experiments) give no clear winner. However, the combinations with a high value of $N_{\text{ref}}$ and a small value of $\alpha$ or vice versa, are clearly inferior. Therefore we choose a pair where both parameters are in the middle of the ranges.
	\item \emph{Refined grid search} (RGS): \textsc{AveRank} statistics mostly hint on the best pair here.
	\item \emph{Random simplices} (RaSi): \textsc{AveRank} statistics, plotted in Figure~\ref{fig:sim1RaSiAveRank}, do not help in choosing the best parameter. One observes that the concentration parameter of the Dirichlet seems not to have much influence on the performance of the algorithm. We decided to allow for most freedom when drawing the direction.
	\item \emph{Simulated annealing} (SA): Substantial number of the considered alternatives complicates the choice of the best performing parameter combination. The parameter choice is thus made based on the summary statistics.
	\item \emph{Coordinate descent} (CD): \textsc{AveRank} of the four considered settings is depicted in Figure~\ref{fig:sim1CDAveRank}. One clearly distinguishes superiority of using great circles' coordinate system and the golden section technique for the line search.
	\item \emph{Nelder-Mead} (NM): First, running usual NM in the Euclidean space performs worse than when sticking to the geometry of the hyper-sphere. One can further notice that starting with the spherical cap around the direction from the data mean to $\bmz$, drawing the initial simplex from the entire hemisphere, as well as forcing the simplex to be always contained in a hemisphere gives on average better results.
\end{itemize}

Several limitations of the above fine-tuning need to be mentioned, which hold true for the simulation comparison in Section~\ref{ssec:comparison} as well:
\begin{itemize}
	\item Strictly speaking, the tuning is subject to the chosen distributions, dimensions and ranges of parameters. Further parameters of the algorithms, not mentioned here (e.g. constants of the Nelder-Mead algorithm, see Section~\ref{ssec:NM}) were kept unchanged on their default values.
	\item The number of random directions is fixed to $N\approx 1000$, which is a budget constraint. Thus, it is possible that one algorithm would approximate better with a few more directions, while another one would not make use of these additional directions. On contrary, with less directions the results could look differently. On the other hand, the simulation study of Section~\ref{ssec:comparison} illustrates only weak dependence of our conclusions on the change of the number of random directions.
\end{itemize}

\subsection{Results of the simulation study}\label{ssec:comparison}



To compare the performance of the algorithms, we run a simulation study for the distributional settings from Section~\ref{ssec:dists}, fixing parameters to the values chosen in Section~\ref{ssec:tuning}. We take $N\approx 100$, $1000$, and $10000$ projections. In this section, only results for $N\approx 1000$ directions are analyzed, since those for the other numbers of directions are similar. Complete results can be found in the Supplementary Materials.

\begin{figure}[h!]
	\includegraphics[width=\textwidth]{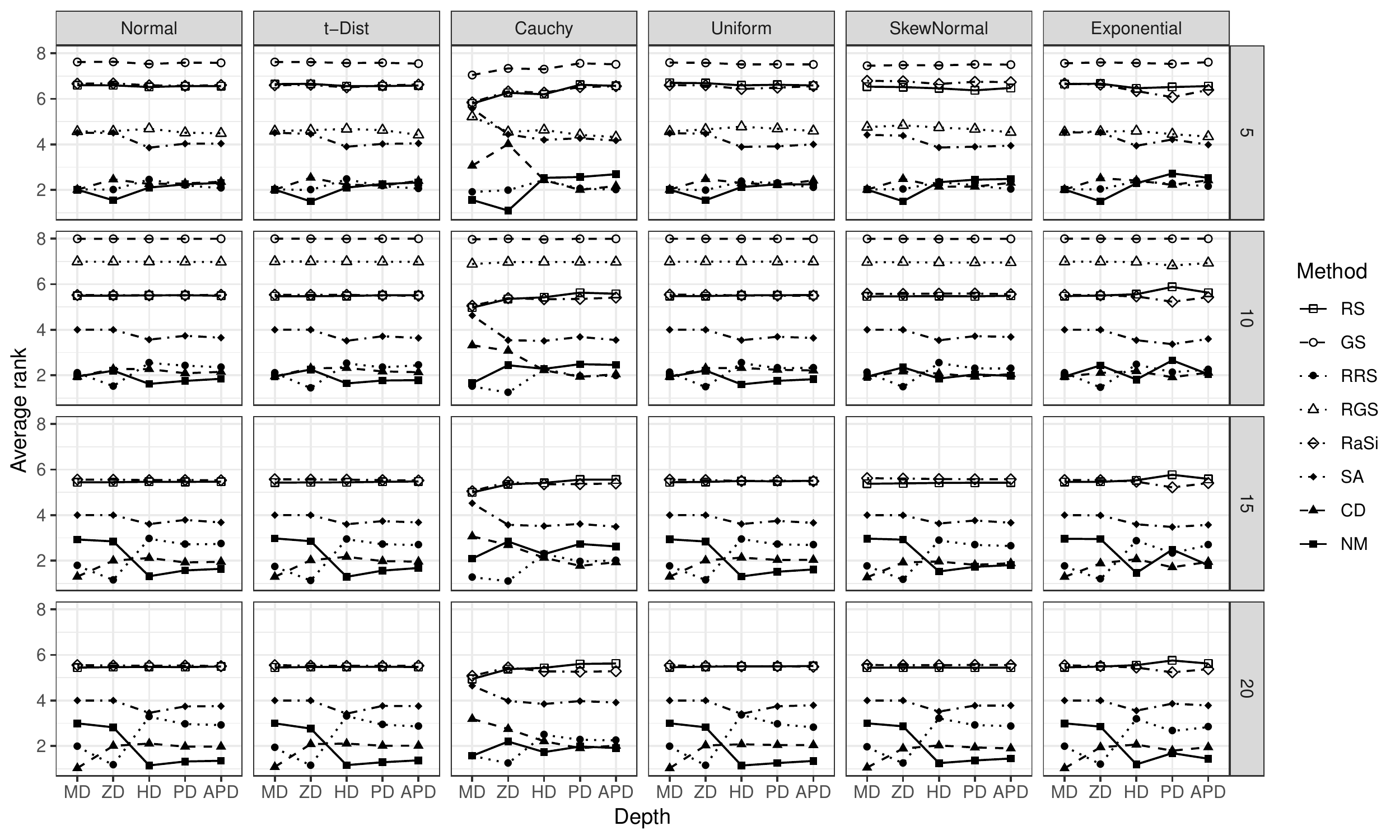}
	\caption{\textsc{AveRank} statistics for the eight considered approximation methods when using the parameter settings from Section~\ref{ssec:tuning} (see also Table~\ref{tab:parsChosen}) and $N\approx1000$ projections.}
	\label{fig:sim2averank1000}
\end{figure}

Figure~\ref{fig:sim2averank1000} exhibits \textsc{AveRank} for each of the eight considered algorithms for different depth notions, distributions, and dimensions. Several observations can be made:
\begin{itemize}
	\item There is a group of methods which have poor performance that further degrades with increasing dimension: random search (RS), grid search (GS), refined grid search (RGS), and random simplices (RaSi). Moreover, GS and RGS are not considered in dimension $d>10$ (as explained in Sections~\ref{ssec:GS} and~\ref{ssec:RGS}), because $1000$ directions are not sufficient to generate even a very sparse grid in such a high dimension.
	\item Refined random search (RRS), coordinate descent (CD), and Nelder-Mead (NM) show rather good performance.
	\item NM shows superior behavior in this latter group, since it possesses almost always lower \textsc{AveRank} compared to the two other methods for the halfspace, projection, and asymmetric projection depths. Thus it can be seen as a general winner. However, it is closely followed by CD.
\end{itemize}

Additionally, Table~\ref{tab:bestm_HDa} illustrates concordance of \textsc{AveRank} and \textsc{PercBest} statistics for the halfspace depth (in most of the cases); see the Supplementary Materials for all depths.

\begin{table}[ht]\footnotesize
\begin{center}
\begin{tabular}{llcccccccccccc}
\toprule
 & & \mc{12}{c}{Number of projections} \\
\cmidrule{3-14}
HD & & \mc{4}{c}{100} & \mc{4}{c}{1000} & \mc{4}{c}{10000} \\
\cmidrule(rl){3-6} \cmidrule(rl){7-10} \cmidrule(rl){11-14} \\
$d$ & Distribution & \rotatebox{90}{RRS} & \rotatebox{90}{SA} & \rotatebox{90}{CD} & \rotatebox{90}{NM} & \rotatebox{90}{RRS} & \rotatebox{90}{SA} & \rotatebox{90}{CD} & \rotatebox{90}{NM} & \rotatebox{90}{RRS} & \rotatebox{90}{SA} & \rotatebox{90}{CD} & \rotatebox{90}{NM} \\
\midrule
5 & Normal &   &   &   & $\bestRankPerc$ &   &   &   & $\bestRankPerc$ &   &   & $\bestPerc$ & $\bestRank$ \\
 & t-Dist &   &   &   & $\bestRankPerc$ &   &   &   & $\bestRankPerc$ &   &   &   & $\bestRankPerc$ \\
 & Cauchy & $\bestPerc$ &   & $\bestRank$ &   & $\bestPerc$ &   & $\bestRank$ &   & $\bestPerc$ &   &   & $\bestRank$ \\
 & Uniform &   &   &   & $\bestRankPerc$ & $\bestPerc$ &   &   & $\bestRank$ &   &   &   & $\bestRankPerc$ \\
 & SkewNormal &   &   & $\bestRankPerc$ &   &   &   & $\bestRankPerc$ &   &   &   & $\bestRankPerc$ &   \\
 & Exponential &   &   & $\bestRankPerc$ &   & $\bestPerc$ &   &   & $\bestRank$ &   &   &   & $\bestRankPerc$ \\
\midrule
10 & Normal &   &   &   & $\bestRankPerc$ &   &   &   & $\bestRankPerc$ &   &   &   & $\bestRankPerc$ \\
 & t-Dist &   &   &   & $\bestRankPerc$ &   &   &   & $\bestRankPerc$ &   &   &   & $\bestRankPerc$ \\
 & Cauchy &   &   &   & $\bestRankPerc$ & $\bestPerc$ &   & $\bestRank$ &   &   &   & $\bestRankPerc$ &   \\
 & Uniform &   &   &   & $\bestRankPerc$ &   &   &   & $\bestRankPerc$ &   &   &   & $\bestRankPerc$ \\
 & SkewNormal &   &   &   & $\bestRankPerc$ &   &   &   & $\bestRankPerc$ &   &   & $\bestPerc$ & $\bestRank$ \\
 & Exponential &   &   &   & $\bestRankPerc$ &   &   &   & $\bestRankPerc$ &   &   &   & $\bestRankPerc$ \\
\midrule
15 & Normal &   &   &   & $\bestRankPerc$ &   &   &   & $\bestRankPerc$ &   &   &   & $\bestRankPerc$ \\
 & t-Dist &   &   &   & $\bestRankPerc$ &   &   &   & $\bestRankPerc$ &   &   &   & $\bestRankPerc$ \\
 & Cauchy &   &   &   & $\bestRankPerc$ &   &   & $\bestRank$ & $\bestPerc$ &   &   &   & $\bestRankPerc$ \\
 & Uniform &   &   &   & $\bestRankPerc$ &   &   &   & $\bestRankPerc$ &   &   &   & $\bestRankPerc$ \\
 & SkewNormal &   &   &   & $\bestRankPerc$ &   &   &   & $\bestRankPerc$ &   &   &   & $\bestRankPerc$ \\
 & Exponential &   &   &   & $\bestRankPerc$ &   &   &   & $\bestRankPerc$ &   &   &   & $\bestRankPerc$ \\
\midrule
20 & Normal &   &   & $\bestRankPerc$ &   &   &   &   & $\bestRankPerc$ &   &   &   & $\bestRankPerc$ \\
 & t-Dist &   &   & $\bestRankPerc$ &   &   &   &   & $\bestRankPerc$ &   &   &   & $\bestRankPerc$ \\
 & Cauchy &   &   &   & $\bestRankPerc$ &   &   &   & $\bestRankPerc$ &   &   &   & $\bestRankPerc$ \\
 & Uniform &   &   & $\bestRankPerc$ &   &   &   &   & $\bestRankPerc$ &   &   &   & $\bestRankPerc$ \\
 & SkewNormal &   &   & $\bestRankPerc$ &   &   &   &   & $\bestRankPerc$ &   &   &   & $\bestRankPerc$ \\
 & Exponential &   &   & $\bestRankPerc$ &   &   &   &   & $\bestRankPerc$ &   &   &   & $\bestRankPerc$ \\
\bottomrule
\end{tabular}
\end{center}
\caption{Best performing methods in sense of \textsc{AveRank} (filled circle) for the halfspace depth (HD). If the best method in view of \textsc{PercBest} differs, it is indicated by an empty circle.}\label{tab:bestm_HDa}
\end{table}

To get more insights into the dynamic of the optimization process, we regard the flow of the minimal reached depth with the number of random directions. A typical behavior of the optimization, on the example of the normal distribution in dimension $20$, is indicated in Figure~\ref{fig:sim2flow20Normal1000}. Similar figures for all six considered distributions are gathered in the Supplementary Materials.

\begin{figure}[h!]
	\includegraphics[width=\textwidth]{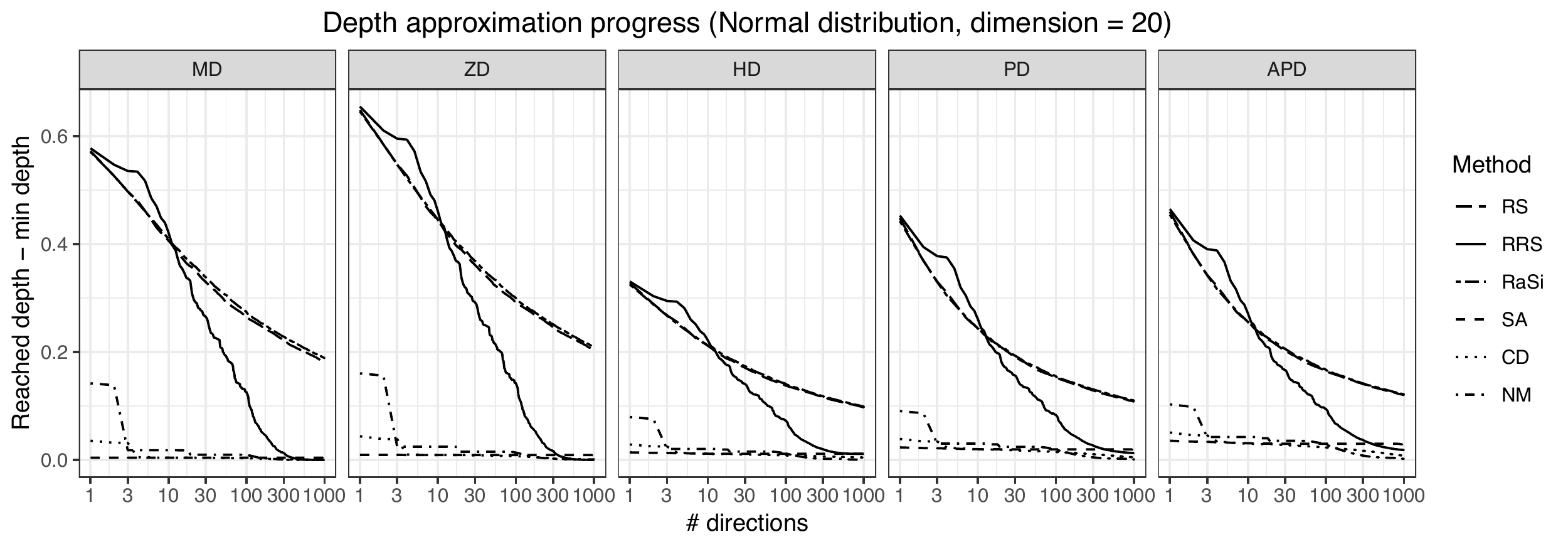}
	\caption{Average (over $1000$ runs) difference between the reached depth and the minimally achieved depth (by all methods for the current triplet distribution---depth---dimension) during the optimization process (normal distribution, dimension $d=20$). (The lines of RS and RaSi almost coincide in the graphs.)}
	\label{fig:sim2flow20Normal1000}
\end{figure}

Our most important observation is a high performance of the optimization techniques (SA, RRS, CD, and NM) compared with the random methods (RS, RaSi). The two latter ones seem to (approximately) follow the bounds derived in \cite{NagyDM19} and are outperformed already before reaching $100$ random directions. Further inspection shows that the improvement of simulated annealing (SA) is very weak, and minor improvement can be expected for even higher number of directions. A possible explanation would be that the parameters of SA should be tuned separately for each setting.


\section{Results for the approximation error}\label{sec:exact}
Apart from knowing which of the discussed approximation methods gives the best
approximation, it is also of interest to have information on the approximation
error. To calculate the approximation error, the exact values of the depths
have to be known. From the considered depths only the Mahalanobis depth
and the zonoid depth can be exactly computed in high dimensions in reasonable
time. For the halfspace depth there exists an exact algorithm 
\citep{DyckerhoffM16} to compute the depth in arbitrary dimension which has
a complexity of $O(n^{d-1}\log n)$. For the considered sample size of $n=1000$
the exact computation of the halfspace depth is (in reasonable time) only 
possible when  $d\le 5$. Although there is an exact algorithm for the projection
depth \citep{LiuZ14}, the considered sample size of $n=1000$ in dimension $d=5$
is already too large to have the value of the depth computed in reasonable time.
For the asymmetric projection depth no exact algorithm exists. Therefore we
decided to have a closer look at two situations. 
First, we choose
to examine the approximation of the halfspace depth since it 
probably is the most prominent depth. Because of the high computational cost
we computed approximation errors for the halfspace depth only in dimension
$d=5$.  
Second, to get some intuition
on approximation errors in high dimensions we choose to compute approximation
errors for the zonoid depth in all dimension $d=5,10,15,20$. The zonoid depth was chosen
since apart from the trivial case of the Mahalanobis depth, it is the only depth 
considered in this study for which exact computation is possible when $d=20$.
The time for computing the zonoid depth of a single point w.r.t. a sample of
size $n=1000$ in dimension $d=20$ is still under one second. 
For both setups (halfspace depth, $d=5$, and zonoid depth, $d\le 20$)
we used the same simulated datasets that were already used in Section~\ref{ssec:comparison}.
For the halfspace depth we only used the first fifteen simulated datasets
to compute the approximation errors. 
The approximated depth values as well as the exact depth values for the fifteen datasets simulated from the normal distribution are shown for the halfspace depth in Table~\ref{tab.exact_HD_Normal_1000a}.
The last two lines of the table show the mean absolute error (MAE) and the mean relative error (MRE) of the considered approximation methods.
The respective tables for all six distributions and $N\approx100,1000,10000$ projections are given in the Supplementary Materials.

\begin{table}[ht]\footnotesize
\centering
\begin{tabular}{r*{9}{c}}
\toprule
    &    RS &    GS &       RRS &   RGS &  RaSi &        SA &        CD &        NM & Exact\\
\midrule
  1 & 0.279 & 0.286 & \md 0.267 & 0.273 & 0.279 &     0.269 &     0.268 &     0.269 & 0.265\\
  2 & 0.143 & 0.160 &     0.132 & 0.139 & 0.148 &     0.136 &     0.132 & \md 0.130 & 0.128\\
  3 & 0.244 & 0.259 & \md 0.238 & 0.245 & 0.247 &     0.241 & \md 0.238 & \md 0.238 & 0.236\\
  4 & 0.329 & 0.340 & \md 0.310 & 0.336 & 0.324 &     0.316 &     0.311 &     0.311 & 0.309\\
  5 & 0.220 & 0.235 &     0.200 & 0.208 & 0.225 &     0.204 & \md 0.199 &     0.200 & 0.197\\
  6 & 0.236 & 0.259 & \md 0.214 & 0.229 & 0.227 &     0.219 &     0.216 &     0.217 & 0.213\\
  7 & 0.238 & 0.242 &     0.230 & 0.235 & 0.244 &     0.233 & \md 0.227 &     0.229 & 0.226\\
  8 & 0.228 & 0.229 & \md 0.218 & 0.225 & 0.229 &     0.223 &     0.219 & \md 0.218 & 0.215\\
  9 & 0.171 & 0.169 &     0.152 & 0.153 & 0.164 &     0.154 & \md 0.151 &     0.152 & 0.149\\
 10 & 0.241 & 0.248 &     0.224 & 0.230 & 0.228 &     0.225 & \md 0.223 & \md 0.223 & 0.221\\
 11 & 0.187 & 0.205 &     0.170 & 0.182 & 0.185 & \md 0.169 & \md 0.169 & \md 0.169 & 0.166\\
 12 & 0.284 & 0.280 & \md 0.269 & 0.273 & 0.288 &     0.274 &     0.272 &     0.271 & 0.268\\
 13 & 0.221 & 0.228 & \md 0.201 & 0.218 & 0.212 &     0.206 &     0.206 &     0.202 & 0.200\\
 14 & 0.171 & 0.182 &     0.161 & 0.161 & 0.176 &     0.160 &     0.158 & \md 0.157 & 0.154\\
 15 & 0.168 & 0.188 & \md 0.157 & 0.158 & 0.168 &     0.161 &     0.161 & \md 0.157 & 0.157\\
\midrule
MRE & 0.105 & 0.169 &     0.017 & 0.063 & 0.099 &     0.035 &     0.019 & \md 0.016 & 0.000\\
MAE & 0.017 & 0.027 & \md 0.003 & 0.011 & 0.016 &     0.006 &     0.003 & \md 0.003 & 0.000\\
\bottomrule
\end{tabular}
\caption{Exact and approximate values of the halfspace depth for $15$ points together with the corresponding MAE and MRE: normal distribution, $n = 1000$, $d = 5$, $N \approx 1000$ projections.}
\label{tab.exact_HD_Normal_1000a}
\end{table}

Because of its good computability, for the zonoid depth we used all 1000
simulated datasets for each combination of dimension and distribution.
In Table~\ref{tab.MRE_ZD_1000a} the mean relative errors (MRE) are shown for the case where $1000$ projections were used.
The respective tables for both the mean relative errors (MRE) and the mean absolute errors (MAE) for all six distributions and $N\approx100,1000,10000$ projections and  are shown in the Supplementary Materials. 

\begin{table}[ht]\footnotesize
\centering
\resizebox{0.99\linewidth}{!}{%
\begin{tabular}{ll*{8}{c}}
\toprule
&&\multicolumn{8}{c}{Approximation algorithm}\\
\cmidrule{3-10}
$d$ & Distribution &           RS &           GS &          RRS &          RGS &         RaSi &           SA &           CD &           NM\\
\midrule
 5 &        Normal &     0.018782 &     0.040681 &     0.000002 &     0.002805 &     0.019107 &     0.000535 &     0.000004 & \md 0.000001\\
   &        Cauchy &     0.043137 &     0.080894 &     0.000465 &     0.016957 &     0.045022 &     0.010639 &     0.017534 & \md 0.000056\\
   &       Uniform &     0.021093 &     0.047018 &     0.000002 &     0.003281 &     0.021418 &     0.000597 &     0.000008 & \md 0.000001\\
   &        t-Dist &     0.018799 &     0.041453 &     0.000002 &     0.003367 &     0.018021 &     0.000537 &     0.000004 & \md 0.000001\\
   &    SkewNormal &     0.021533 &     0.046762 &     0.000002 &     0.005977 &     0.029959 &     0.000643 &     0.000005 & \md 0.000001\\
   &   Exponential &     0.021569 &     0.047682 &     0.000002 &     0.003068 &     0.020915 &     0.000653 &     0.000007 & \md 0.000001\\
\midrule
10 &        Normal &     0.217197 &     1.118694 & \md 0.000025 &     0.942766 &     0.219397 &     0.007122 &     0.000057 &     0.000094\\
   &        Cauchy &     0.303627 &     2.212219 & \md 0.006813 &     1.661869 &     0.307660 &     0.064152 &     0.052861 &     0.034096\\
   &       Uniform &     0.222892 &     1.289776 & \md 0.000036 &     1.094230 &     0.228138 &     0.008009 &     0.000094 &     0.000168\\
   &        t-Dist &     0.215653 &     1.090465 & \md 0.000023 &     0.937243 &     0.219798 &     0.006999 &     0.000054 &     0.000098\\
   &    SkewNormal &     0.218591 &     1.081675 & \md 0.000028 &     0.923719 &     0.237190 &     0.008551 &     0.000062 &     0.000228\\
   &   Exponential &     0.229525 &     1.201967 & \md 0.000037 &     1.001438 &     0.232001 &     0.009666 &     0.000090 &     0.000489\\
\midrule
15 &        Normal &     0.585529 &          --- & \md 0.000182 &          --- &     0.609475 &     0.022943 &     0.000392 &     0.001717\\
   &        Cauchy &     0.676690 &          --- & \md 0.020609 &          --- &     0.692426 &     0.166416 &     0.084469 &     0.097217\\
   &       Uniform &     0.575809 &          --- & \md 0.000277 &          --- &     0.596498 &     0.027329 &     0.000737 &     0.002582\\
   &        t-Dist &     0.579684 &          --- & \md 0.000160 &          --- &     0.588454 &     0.022043 &     0.000365 &     0.001557\\
   &    SkewNormal &     0.581892 &          --- & \md 0.000200 &          --- &     0.619134 &     0.027594 &     0.000437 &     0.003568\\
   &   Exponential &     0.607952 &          --- & \md 0.000343 &          --- &     0.622455 &     0.033453 &     0.000909 &     0.006960\\
\midrule
20 &        Normal &     1.143799 &          --- & \md 0.001079 &          --- &     1.177983 &     0.052060 &     0.002086 &     0.006032\\
   &        Cauchy &     1.110807 &          --- & \md 0.049412 &          --- &     1.125693 &     0.369281 &     0.158725 &     0.099501\\
   &       Uniform &     1.024786 &          --- & \md 0.001421 &          --- &     1.052585 &     0.060773 &     0.003030 &     0.007248\\
   &        t-Dist &     1.131364 &          --- & \md 0.001026 &          --- &     1.148654 &     0.050696 &     0.002042 &     0.005657\\
   &    SkewNormal &     1.141836 &          --- & \md 0.001293 &          --- &     1.178756 &     0.070894 &     0.002138 &     0.008604\\
   &   Exponential &     1.151532 &          --- & \md 0.002333 &          --- &     1.180201 &     0.126961 &     0.004631 &     0.014133\\
\bottomrule
\end{tabular}}
\caption{Mean relative error (MRE) for the approximation of the zonoid depth, $n = 1000$ data points, $N \approx 1000$ projections.}
\label{tab.MRE_ZD_1000a}
\end{table}



For the halfspace depth and normally distributed data in dimension $d=5$, when $N\approx1000$ projections are used, the best methods are NM and RRS, followed be CD, whereas the worst methods are RS, GS and RaSi. Even though there is one case (dataset 15) where RRS and NM found the exact halfspace depth, MAE suggests that on average the best halfspace found by RRS or NM contains three points more than the optimal halfspace. Furthermore, relative depth approximation error remains (again on an average) below $2\%$ of the exact depth value.




For the zonoid depth, when the approximation was done using $N\approx1000$ projections, RRS gets the first place (when $d>5$) followed by CD, NM and SA. An important point to note is that the very basic methods like RS and RaSi are unusable when dimension is high with relative error rates beyond $50\%$ ($d=15$) or even beyond $100\%$ ($d=20$). It is noteworthy that the same holds for RRS when the number of projections is low (see the Supplementary Materials), which suggests that RRS needs a substantial number of projections to work well. However, the more elaborate methods like CD, NM and SA perform well regardless of the number of projections. With these methods, relative errors can be kept reasonably low, even far below $1\%$ (except the Cauchy distribution) of the exact value.

\section{Guidelines for practitioners}

The current article shows that even depths that require substantial burden for exact computation can be (potentially) well approximated even in higher dimensions in reasonable time. For this, exploiting the geometry of the unit hyper-sphere is definitely advantageous (see, e.g., parameter tuning of CD or NM in Section~\ref{ssec:tuning}). Further, methods based on random projections are clearly outperformed by those launching optimization over the surface of the hyper-sphere. Among the latter ones, NM performs the best, closely followed by CD and RRS. For the optimization techniques, the direction from the sample average to the point of interest seems to be a good initial argument. For the considered depth notions, the time complexity of all algorithms is $O(Nn)$ only, i.e. linear in both number of sample points and random directions, while the running time of the algorithms is very small (on average always below $0.05$ second for $1000$ directions), see Table~\ref{tab:times} for the run time of the algorithms. At the same time, approximation precision seems to be high as well.

\begin{table}[ht]\footnotesize
\centering
\begin{tabular}{rccccccccc}
\toprule
\multicolumn{2}{c}{} & \multicolumn{8}{c}{Approximation algorithm} \\
\cmidrule{3-10}
$d$ & Depth & RS & GS & RRS & RGS & RaSi & SA & CD & NM \\ 
\midrule
 & MD & 0.007 & 0.004 & 0.007 & 0.003 & 0.008 & 0.007 & 0.006 & 0.007 \\ 
 & ZD & 0.014 & 0.007 & 0.015 & 0.006 & 0.016 & 0.015 & 0.015 & 0.014 \\ 
$5$ & HD & 0.009 & 0.004 & 0.007 & 0.003 & 0.010 & 0.007 & 0.006 & 0.007 \\ 
 & PD & 0.027 & 0.015 & 0.026 & 0.011 & 0.029 & 0.027 & 0.027 & 0.026 \\ 
 & APD & 0.020 & 0.011 & 0.020 & 0.008 & 0.022 & 0.020 & 0.019 & 0.019 \\
\midrule
 & MD & 0.011 & 0.001 & 0.010 & 0.001 & 0.014 & 0.011 & 0.010 & 0.012 \\ 
 & ZD & 0.018 & 0.001 & 0.019 & 0.002 & 0.021 & 0.020 & 0.018 & 0.018 \\ 
$10$ & HD & 0.013 & 0.001 & 0.010 & 0.001 & 0.017 & 0.011 & 0.010 & 0.010 \\ 
 & PD & 0.031 & 0.001 & 0.030 & 0.003 & 0.035 & 0.031 & 0.030 & 0.030 \\ 
 & APD & 0.024 & 0.001 & 0.024 & 0.002 & 0.029 & 0.024 & 0.023 & 0.025 \\
\midrule
 & MD & 0.012 & --- & 0.012 & --- & 0.016 & 0.012 & 0.012 & 0.012 \\ 
 & ZD & 0.020 & --- & 0.020 & --- & 0.025 & 0.021 & 0.020 & 0.021 \\ 
$15$ & HD & 0.015 & --- & 0.012 & --- & 0.019 & 0.012 & 0.011 & 0.012 \\ 
 & PD & 0.033 & --- & 0.032 & --- & 0.037 & 0.033 & 0.032 & 0.032 \\ 
 & APD & 0.026 & --- & 0.025 & --- & 0.030 & 0.026 & 0.025 & 0.024 \\
\midrule
 & MD & 0.014 & --- & 0.014 & --- & 0.020 & 0.014 & 0.013 & 0.014 \\ 
 & ZD & 0.022 & --- & 0.022 & --- & 0.028 & 0.023 & 0.021 & 0.022 \\ 
$20$ & HD & 0.017 & --- & 0.014 & --- & 0.022 & 0.015 & 0.013 & 0.014 \\ 
 & PD & 0.035 & --- & 0.034 & --- & 0.041 & 0.035 & 0.033 & 0.034 \\ 
 & APD & 0.028 & --- & 0.027 & --- & 0.034 & 0.029 & 0.026 & 0.026 \\
\bottomrule  
\end{tabular}
\caption{Average time (in seconds) of the algorithms for the budget of $1000$ directions, over $1000$ repetitions. Averaging is also performed over all the six distributions, since running times are independent of distributions. GS and RGS cannot be run in dimensions $d=15$ and $d=20$, while in dimension $d=10$ only a very sparse grid (11 and 110 directions for GS and RGS, respectively) satisfies the condition $N\le 1000$.}\label{tab:times}
\end{table}

In applications, if possible, it is recommended to first fine-tune the method using either available real data or similar simulated ones. If computational budget allows, it is further advised to benchmark several methods, since they are comparable while having the upward bias, as in Section~\ref{ssec:tuning}.

It is necessary to emphasize the limitations of the entire simulation study. First of all, the presented results and the accompanying analysis is --- strictly speaking --- valid only for statistical processes which are similar enough to the six considered distributions. Further, the behavior of the explored performance indicators is unpredictable outside the considered parameter ranges since it can have a non-linear character; this holds for the tuning procedure in Section~\ref{ssec:tuning} as well. Also, the study is restricted to a sample size of $n=1000$ observations, while considered dimensions $d=5,10,15,20$ and numbers of random directions $N=100,1000,10000$ are somewhat liming as well. Likewise, one should be careful when interpreting results of the aggregated (averaging) statistics since they might hide information that could be of use in particular cases. Finally, the approximation algorithms are compared with each other with respect to minimum achieved depth, while exact depth values --- being unknown (in most of the cases) --- are not addressed, and are only studied for the halfspace depth (in $d=5$) and the zonoid depth in Section~\ref{sec:exact}.

The source codes of implementations of the methods described in Section~\ref{sec:Algorithms} and the reproducing scripts of the experiments, as well as results of the fine-tuning simulation study from Section~\ref{ssec:tuning} are gathered in the Supplementary Materials. The disk space occupied by the complete results of the main simulation study of Section~\ref{ssec:comparison} is $79.3$ \texttt{GB} (with $3.7$ \texttt{GB} for the generated data, and $0.7$ \texttt{GB}, $6.8$ \texttt{GB} and $68.1$ \texttt{GB} for the simulation results with $100$, $1000$ and $10000$ directions, respectively) and thus cannot be uploaded online. Nevertheless, these results in the compact form sufficient to reproduce all the illustrations and tables of the article (including the Supplementary Materials) can be obtained upon request from the authors.

\section*{Supplementary materials}
Supplementary materials to this article include:
\begin{itemize}
	\item {\bf Additional figures and tables:} Additional figures and tables illustrating more comprehensive results of the experimental study, i.e. illustrations to the fine-tuning, simulation results and approximation error. 
	\item {\bf Reproducing codes:} \texttt{C++} codes of all the methods from Section~\ref{sec:Algorithms} as well as the scripts for reproduction of experiments. 
	\item {\bf Experimental results:} Results of the fine-tuning simulation study of Section~\ref{ssec:tuning}. 
\end{itemize}

\section*{Acknowledgements}
The research of Stanislav Nagy was supported by the Czech Science Foundation
[grant number 19-16097Y] and by the PRIMUS/17/SCI/3 project of Charles University. 




\clearpage

\setcounter{section}{0}
\setcounter{figure}{0}
\setcounter{table}{0}
\renewcommand{\thesection}{S.\arabic{section}}
\renewcommand{\thesubsection}{S.\arabic{section}.\arabic{subsection}}
\renewcommand{\thefigure}{S.\arabic{figure}}
\renewcommand{\thetable}{S.\arabic{table}}

\begin{center}
	{\LARGE Supplementary Materials to the article \\
	\indent\\
	``Approximate computation of projection depths''}
	
	\indent\\	 \indent\\
	
	{\Large Rainer Dyckerhoff, Pavlo Mozharovskyi, Stanislav Nagy}
\end{center}

First, these Supplementary Materials include a visualization of the average depth ranks (\textsc{AveRank}) and the percentages of attaining the lowest depth (\textsc{PercBest}) over the considered parameter sets in Section~\ref{asec:finetuning}, which also includes Table~\ref{atab:parsChosen} with description of the considered parameters of the algorithms. Further, results of the main simulation study are gathered in Section~\ref{asec:simresults}: \textsc{AveRank} and \textsc{PercBest} statistics (Section~\ref{assec:simcharts}), tables indicating the best performing methods for different depths/dimensions/distributions in sense of both \textsc{AveRank} and \textsc{PercBest} (Section~\ref{assec:simtables}), and graphs of the development of the average difference of the approximated depth and the lowest achieved depth (Section~\ref{assec:simgraphs}). Finally, Section~\ref{asec:error} contains error analysis in comparison with the exact depth, for the halfspace and zonoid depths.

\section{Illustrations to the fine-tuning}\label{asec:finetuning}

\begin{table}[h!]\small
	\begin{center}
	\begin{tabular}{llll}
	\toprule
		Method & Parameters & Values & Description\\
        \midrule
		RRS  & $N_{\text{ref}}$ & $5$, $\boldsymbol{10}$, $15$    & Number of refinement steps\\
		     & $\alpha$ & $0.2$, $\boldsymbol{0.5}$, $0.8$        & Shrinking factor of the spherical cap\\ 
        \midrule                                            
		RGS  & $N_{\text{ref}}$ & $5$, $\boldsymbol{10}$, $15$    & Number of refinement steps\\                
		     & $\alpha$ & $0.2$, $\boldsymbol{0.5}$, $0.8$        & Shrinking factor of the spherical cap\\    
        \midrule                                            
		RaSi & $\alpha$ & $\boldsymbol{1.25}$, $1.5$, $1.8$       & Parameter of the Dirichlet distribution\\
        \midrule
		SA   & $\alpha$ & $0.5$, $0.8$, $\boldsymbol{0.95}$       & Cooling factor\\
		     & $\beta$ & $5$, $\boldsymbol{10}$, $15$             & Size of the spherical cap\\
		     & $\texttt{Start}$ & \texttt{\bf Mn}, $\texttt{Rn}$  & Starting value (mean, random)\\
        \midrule
		CD   & $\texttt{Space}$ & $\texttt{Ec}$, \texttt{\bf Sp}  & Euclidean space or sphere\\
		     & $\texttt{LS}$ & $\texttt{Eq}$, \texttt{\bf GS}     & Line search: equally spaced or golden section\\
        \midrule
		NM   & $\texttt{Space}$ & $\texttt{Ec}$, \texttt{\bf Sp}  & Euclidean space or sphere\\ 
		     & $\texttt{Start}$ & \texttt{\bf Mn}, $\texttt{Rn}$  & Starting value (mean, random)\\
		     & $\beta$ & $\boldsymbol{1}$, $2$, $4$               & Size of the spherical cap\\
		     & $\texttt{Bound}$ & \texttt{\bf y}, $\texttt{n}$    & Bound movement on great circles (yes/no)\\
        \bottomrule
	\end{tabular}
	\end{center}
	\caption{Considered and selected (in bold) parameters, obtained by the fine-tuning.}
	\label{atab:parsChosen}
\end{table}

\clearpage

\begin{figure}[h!]\center
	\includegraphics[width=0.85\textwidth]{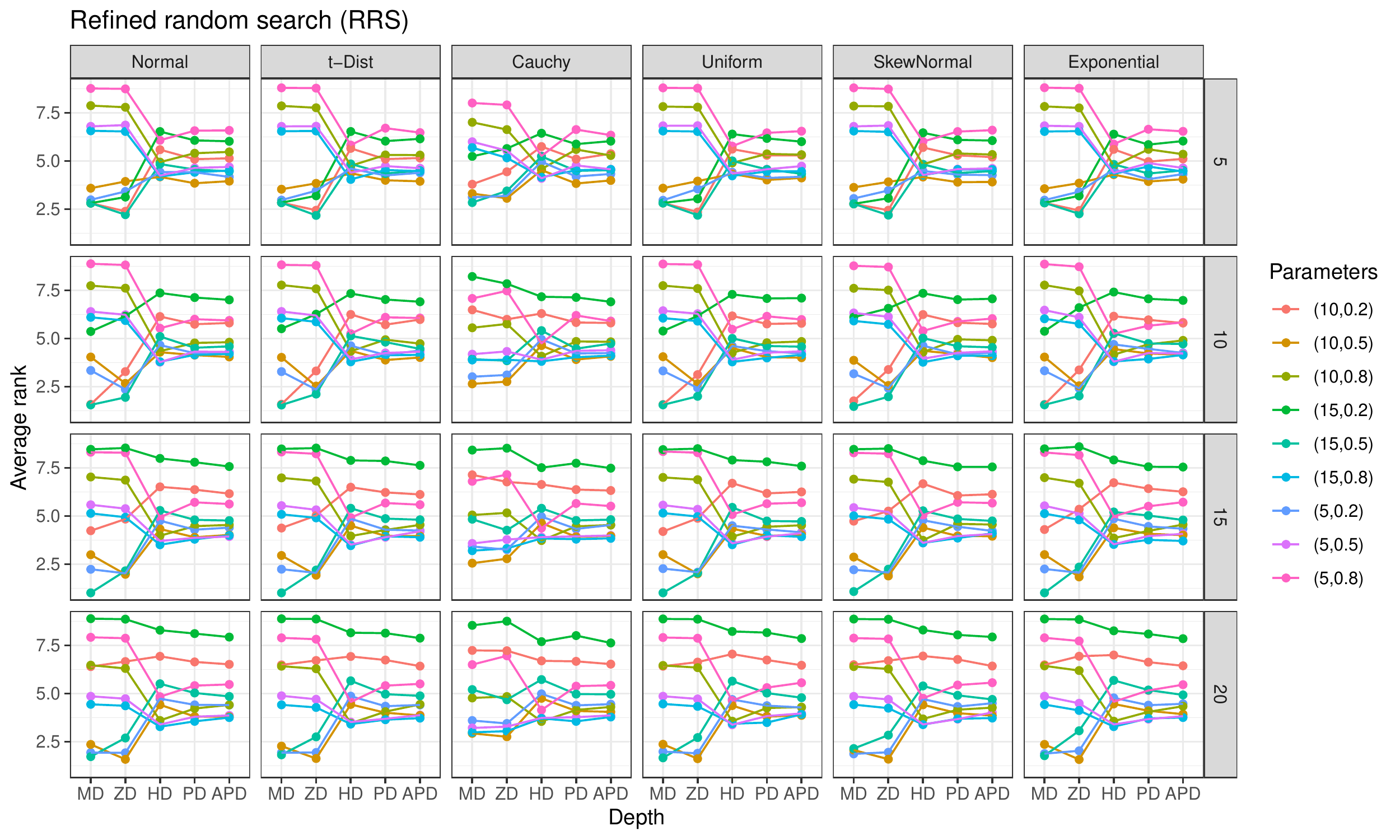}
	\caption{\textsc{AveRank} for the refined random search (RRS) algorithm. Parameters are: Number of refinement steps ($N_{\text{ref}}\in\{5, 10, 15\}$); Shrinking factor of the spherical cap ($\alpha\in\{0.2, 0.5, 0.8\}$).}
\end{figure}

\begin{figure}[h!]\center
	\includegraphics[width=0.85\textwidth]{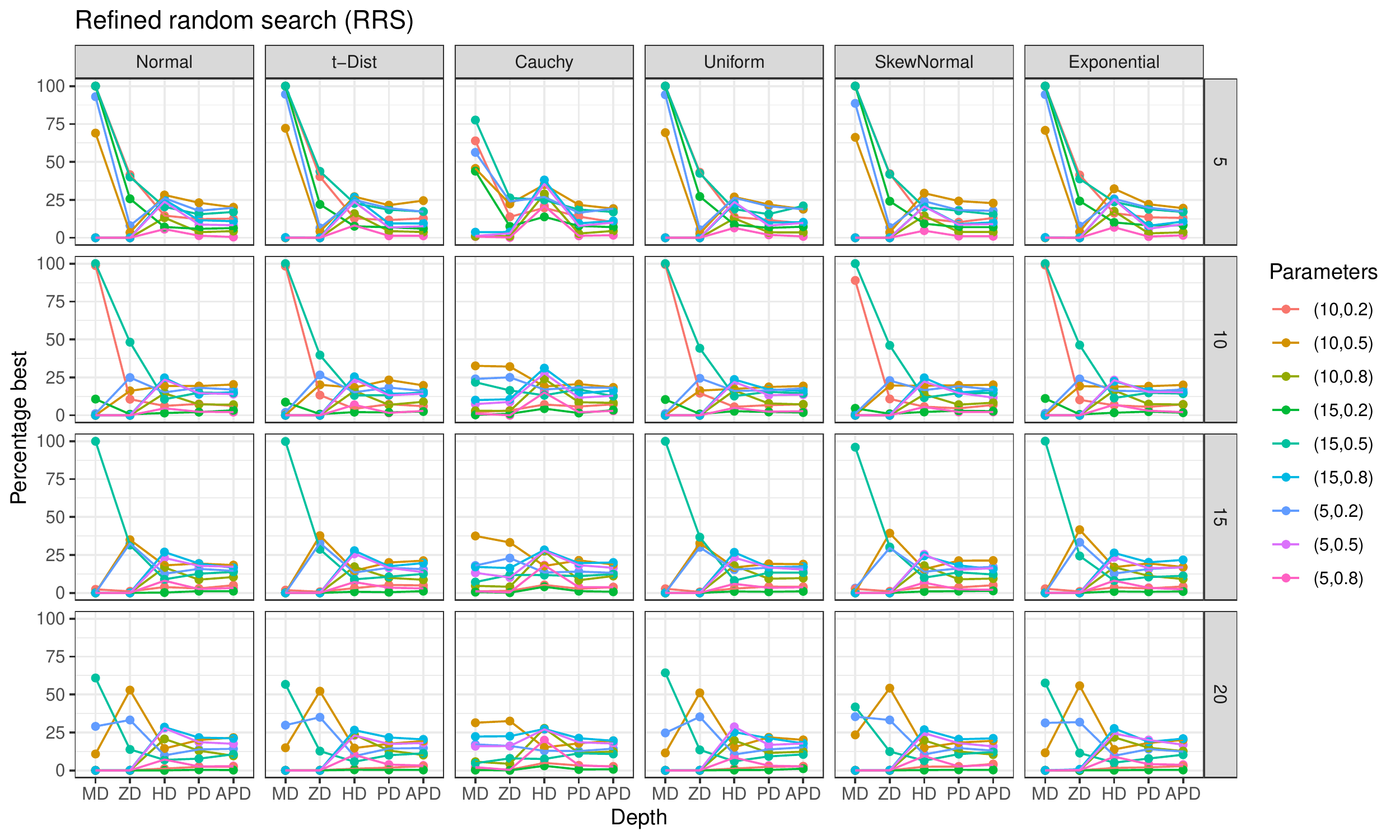}
	\caption{\textsc{PercBest} for the refined random search (RRS) algorithm. Parameters are: Number of refinement steps ($N_{\text{ref}}\in\{5, 10, 15\}$); Shrinking factor of the spherical cap ($\alpha\in\{0.2, 0.5, 0.8\}$).}
\end{figure}

\begin{figure}[h!]\center
	\includegraphics[width=0.9\textwidth]{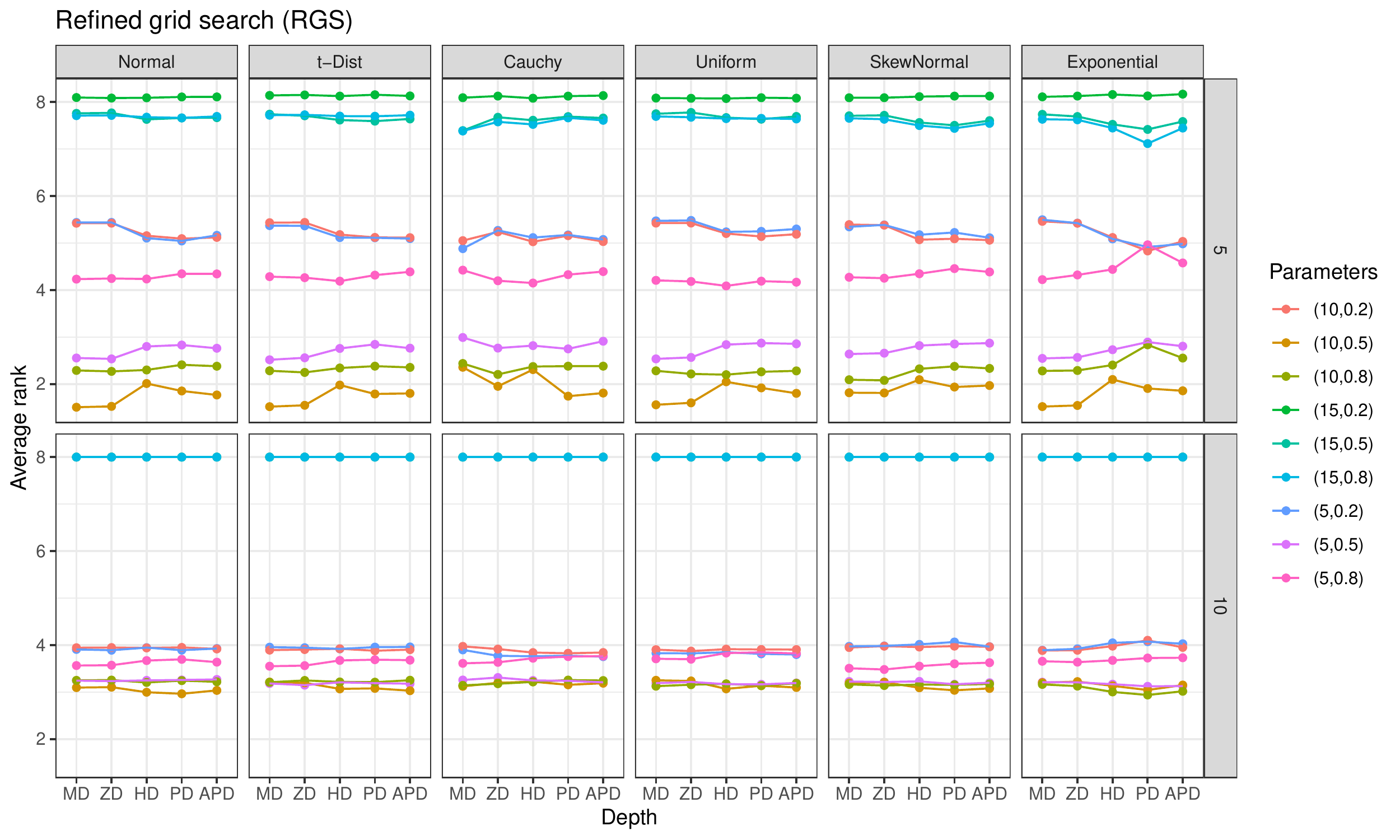}
	\caption{\textsc{AveRank} for the refined grid search (RGS) algorithm. Parameters are: Number of refinement steps ($N_{\text{ref}}\in\{5, 10, 15\}$); Shrinking factor of the spherical cap ($\alpha\in\{0.2, 0.5, 0.8\}$).}
\end{figure}

\begin{figure}[h!]\center
	\includegraphics[width=0.9\textwidth]{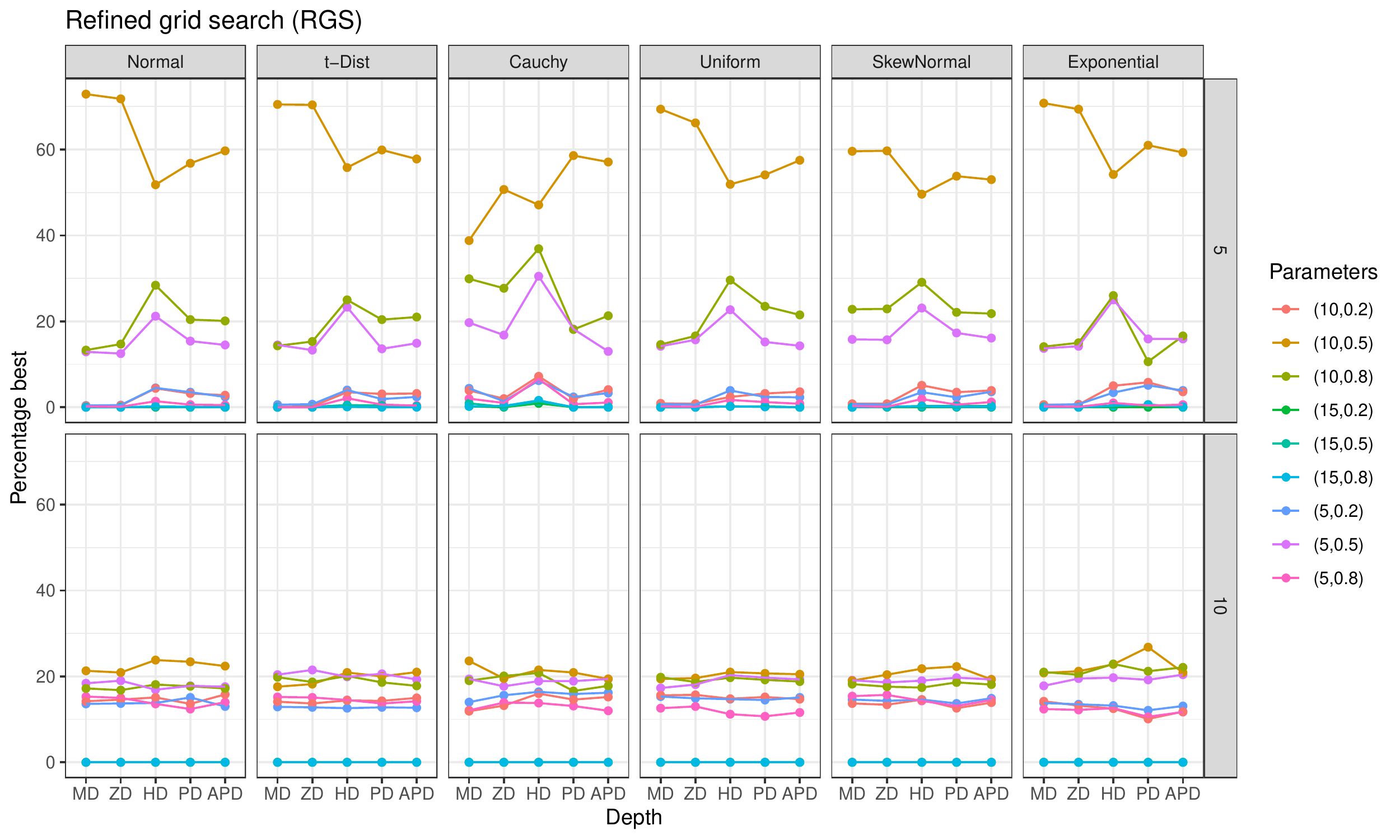}
	\caption{\textsc{PercBest} for the refined grid search (RGS) algorithm. Parameters are: Number of refinement steps ($N_{\text{ref}}\in\{5, 10, 15\}$); Shrinking factor of the spherical cap ($\alpha\in\{0.2, 0.5, 0.8\}$).}
\end{figure}

\begin{figure}[h!]\center
	\includegraphics[width=0.95\textwidth]{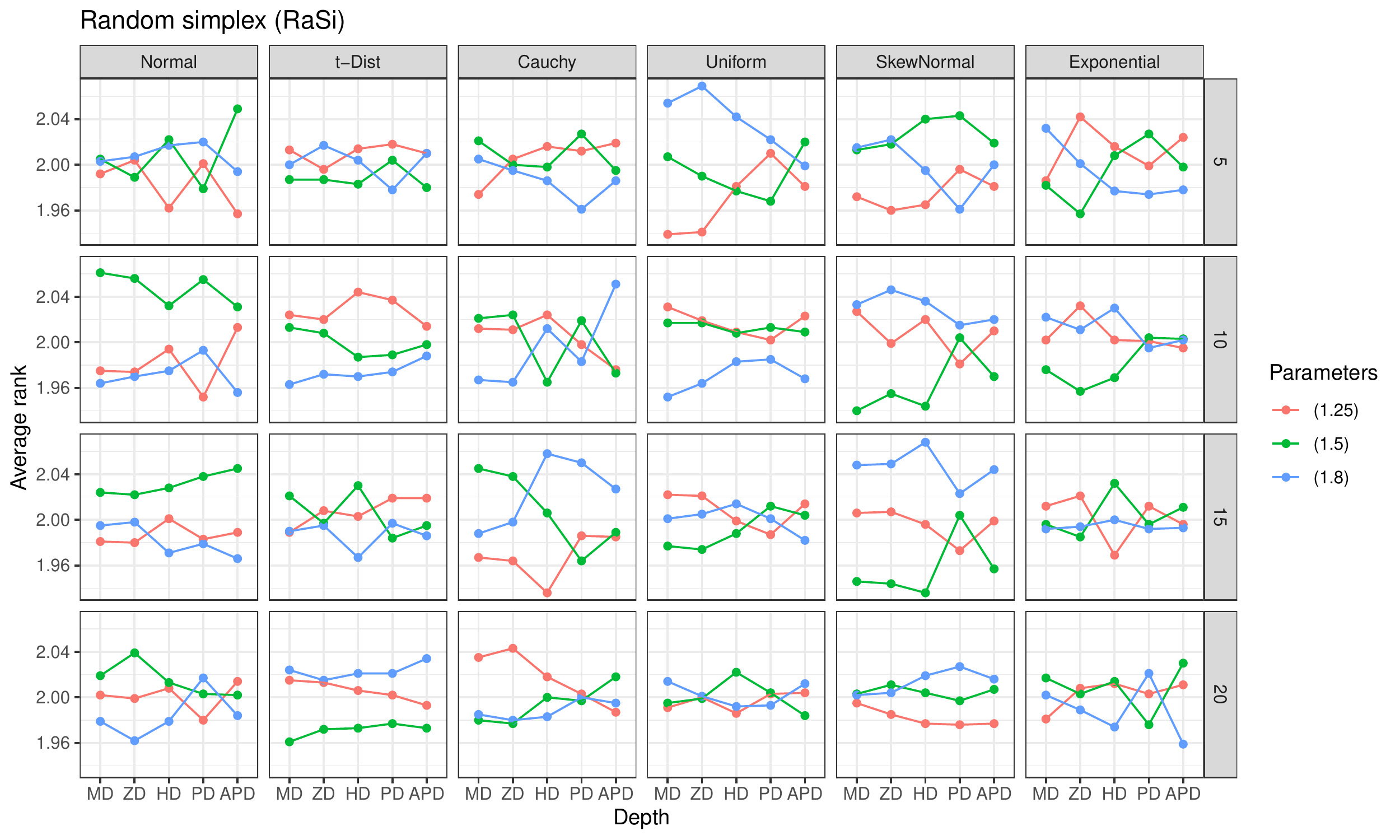}
	\caption{\textsc{AveRank} for the random simplices (RaSi) algorithm. Parameter is: Parameter of the Dirichlet distribution ($\alpha\in\{1.25, 1.5, 1.8\}$).}
\end{figure}

\begin{figure}[h!]\center
	\includegraphics[width=0.95\textwidth]{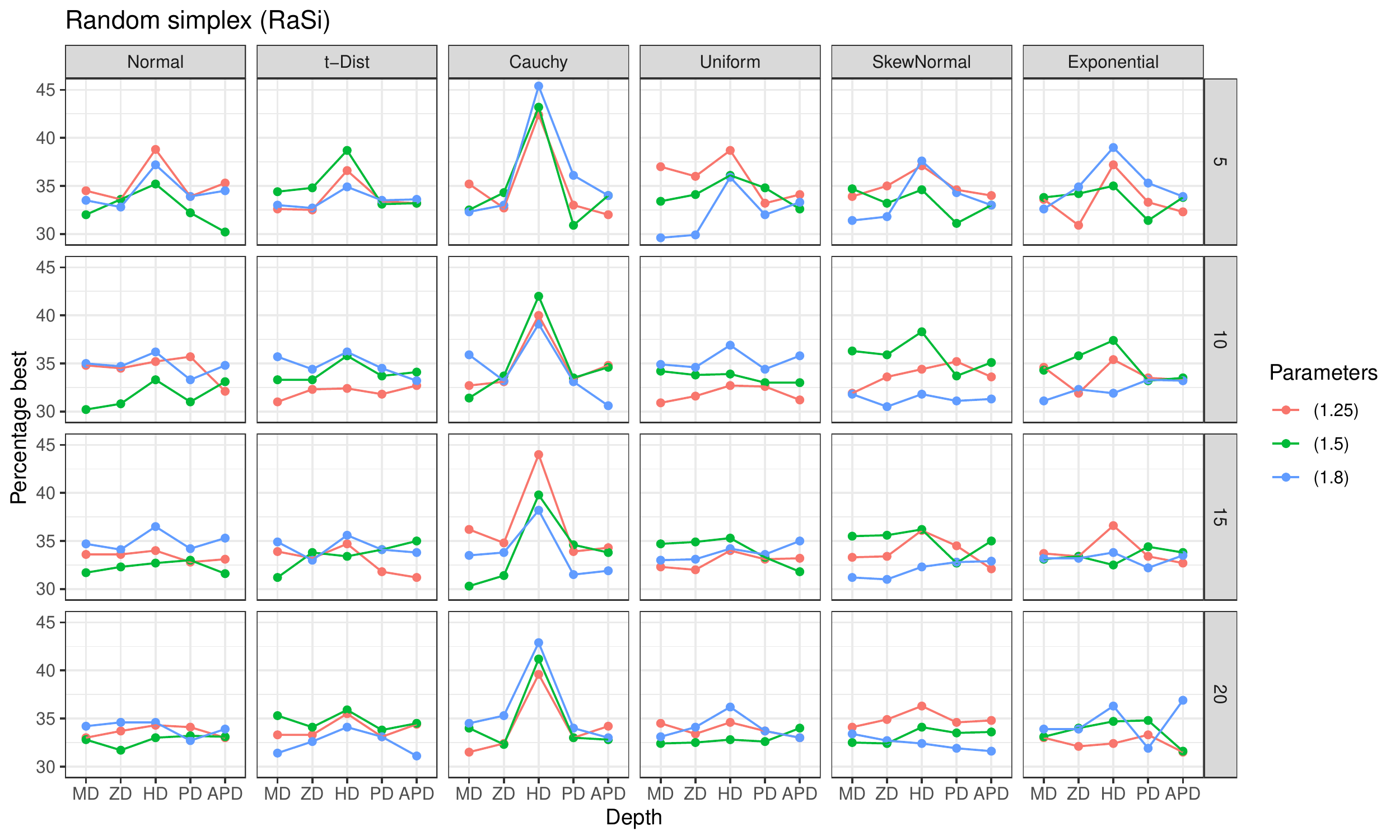}
	\caption{\textsc{PercBest} for the random simplices (RaSi) algorithm. Parameter is: Parameter of the Dirichlet distribution ($\alpha\in\{1.25, 1.5, 1.8\}$).}
\end{figure}

\begin{figure}[h!]\center
	\includegraphics[width=0.9\textwidth]{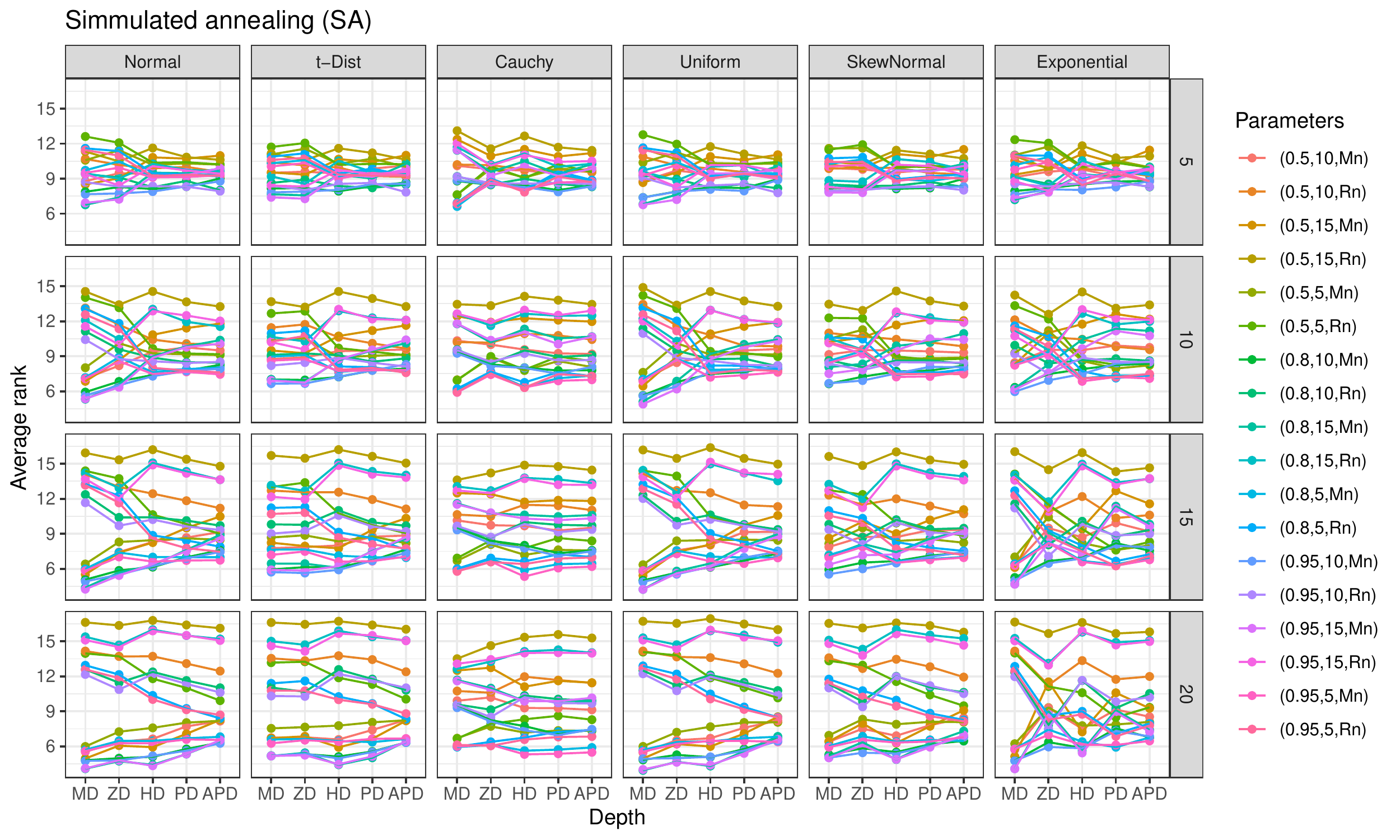}
	\caption{\textsc{AveRank} for the simulated annealing (SA) algorithm. Parameters are: Cooling factor ($\alpha\in\{0.5, 0.8, 0.95\}$); Size of the spherical cap ($\beta\in\{5, 10, 15\}$); Starting value (mean, random) ($\texttt{Start}\in\{\text{Mn}, \text{Rn}\}$).}
\end{figure}

\begin{figure}[h!]\center
	\includegraphics[width=0.9\textwidth]{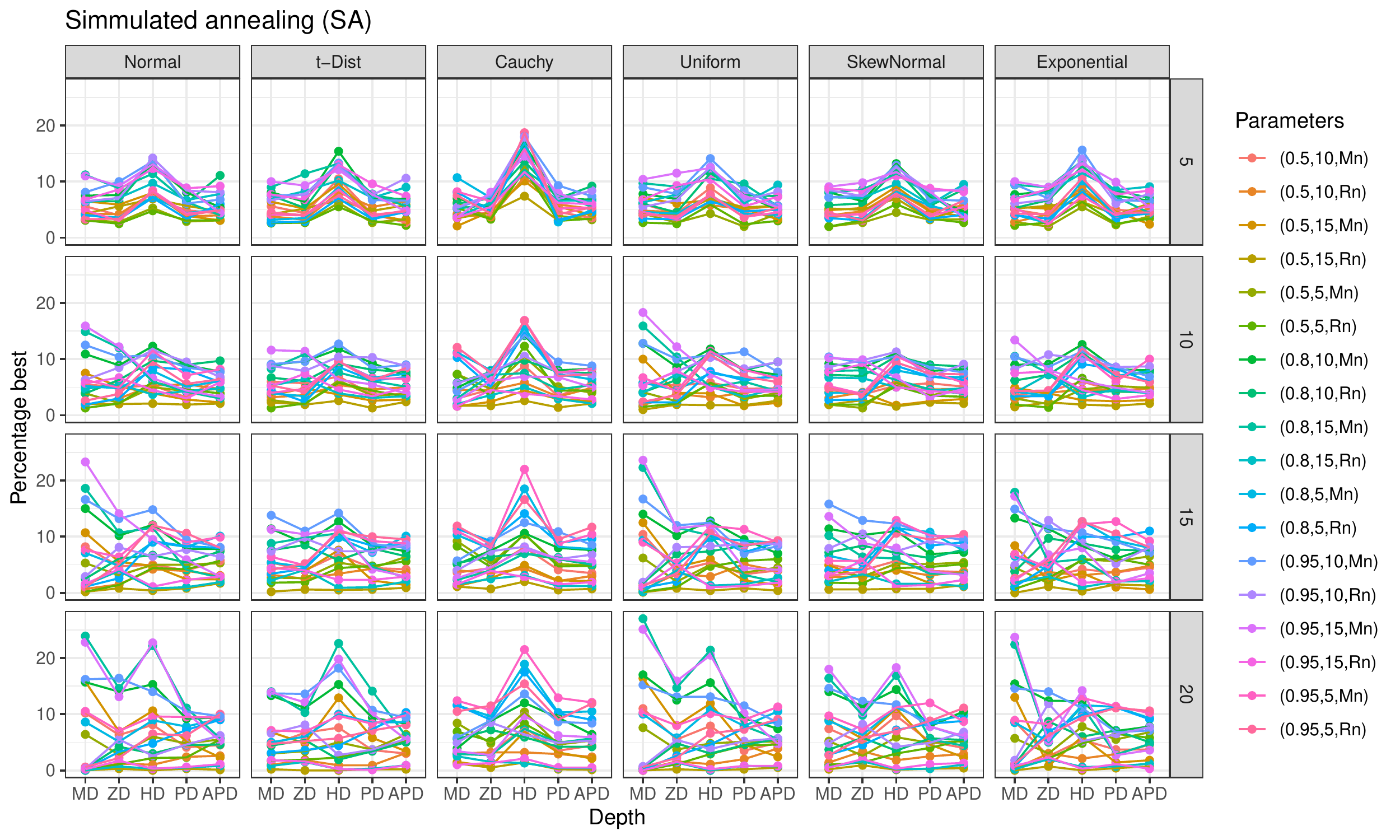}
	\caption{\textsc{PercBest} for the simulated annealing (SA) algorithm. Parameters are: Cooling factor ($\alpha\in\{0.5, 0.8, 0.95\}$); Size of the spherical cap ($\beta\in\{5, 10, 15\}$); Starting value (mean, random) ($\texttt{Start}\in\{\text{Mn}, \text{Rn}\}$).}
\end{figure}

\begin{figure}[h!]\center
	\includegraphics[width=0.9\textwidth]{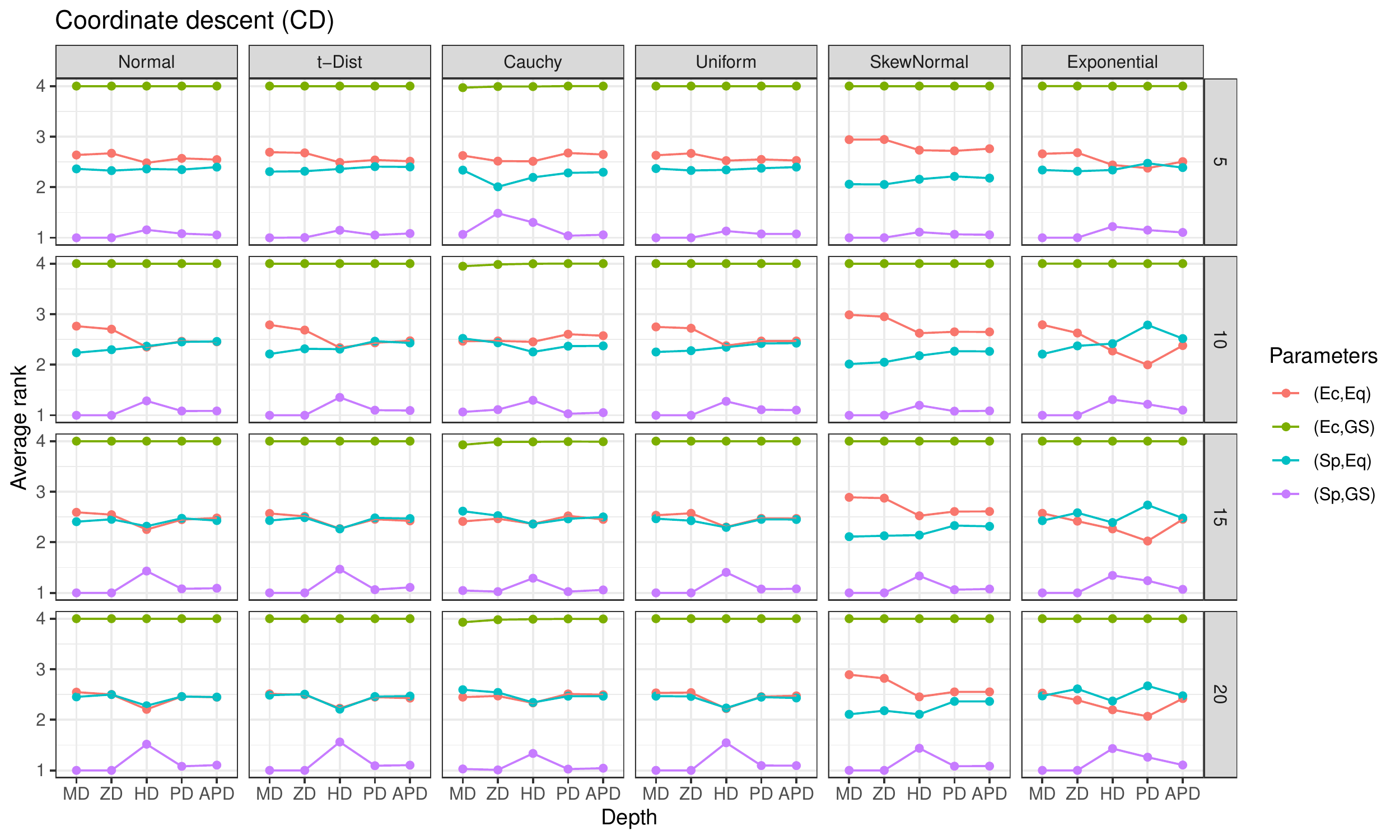}
	\caption{\textsc{AveRank} for the coordinate descent (CD) algorithm. Parameters are: Euclidean space or sphere ($\texttt{Space}\in\{\text{Ec}, \text{Sp}\}$); Line search: equally spaced or golden section ($\texttt{LS}\in\{\text{Eq}, \text{GS}\}$).}
\end{figure}

\begin{figure}[h!]\center
	\includegraphics[width=0.9\textwidth]{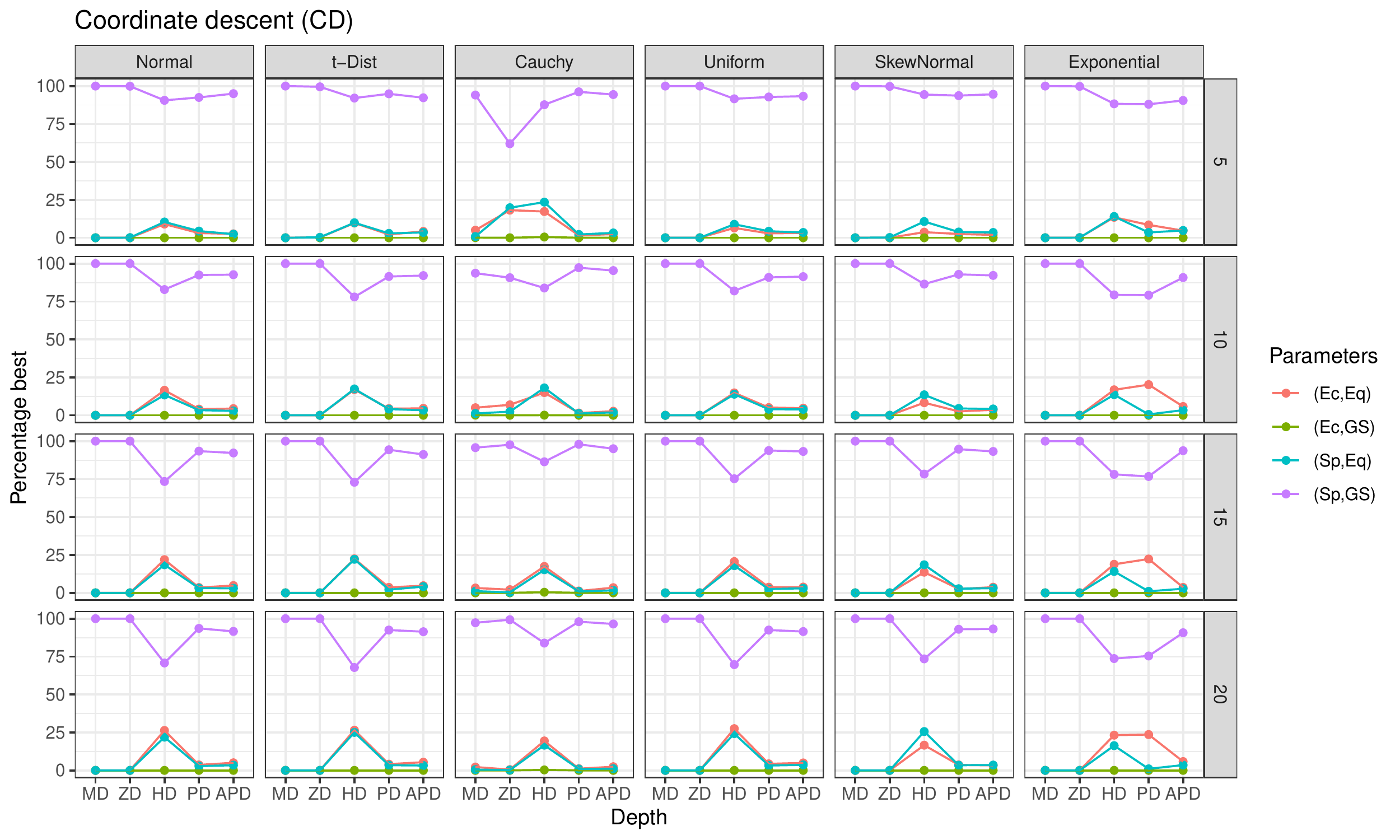}
	\caption{\textsc{PercBest} for the coordinate descent (CD) algorithm. Parameters are: Euclidean space or sphere ($\texttt{Space}\in\{\text{Ec}, \text{Sp}\}$); Line search: equally spaced or golden section ($\texttt{LS}\in\{\text{Eq}, \text{GS}\}$).}
\end{figure}

\begin{figure}[h!]\center
	\includegraphics[width=0.85\textwidth]{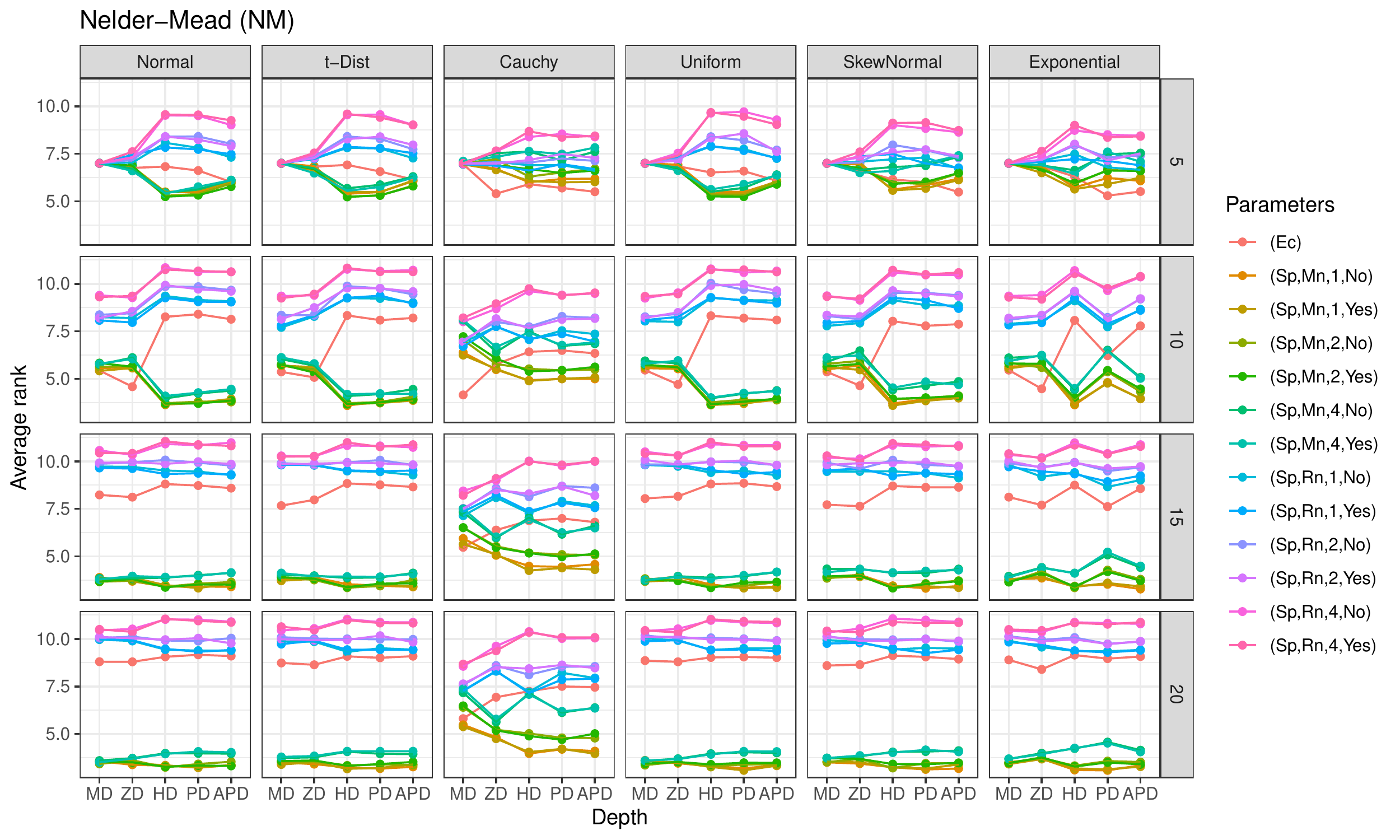}
	\caption{\textsc{AveRank} for the Nelder-Mead (NM) algorithm. Parameters are: Euclidean space or sphere ($\texttt{Space}\in\{\text{Ec}, \text{Sp}\}$); Starting value (mean, random) ($\texttt{Start}\in\{\text{Mn}, \text{Rn}\}$); Size of the spherical cap ($\beta\in\{1, 2, 4\}$); Bound movement on great circles (yes/no) ($\texttt{Bound}\in\{\text{y}, \text{n}\}$).}
\end{figure}

\begin{figure}[h!]\center
	\includegraphics[width=0.85\textwidth]{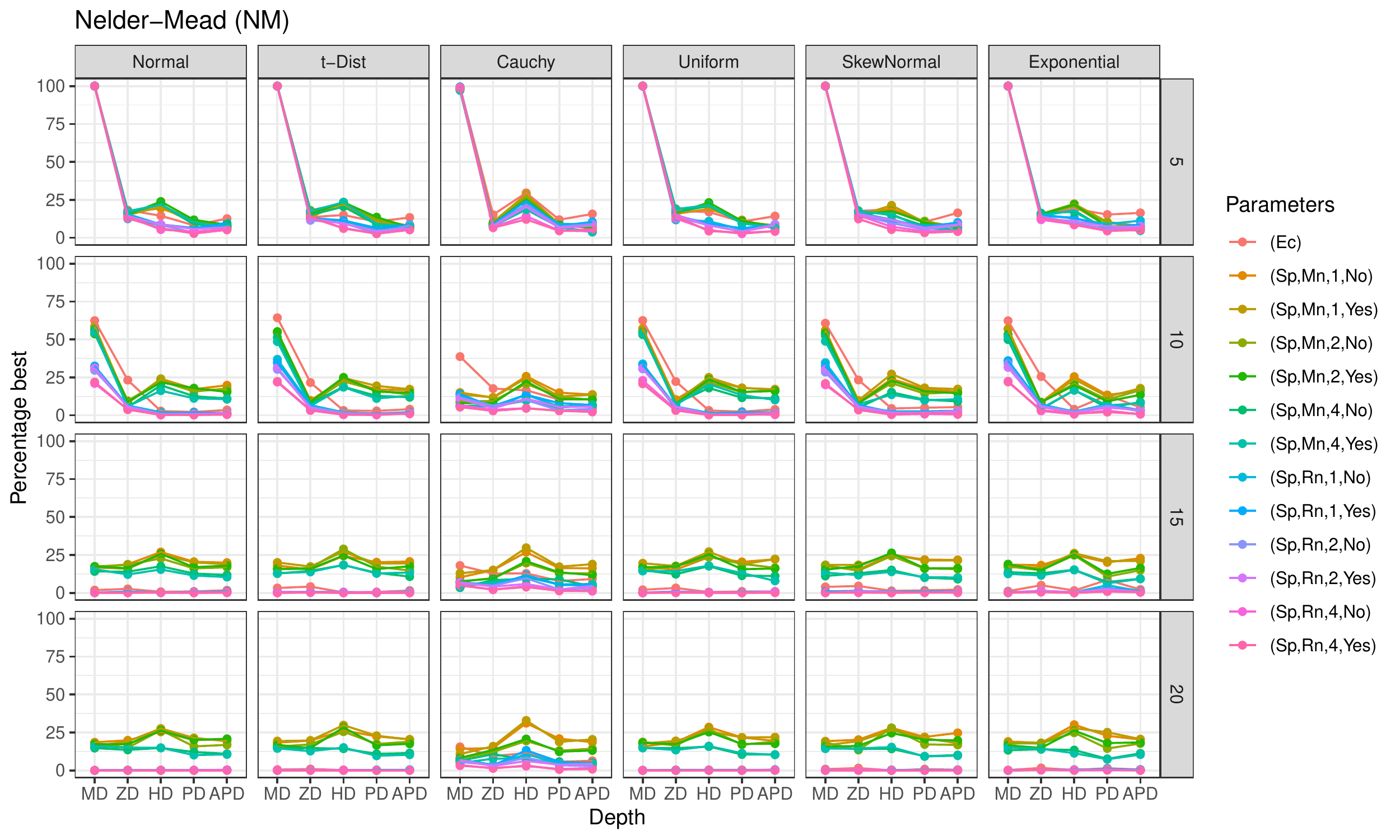}
	\caption{\textsc{PercBest} for the Nelder-Mead (NM) algorithm. Parameters are: Euclidean space or sphere ($\texttt{Space}\in\{\text{Ec}, \text{Sp}\}$); Starting value (mean, random) ($\texttt{Start}\in\{\text{Mn}, \text{Rn}\}$); Size of the spherical cap ($\beta\in\{1, 2, 4\}$); Bound movement on great circles (yes/no) ($\texttt{Bound}\in\{\text{y}, \text{n}\}$).}
\end{figure}

\clearpage

\section{Illustration on the simulation results}\label{asec:simresults}

\subsection{Figures of average ranks (\textsc{AveRank}) and percentage of achieving lowest depth (\textsc{PercBest})}\label{assec:simcharts}

\begin{figure}[h!]\center
	\includegraphics[width=0.775\textwidth]{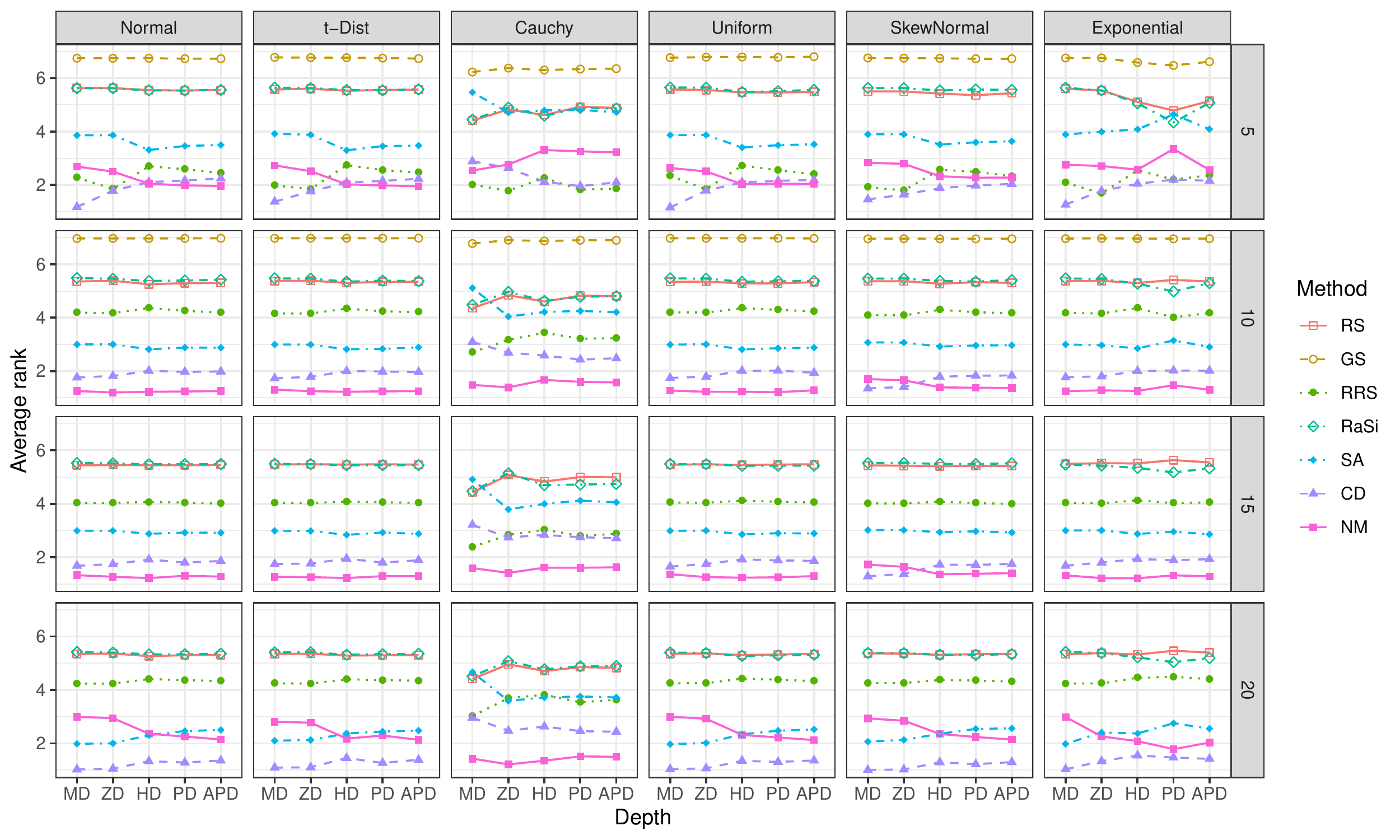}
	\caption{\textsc{AveRank} statistics for the eight considered approximation methods when using the parameter settings from Section~\ref{ssec:tuning} (see also Table~\ref{tab:parsChosen}) and $N\approx100$ projections.}
\end{figure}

\begin{figure}[h!]\center
	\includegraphics[width=0.775\textwidth]{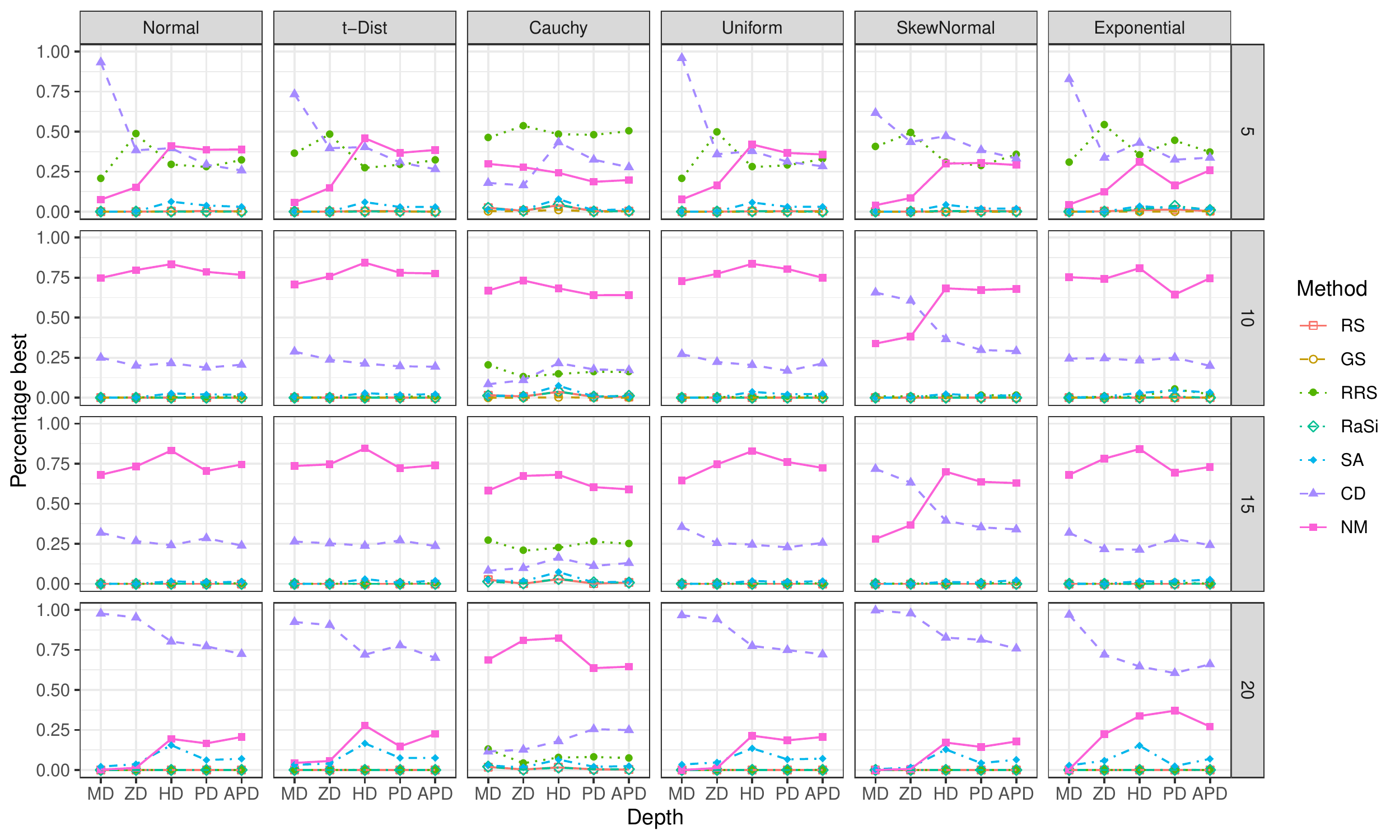}
	\caption{\textsc{PercBest} statistics for the eight considered approximation methods when using the parameter settings from Section~\ref{ssec:tuning} (see also Table~\ref{tab:parsChosen}) and $N\approx100$ projections.}
\end{figure}

\begin{figure}[h!]\center
	\includegraphics[width=0.95\textwidth]{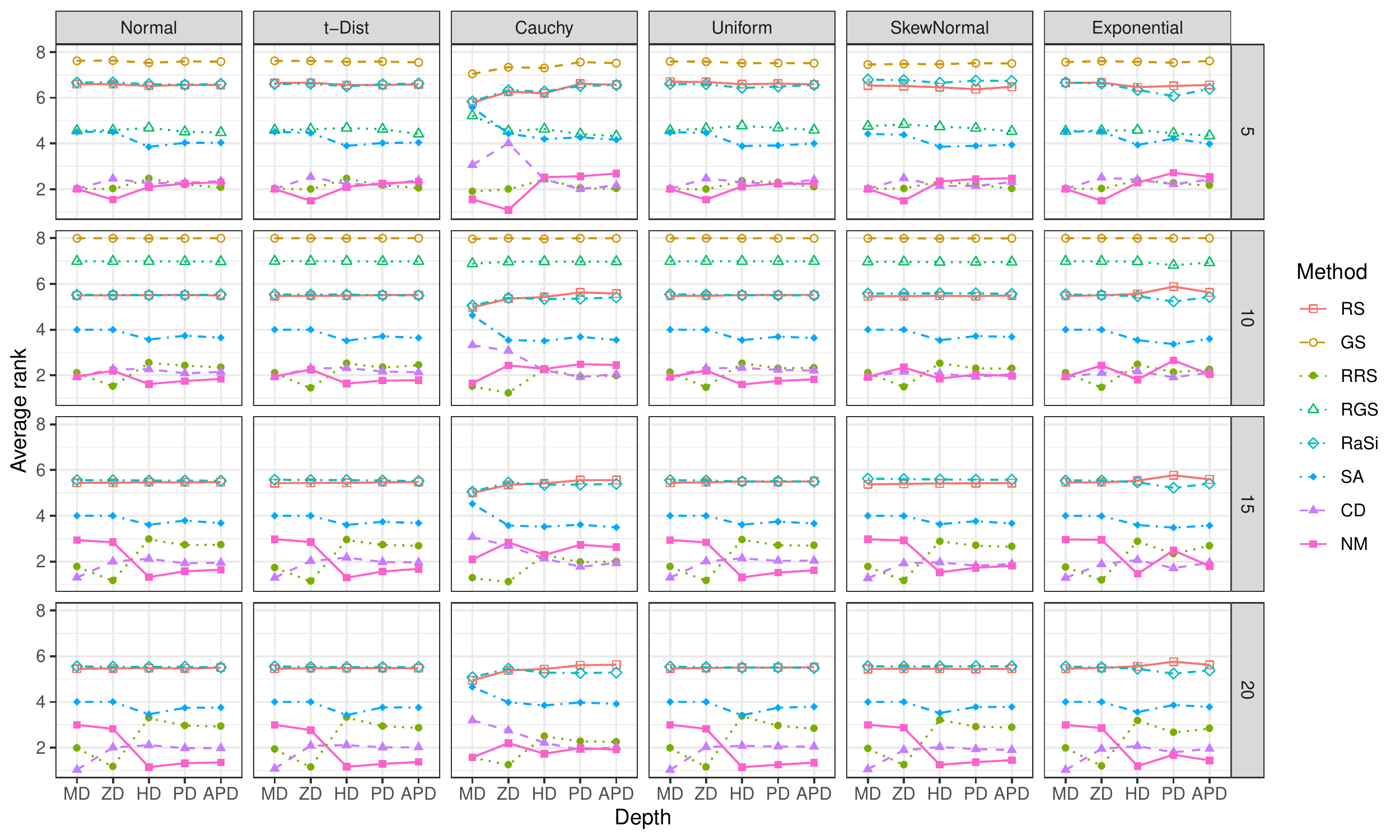}
	\caption{\textsc{AveRank} statistics for the eight considered approximation methods when using the parameter settings from Section~\ref{ssec:tuning} (see also Table~\ref{tab:parsChosen}) and $N\approx1000$ projections.}
\end{figure}

\begin{figure}[h!]\center
	\includegraphics[width=0.95\textwidth]{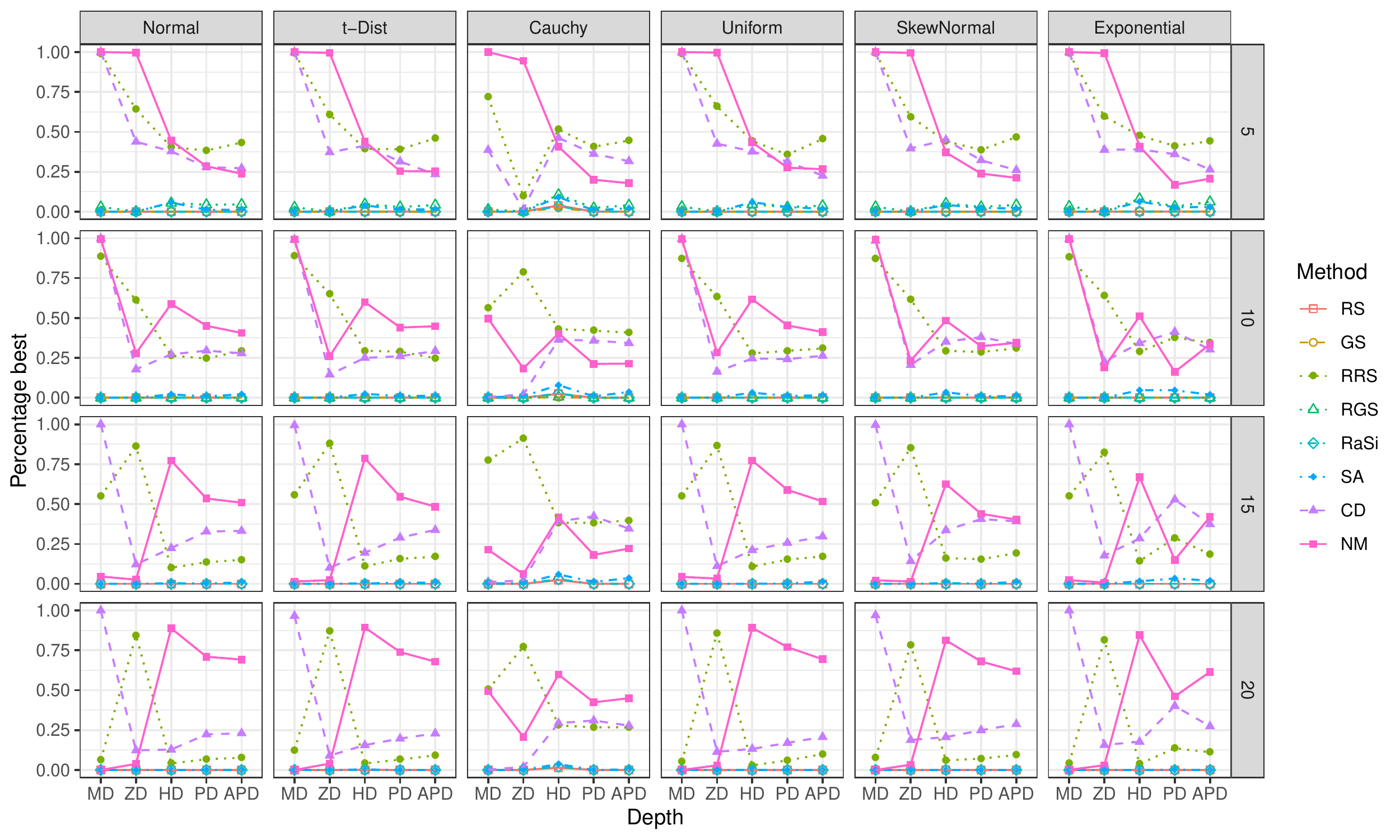}
	\caption{\textsc{PercBest} statistics for the eight considered approximation methods when using the parameter settings from Section~\ref{ssec:tuning} (see also Table~\ref{tab:parsChosen}) and $N\approx1000$ projections.}
\end{figure}

\begin{figure}[h!]\center
	\includegraphics[width=0.95\textwidth]{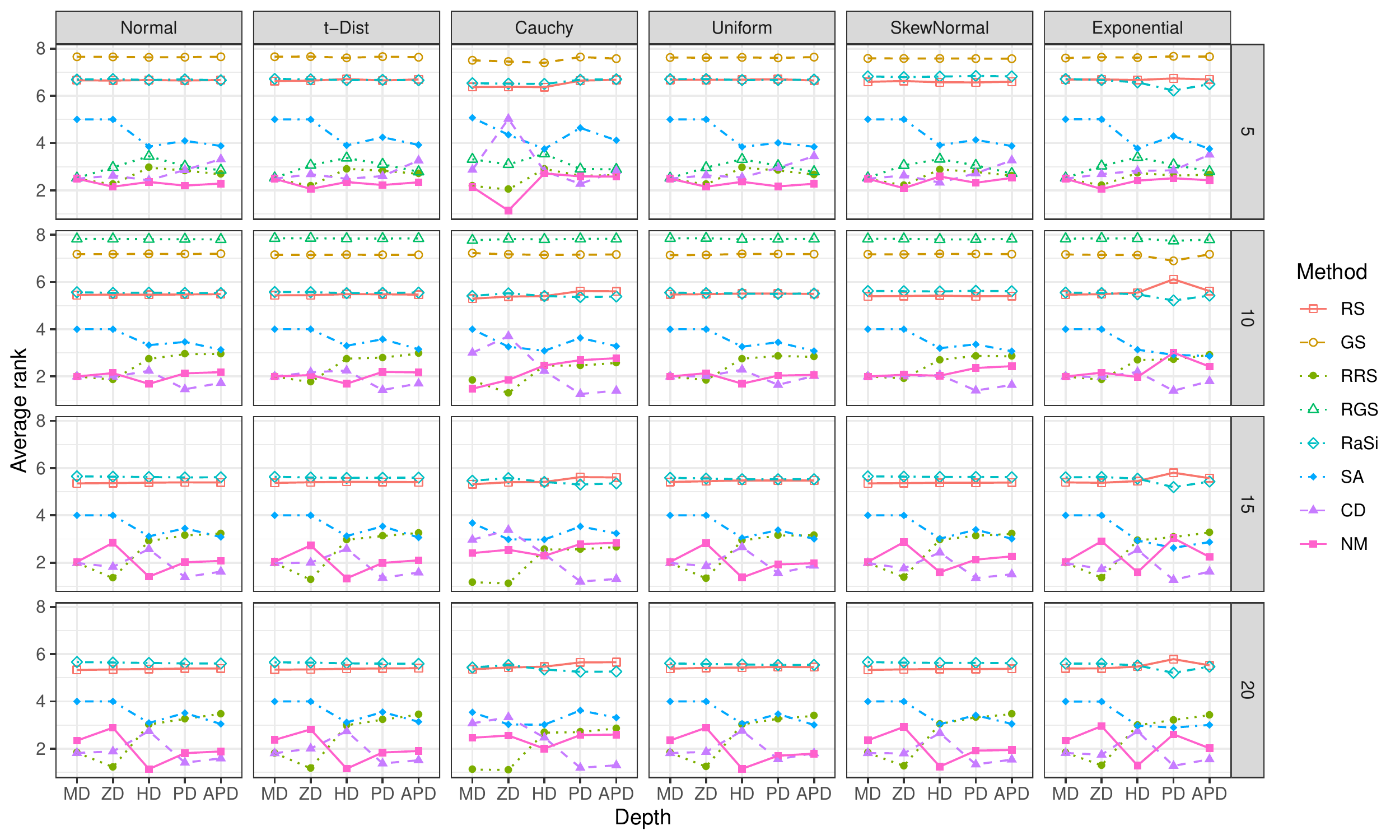}
	\caption{\textsc{AveRank} statistics for the eight considered approximation methods when using the parameter settings from Section~\ref{ssec:tuning} (see also Table~\ref{tab:parsChosen}) and $N\approx10000$ projections.}
\end{figure}

\begin{figure}[h!]\center
	\includegraphics[width=0.95\textwidth]{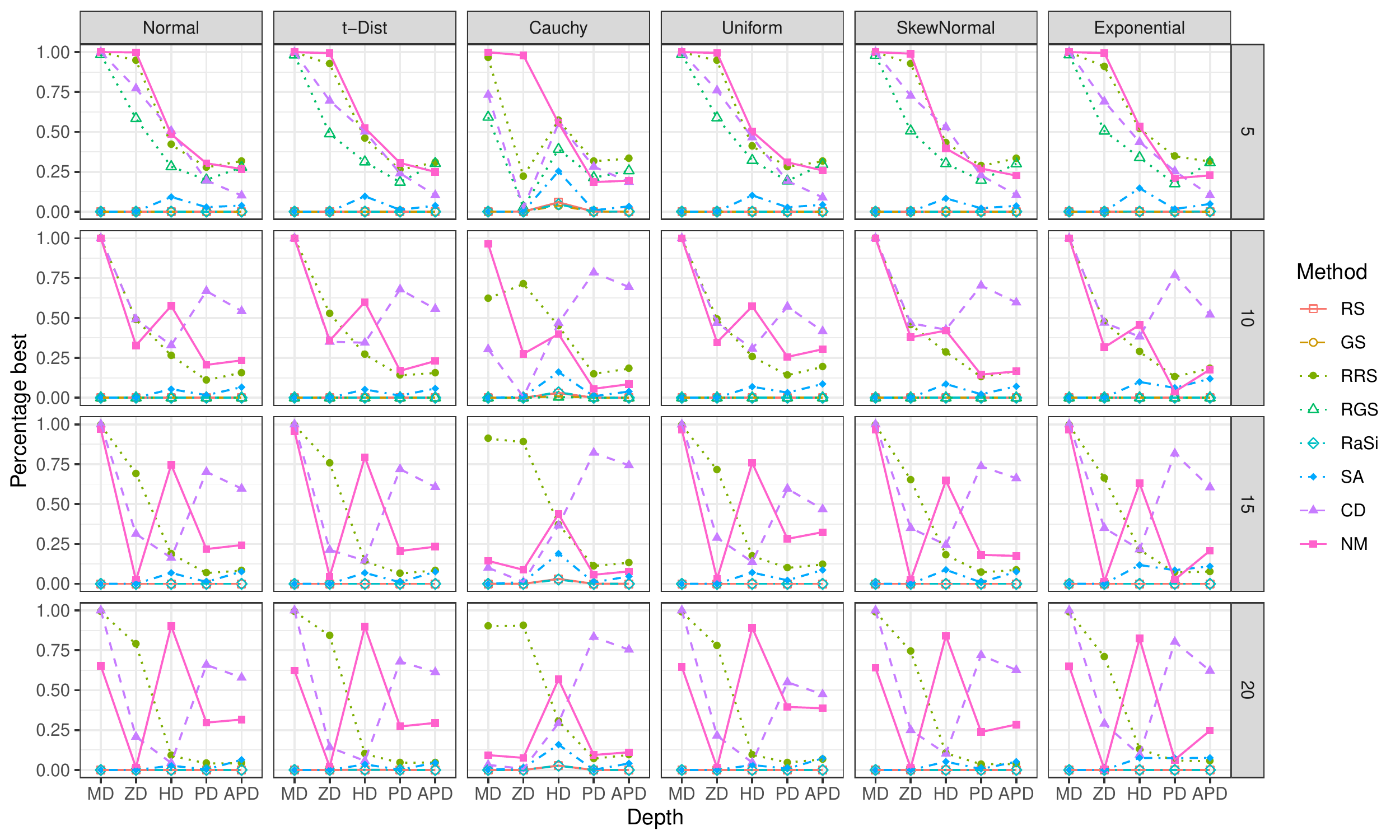}
	\caption{\textsc{PercBest} statistics for the eight considered approximation methods when using the parameter settings from Section~\ref{ssec:tuning} (see also Table~\ref{tab:parsChosen}) and $N\approx10000$ projections.}
\end{figure}

\clearpage

\subsection{Tables of best methods 
in terms of average ranks (\textsc{AveRank}) and percentage of achieving lowest depth (\textsc{PercBest})}\label{assec:simtables}

\begin{table}[ht]\small
\begin{center}
\begin{tabular}{llcccccccccccc}
\toprule
 & & \mc{12}{c}{Number of projections} \\
\cmidrule{3-14}
MD & & \mc{4}{c}{100} & \mc{4}{c}{1000} & \mc{4}{c}{10000} \\
\cmidrule(rl){3-6} \cmidrule(rl){7-10} \cmidrule(rl){11-14} \\
$d$ & Distribution & \rotatebox{90}{RRS} & \rotatebox{90}{SA} & \rotatebox{90}{CD} & \rotatebox{90}{NM} & \rotatebox{90}{RRS} & \rotatebox{90}{SA} & \rotatebox{90}{CD} & \rotatebox{90}{NM} & \rotatebox{90}{RRS} & \rotatebox{90}{SA} & \rotatebox{90}{CD} & \rotatebox{90}{NM} \\
\midrule
5 & Normal &   &   & $\bestRankPerc$ &   &   &   & $\bestRankPerc$ &   & $\bestRankPerc$ &   &   &   \\
 & t-Dist &   &   & $\bestRankPerc$ &   &   &   & $\bestRankPerc$ &   & $\bestRankPerc$ &   &   &   \\
 & Cauchy & $\bestRankPerc$ &   &   &   &   &   &   & $\bestRankPerc$ &   &   &   & $\bestRankPerc$ \\
 & Uniform &   &   & $\bestRankPerc$ &   &   &   &   & $\bestRankPerc$ & $\bestRankPerc$ &   &   &   \\
 & SkewNormal &   &   & $\bestRankPerc$ &   &   &   & $\bestRankPerc$ &   & $\bestRankPerc$ &   &   &   \\
 & Exponential &   &   & $\bestRankPerc$ &   &   &   & $\bestRankPerc$ &   & $\bestRankPerc$ &   &   &   \\
\midrule
10 & Normal &   &   &   & $\bestRankPerc$ &   &   & $\bestRankPerc$ &   & $\bestRankPerc$ &   &   &   \\
 & t-Dist &   &   &   & $\bestRankPerc$ &   &   &   & $\bestRankPerc$ & $\bestRankPerc$ &   &   &   \\
 & Cauchy &   &   &   & $\bestRankPerc$ & $\bestRankPerc$ &   &   &   &   &   &   & $\bestRankPerc$ \\
 & Uniform &   &   &   & $\bestRankPerc$ &   &   & $\bestRankPerc$ &   &   &   & $\bestRankPerc$ &   \\
 & SkewNormal &   &   & $\bestRankPerc$ &   &   &   &   & $\bestRankPerc$ & $\bestRankPerc$ &   &   &   \\
 & Exponential &   &   &   & $\bestRankPerc$ &   &   & $\bestRankPerc$ &   &   &   & $\bestRankPerc$ &   \\
\midrule
15 & Normal &   &   &   & $\bestRankPerc$ &   &   & $\bestRankPerc$ &   &   &   & $\bestRankPerc$ &   \\
 & t-Dist &   &   &   & $\bestRankPerc$ &   &   & $\bestRankPerc$ &   &   &   & $\bestRankPerc$ &   \\
 & Cauchy &   &   &   & $\bestRankPerc$ & $\bestRankPerc$ &   &   &   & $\bestRankPerc$ &   &   &   \\
 & Uniform &   &   &   & $\bestRankPerc$ &   &   & $\bestRankPerc$ &   &   &   & $\bestRankPerc$ &   \\
 & SkewNormal &   &   & $\bestRankPerc$ &   &   &   & $\bestRankPerc$ &   &   &   & $\bestRankPerc$ &   \\
 & Exponential &   &   &   & $\bestRankPerc$ &   &   & $\bestRankPerc$ &   &   &   & $\bestRankPerc$ &   \\
\midrule
20 & Normal &   &   & $\bestRankPerc$ &   &   &   & $\bestRankPerc$ &   &   &   & $\bestRankPerc$ &   \\
 & t-Dist &   &   & $\bestRankPerc$ &   &   &   & $\bestRankPerc$ &   &   &   & $\bestRankPerc$ &   \\
 & Cauchy &   &   &   & $\bestRankPerc$ & $\bestRankPerc$ &   &   &   & $\bestRankPerc$ &   &   &   \\
 & Uniform &   &   & $\bestRankPerc$ &   &   &   & $\bestRankPerc$ &   &   &   & $\bestRankPerc$ &   \\
 & SkewNormal &   &   & $\bestRankPerc$ &   &   &   & $\bestRankPerc$ &   &   &   & $\bestRankPerc$ &   \\
 & Exponential &   &   & $\bestRankPerc$ &   &   &   & $\bestRankPerc$ &   &   &   & $\bestRankPerc$ &   \\
\bottomrule
\end{tabular}
\end{center}
\caption{Best performing methods in sense of \textsc{AveRank} (filled circle) for the Mahalanobis depth (MD). If the best method in view of \textsc{PercBest} differs, it is indicated by an empty circle.}\label{tab:bestm_MD}
\end{table}

\begin{table}[ht]\small
\begin{center}
\begin{tabular}{llcccccccccccc}
\toprule
 & & \mc{12}{c}{Number of projections} \\
\cmidrule{3-14}
ZD & & \mc{4}{c}{100} & \mc{4}{c}{1000} & \mc{4}{c}{10000} \\
\cmidrule(rl){3-6} \cmidrule(rl){7-10} \cmidrule(rl){11-14} \\
$d$ & Distribution & \rotatebox{90}{RRS} & \rotatebox{90}{SA} & \rotatebox{90}{CD} & \rotatebox{90}{NM} & \rotatebox{90}{RRS} & \rotatebox{90}{SA} & \rotatebox{90}{CD} & \rotatebox{90}{NM} & \rotatebox{90}{RRS} & \rotatebox{90}{SA} & \rotatebox{90}{CD} & \rotatebox{90}{NM} \\
\midrule
5 & Normal & $\bestPerc$ &   & $\bestRank$ &   &   &   &   & $\bestRankPerc$ &   &   &   & $\bestRankPerc$ \\
 & t-Dist & $\bestPerc$ &   & $\bestRank$ &   &   &   &   & $\bestRankPerc$ &   &   &   & $\bestRankPerc$ \\
 & Cauchy & $\bestRankPerc$ &   &   &   &   &   &   & $\bestRankPerc$ &   &   &   & $\bestRankPerc$ \\
 & Uniform & $\bestPerc$ &   & $\bestRank$ &   &   &   &   & $\bestRankPerc$ &   &   &   & $\bestRankPerc$ \\
 & SkewNormal & $\bestPerc$ &   & $\bestRank$ &   &   &   &   & $\bestRankPerc$ &   &   &   & $\bestRankPerc$ \\
 & Exponential & $\bestRankPerc$ &   &   &   &   &   &   & $\bestRankPerc$ &   &   &   & $\bestRankPerc$ \\
\midrule
10 & Normal &   &   &   & $\bestRankPerc$ & $\bestRankPerc$ &   &   &   & $\bestRank$ &   & $\bestPerc$ &   \\
 & t-Dist &   &   &   & $\bestRankPerc$ & $\bestRankPerc$ &   &   &   & $\bestRankPerc$ &   &   &   \\
 & Cauchy &   &   &   & $\bestRankPerc$ & $\bestRankPerc$ &   &   &   & $\bestRankPerc$ &   &   &   \\
 & Uniform &   &   &   & $\bestRankPerc$ & $\bestRankPerc$ &   &   &   & $\bestRankPerc$ &   &   &   \\
 & SkewNormal &   &   & $\bestRankPerc$ &   & $\bestRankPerc$ &   &   &   & $\bestRank$ &   & $\bestPerc$ &   \\
 & Exponential &   &   &   & $\bestRankPerc$ & $\bestRankPerc$ &   &   &   & $\bestRankPerc$ &   &   &   \\
\midrule
15 & Normal &   &   &   & $\bestRankPerc$ & $\bestRankPerc$ &   &   &   & $\bestRankPerc$ &   &   &   \\
 & t-Dist &   &   &   & $\bestRankPerc$ & $\bestRankPerc$ &   &   &   & $\bestRankPerc$ &   &   &   \\
 & Cauchy &   &   &   & $\bestRankPerc$ & $\bestRankPerc$ &   &   &   & $\bestRankPerc$ &   &   &   \\
 & Uniform &   &   &   & $\bestRankPerc$ & $\bestRankPerc$ &   &   &   & $\bestRankPerc$ &   &   &   \\
 & SkewNormal &   &   & $\bestRankPerc$ &   & $\bestRankPerc$ &   &   &   & $\bestRankPerc$ &   &   &   \\
 & Exponential &   &   &   & $\bestRankPerc$ & $\bestRankPerc$ &   &   &   & $\bestRankPerc$ &   &   &   \\
\midrule
20 & Normal &   &   & $\bestRankPerc$ &   & $\bestRankPerc$ &   &   &   & $\bestRankPerc$ &   &   &   \\
 & t-Dist &   &   & $\bestRankPerc$ &   & $\bestRankPerc$ &   &   &   & $\bestRankPerc$ &   &   &   \\
 & Cauchy &   &   &   & $\bestRankPerc$ & $\bestRankPerc$ &   &   &   & $\bestRankPerc$ &   &   &   \\
 & Uniform &   &   & $\bestRankPerc$ &   & $\bestRankPerc$ &   &   &   & $\bestRankPerc$ &   &   &   \\
 & SkewNormal &   &   & $\bestRankPerc$ &   & $\bestRankPerc$ &   &   &   & $\bestRankPerc$ &   &   &   \\
 & Exponential &   &   & $\bestRankPerc$ &   & $\bestRankPerc$ &   &   &   & $\bestRankPerc$ &   &   &   \\
\bottomrule
\end{tabular}
\end{center}
\caption{Best performing methods in sense of \textsc{AveRank} (filled circle) for the zonoid depth (ZD). If the best method in view of \textsc{PercBest} differs, it is indicated by an empty circle.}\label{tab:bestm_ZD}
\end{table}

\begin{table}[ht]\small
\begin{center}
\begin{tabular}{llcccccccccccc}
\toprule
 & & \mc{12}{c}{Number of projections} \\
\cmidrule{3-14}
HD & & \mc{4}{c}{100} & \mc{4}{c}{1000} & \mc{4}{c}{10000} \\
\cmidrule(rl){3-6} \cmidrule(rl){7-10} \cmidrule(rl){11-14} \\
$d$ & Distribution & \rotatebox{90}{RRS} & \rotatebox{90}{SA} & \rotatebox{90}{CD} & \rotatebox{90}{NM} & \rotatebox{90}{RRS} & \rotatebox{90}{SA} & \rotatebox{90}{CD} & \rotatebox{90}{NM} & \rotatebox{90}{RRS} & \rotatebox{90}{SA} & \rotatebox{90}{CD} & \rotatebox{90}{NM} \\
\midrule
5 & Normal &   &   &   & $\bestRankPerc$ &   &   &   & $\bestRankPerc$ &   &   & $\bestPerc$ & $\bestRank$ \\
 & t-Dist &   &   &   & $\bestRankPerc$ &   &   &   & $\bestRankPerc$ &   &   &   & $\bestRankPerc$ \\
 & Cauchy & $\bestPerc$ &   & $\bestRank$ &   & $\bestPerc$ &   & $\bestRank$ &   & $\bestPerc$ &   &   & $\bestRank$ \\
 & Uniform &   &   &   & $\bestRankPerc$ & $\bestPerc$ &   &   & $\bestRank$ &   &   &   & $\bestRankPerc$ \\
 & SkewNormal &   &   & $\bestRankPerc$ &   &   &   & $\bestRankPerc$ &   &   &   & $\bestRankPerc$ &   \\
 & Exponential &   &   & $\bestRankPerc$ &   & $\bestPerc$ &   &   & $\bestRank$ &   &   &   & $\bestRankPerc$ \\
\midrule
10 & Normal &   &   &   & $\bestRankPerc$ &   &   &   & $\bestRankPerc$ &   &   &   & $\bestRankPerc$ \\
 & t-Dist &   &   &   & $\bestRankPerc$ &   &   &   & $\bestRankPerc$ &   &   &   & $\bestRankPerc$ \\
 & Cauchy &   &   &   & $\bestRankPerc$ & $\bestPerc$ &   & $\bestRank$ &   &   &   & $\bestRankPerc$ &   \\
 & Uniform &   &   &   & $\bestRankPerc$ &   &   &   & $\bestRankPerc$ &   &   &   & $\bestRankPerc$ \\
 & SkewNormal &   &   &   & $\bestRankPerc$ &   &   &   & $\bestRankPerc$ &   &   & $\bestPerc$ & $\bestRank$ \\
 & Exponential &   &   &   & $\bestRankPerc$ &   &   &   & $\bestRankPerc$ &   &   &   & $\bestRankPerc$ \\
\midrule
15 & Normal &   &   &   & $\bestRankPerc$ &   &   &   & $\bestRankPerc$ &   &   &   & $\bestRankPerc$ \\
 & t-Dist &   &   &   & $\bestRankPerc$ &   &   &   & $\bestRankPerc$ &   &   &   & $\bestRankPerc$ \\
 & Cauchy &   &   &   & $\bestRankPerc$ &   &   & $\bestRank$ & $\bestPerc$ &   &   &   & $\bestRankPerc$ \\
 & Uniform &   &   &   & $\bestRankPerc$ &   &   &   & $\bestRankPerc$ &   &   &   & $\bestRankPerc$ \\
 & SkewNormal &   &   &   & $\bestRankPerc$ &   &   &   & $\bestRankPerc$ &   &   &   & $\bestRankPerc$ \\
 & Exponential &   &   &   & $\bestRankPerc$ &   &   &   & $\bestRankPerc$ &   &   &   & $\bestRankPerc$ \\
\midrule
20 & Normal &   &   & $\bestRankPerc$ &   &   &   &   & $\bestRankPerc$ &   &   &   & $\bestRankPerc$ \\
 & t-Dist &   &   & $\bestRankPerc$ &   &   &   &   & $\bestRankPerc$ &   &   &   & $\bestRankPerc$ \\
 & Cauchy &   &   &   & $\bestRankPerc$ &   &   &   & $\bestRankPerc$ &   &   &   & $\bestRankPerc$ \\
 & Uniform &   &   & $\bestRankPerc$ &   &   &   &   & $\bestRankPerc$ &   &   &   & $\bestRankPerc$ \\
 & SkewNormal &   &   & $\bestRankPerc$ &   &   &   &   & $\bestRankPerc$ &   &   &   & $\bestRankPerc$ \\
 & Exponential &   &   & $\bestRankPerc$ &   &   &   &   & $\bestRankPerc$ &   &   &   & $\bestRankPerc$ \\
\bottomrule
\end{tabular}
\end{center}
\caption{Best performing methods in sense of \textsc{AveRank} (filled circle) for the halfspace depth (HD). If the best method in view of \textsc{PercBest} differs, it is indicated by an empty circle.}\label{tab:bestm_HD}
\end{table}

\begin{table}[ht]\small
\begin{center}
\begin{tabular}{llcccccccccccc}
\toprule
 & & \mc{12}{c}{Number of projections} \\
\cmidrule{3-14}
PD & & \mc{4}{c}{100} & \mc{4}{c}{1000} & \mc{4}{c}{10000} \\
\cmidrule(rl){3-6} \cmidrule(rl){7-10} \cmidrule(rl){11-14} \\
$d$ & Distribution & \rotatebox{90}{RRS} & \rotatebox{90}{SA} & \rotatebox{90}{CD} & \rotatebox{90}{NM} & \rotatebox{90}{RRS} & \rotatebox{90}{SA} & \rotatebox{90}{CD} & \rotatebox{90}{NM} & \rotatebox{90}{RRS} & \rotatebox{90}{SA} & \rotatebox{90}{CD} & \rotatebox{90}{NM} \\
\midrule
5 & Normal &   &   &   & $\bestRankPerc$ & $\bestRankPerc$ &   &   &   &   &   &   & $\bestRankPerc$ \\
 & t-Dist &   &   &   & $\bestRankPerc$ & $\bestRankPerc$ &   &   &   &   &   &   & $\bestRankPerc$ \\
 & Cauchy & $\bestRankPerc$ &   &   &   & $\bestPerc$ &   & $\bestRank$ &   & $\bestPerc$ &   & $\bestRank$ &   \\
 & Uniform &   &   &   & $\bestRankPerc$ & $\bestPerc$ &   & $\bestRank$ &   &   &   &   & $\bestRankPerc$ \\
 & SkewNormal &   &   & $\bestRankPerc$ &   & $\bestPerc$ &   & $\bestRank$ &   & $\bestPerc$ &   &   & $\bestRank$ \\
 & Exponential & $\bestPerc$ &   & $\bestRank$ &   & $\bestPerc$ &   & $\bestRank$ &   & $\bestPerc$ &   &   & $\bestRank$ \\
\midrule
10 & Normal &   &   &   & $\bestRankPerc$ &   &   &   & $\bestRankPerc$ &   &   & $\bestRankPerc$ &   \\
 & t-Dist &   &   &   & $\bestRankPerc$ &   &   &   & $\bestRankPerc$ &   &   & $\bestRankPerc$ &   \\
 & Cauchy &   &   &   & $\bestRankPerc$ & $\bestPerc$ &   & $\bestRank$ &   &   &   & $\bestRankPerc$ &   \\
 & Uniform &   &   &   & $\bestRankPerc$ &   &   &   & $\bestRankPerc$ &   &   & $\bestRankPerc$ &   \\
 & SkewNormal &   &   &   & $\bestRankPerc$ &   &   & $\bestRankPerc$ &   &   &   & $\bestRankPerc$ &   \\
 & Exponential &   &   &   & $\bestRankPerc$ &   &   & $\bestRankPerc$ &   &   &   & $\bestRankPerc$ &   \\
\midrule
15 & Normal &   &   &   & $\bestRankPerc$ &   &   &   & $\bestRankPerc$ &   &   & $\bestRankPerc$ &   \\
 & t-Dist &   &   &   & $\bestRankPerc$ &   &   &   & $\bestRankPerc$ &   &   & $\bestRankPerc$ &   \\
 & Cauchy &   &   &   & $\bestRankPerc$ &   &   & $\bestRankPerc$ &   &   &   & $\bestRankPerc$ &   \\
 & Uniform &   &   &   & $\bestRankPerc$ &   &   &   & $\bestRankPerc$ &   &   & $\bestRankPerc$ &   \\
 & SkewNormal &   &   &   & $\bestRankPerc$ &   &   &   & $\bestRankPerc$ &   &   & $\bestRankPerc$ &   \\
 & Exponential &   &   &   & $\bestRankPerc$ &   &   & $\bestRankPerc$ &   &   &   & $\bestRankPerc$ &   \\
\midrule
20 & Normal &   &   & $\bestRankPerc$ &   &   &   &   & $\bestRankPerc$ &   &   & $\bestRankPerc$ &   \\
 & t-Dist &   &   & $\bestRankPerc$ &   &   &   &   & $\bestRankPerc$ &   &   & $\bestRankPerc$ &   \\
 & Cauchy &   &   &   & $\bestRankPerc$ &   &   & $\bestRank$ & $\bestPerc$ &   &   & $\bestRankPerc$ &   \\
 & Uniform &   &   & $\bestRankPerc$ &   &   &   &   & $\bestRankPerc$ &   &   & $\bestRankPerc$ &   \\
 & SkewNormal &   &   & $\bestRankPerc$ &   &   &   &   & $\bestRankPerc$ &   &   & $\bestRankPerc$ &   \\
 & Exponential &   &   & $\bestRankPerc$ &   &   &   &   & $\bestRankPerc$ &   &   & $\bestRankPerc$ &   \\
\bottomrule
\end{tabular}
\end{center}
\caption{Best performing methods in sense of \textsc{AveRank} (filled circle) for the projection depth (PD). If the best method in view of \textsc{PercBest} differs, it is indicated by an empty circle.}\label{tab:bestm_PD}
\end{table}

\begin{table}[ht]\small
\begin{center}
\begin{tabular}{llcccccccccccc}
\toprule
 & & \mc{12}{c}{Number of projections} \\
\cmidrule{3-14}
APD & & \mc{4}{c}{100} & \mc{4}{c}{1000} & \mc{4}{c}{10000} \\
\cmidrule(rl){3-6} \cmidrule(rl){7-10} \cmidrule(rl){11-14} \\
$d$ & Distribution & \rotatebox{90}{RRS} & \rotatebox{90}{SA} & \rotatebox{90}{CD} & \rotatebox{90}{NM} & \rotatebox{90}{RRS} & \rotatebox{90}{SA} & \rotatebox{90}{CD} & \rotatebox{90}{NM} & \rotatebox{90}{RRS} & \rotatebox{90}{SA} & \rotatebox{90}{CD} & \rotatebox{90}{NM} \\
\midrule
5 & Normal &   &   &   & $\bestRankPerc$ & $\bestRankPerc$ &   &   &   & $\bestPerc$ &   &   & $\bestRank$ \\
 & t-Dist &   &   &   & $\bestRankPerc$ & $\bestRankPerc$ &   &   &   & $\bestPerc$ &   &   & $\bestRank$ \\
 & Cauchy & $\bestRankPerc$ &   &   &   & $\bestRankPerc$ &   &   &   & $\bestPerc$ &   &   & $\bestRank$ \\
 & Uniform &   &   &   & $\bestRankPerc$ & $\bestRankPerc$ &   &   &   & $\bestPerc$ &   &   & $\bestRank$ \\
 & SkewNormal & $\bestPerc$ &   & $\bestRank$ &   & $\bestRankPerc$ &   &   &   & $\bestPerc$ &   &   & $\bestRank$ \\
 & Exponential & $\bestPerc$ &   & $\bestRank$ &   & $\bestRankPerc$ &   &   &   & $\bestPerc$ &   &   & $\bestRank$ \\
\midrule
10 & Normal &   &   &   & $\bestRankPerc$ &   &   &   & $\bestRankPerc$ &   &   & $\bestRankPerc$ &   \\
 & t-Dist &   &   &   & $\bestRankPerc$ &   &   &   & $\bestRankPerc$ &   &   & $\bestRankPerc$ &   \\
 & Cauchy &   &   &   & $\bestRankPerc$ & $\bestRankPerc$ &   &   &   &   &   & $\bestRankPerc$ &   \\
 & Uniform &   &   &   & $\bestRankPerc$ &   &   &   & $\bestRankPerc$ &   &   & $\bestRankPerc$ &   \\
 & SkewNormal &   &   &   & $\bestRankPerc$ &   &   &   & $\bestRankPerc$ &   &   & $\bestRankPerc$ &   \\
 & Exponential &   &   &   & $\bestRankPerc$ & $\bestPerc$ &   &   & $\bestRank$ &   &   & $\bestRankPerc$ &   \\
\midrule
15 & Normal &   &   &   & $\bestRankPerc$ &   &   &   & $\bestRankPerc$ &   &   & $\bestRankPerc$ &   \\
 & t-Dist &   &   &   & $\bestRankPerc$ &   &   &   & $\bestRankPerc$ &   &   & $\bestRankPerc$ &   \\
 & Cauchy &   &   &   & $\bestRankPerc$ & $\bestPerc$ &   & $\bestRank$ &   &   &   & $\bestRankPerc$ &   \\
 & Uniform &   &   &   & $\bestRankPerc$ &   &   &   & $\bestRankPerc$ &   &   & $\bestRankPerc$ &   \\
 & SkewNormal &   &   &   & $\bestRankPerc$ &   &   &   & $\bestRankPerc$ &   &   & $\bestRankPerc$ &   \\
 & Exponential &   &   &   & $\bestRankPerc$ &   &   &   & $\bestRankPerc$ &   &   & $\bestRankPerc$ &   \\
\midrule
20 & Normal &   &   & $\bestRankPerc$ &   &   &   &   & $\bestRankPerc$ &   &   & $\bestRankPerc$ &   \\
 & t-Dist &   &   & $\bestRankPerc$ &   &   &   &   & $\bestRankPerc$ &   &   & $\bestRankPerc$ &   \\
 & Cauchy &   &   &   & $\bestRankPerc$ &   &   &   & $\bestRankPerc$ &   &   & $\bestRankPerc$ &   \\
 & Uniform &   &   & $\bestRankPerc$ &   &   &   &   & $\bestRankPerc$ &   &   & $\bestPerc$ & $\bestRank$ \\
 & SkewNormal &   &   & $\bestRankPerc$ &   &   &   &   & $\bestRankPerc$ &   &   & $\bestRankPerc$ &   \\
 & Exponential &   &   & $\bestRankPerc$ &   &   &   &   & $\bestRankPerc$ &   &   & $\bestRankPerc$ &   \\
\bottomrule
\end{tabular}
\end{center}
\caption{Best performing methods in sense of \textsc{AveRank} (filled circle) for the asymmetric projection depth (APD). If the best method in view of \textsc{PercBest} differs, it is indicated by an empty circle.}\label{tab:bestm_APD}
\end{table}

\clearpage

\subsection{Graphs of the development with the number of directions of the average difference between the approximated depth and the lowest achieved depth}\label{assec:simgraphs}

\begin{figure}[h!]\center
	\includegraphics[width=0.775\textwidth,trim=0 0 0 0.75cm,clip=true]{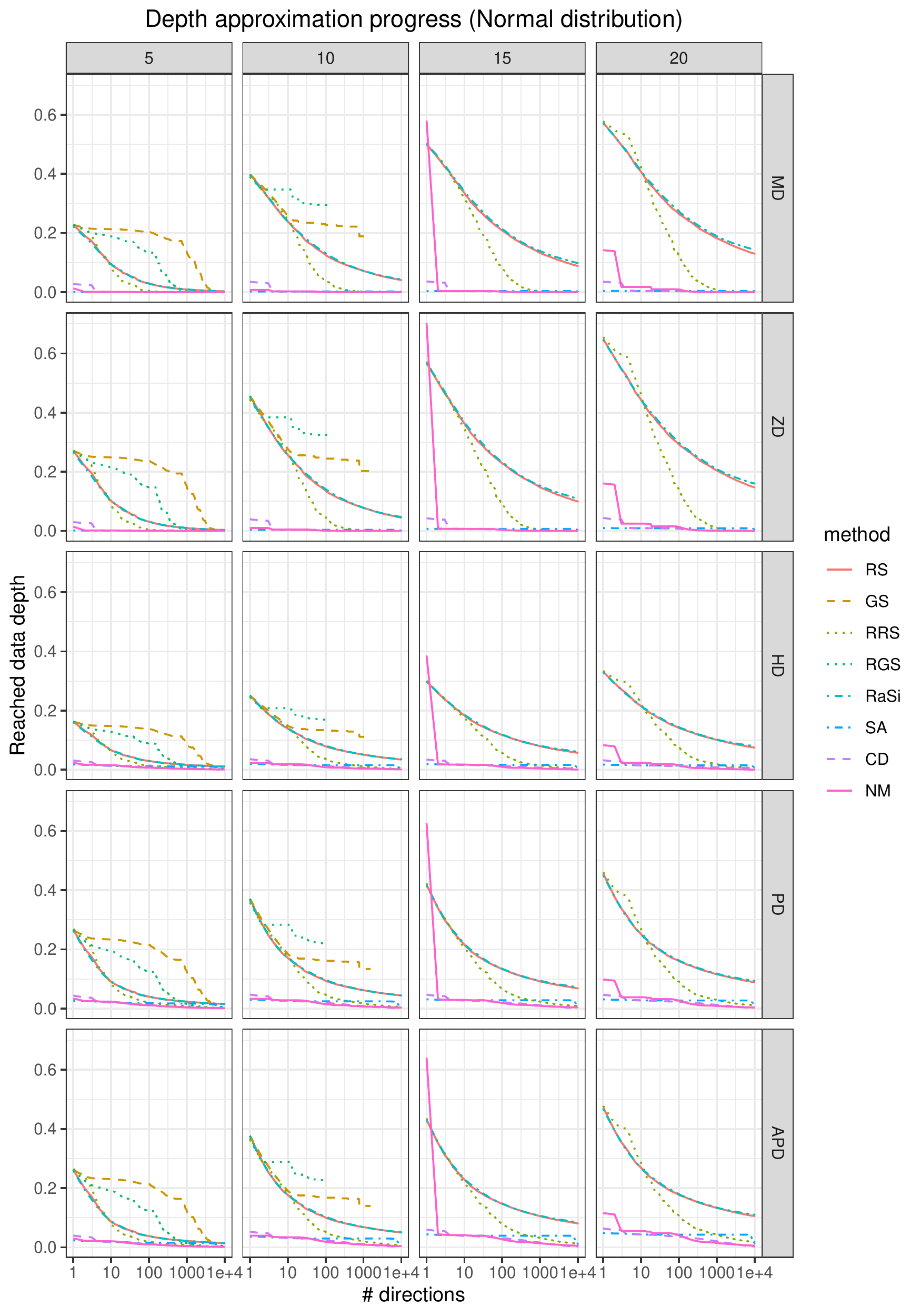}
	\caption{Development of the average minimal depth with the number of directions, using (up to) $N=10000$ random directions for normal distribution.}
\end{figure}

\begin{figure}[h!]\center
	\includegraphics[width=0.9\textwidth,trim=0 0 0 0.75cm,clip=true]{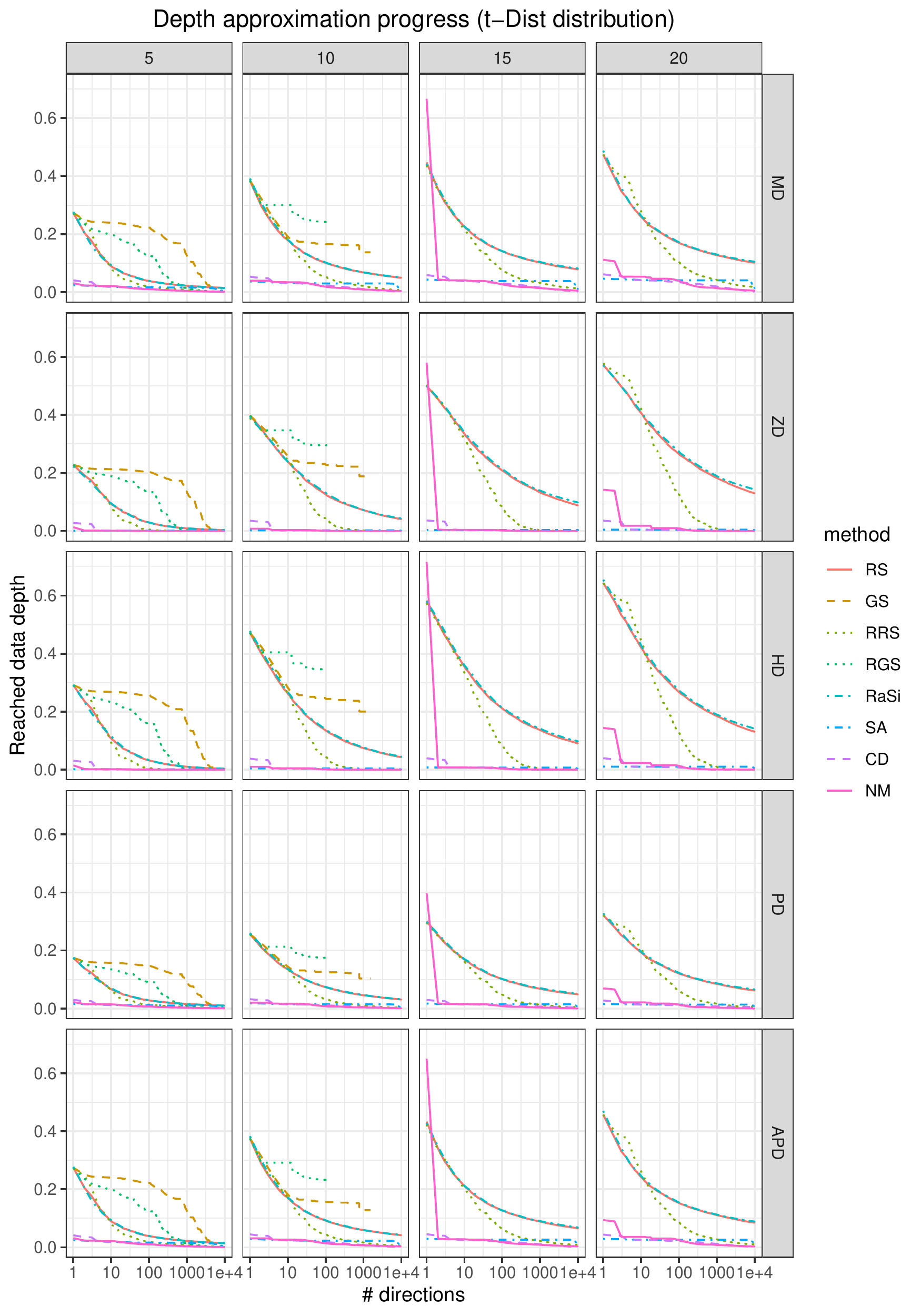}
	\caption{Development of the average minimal depth with the number of directions, using (up to) $N=10000$ random directions for Student $t_5$ distribution.}
\end{figure}

\begin{figure}[h!]\center
	\includegraphics[width=0.9\textwidth,trim=0 0 0 0.75cm,clip=true]{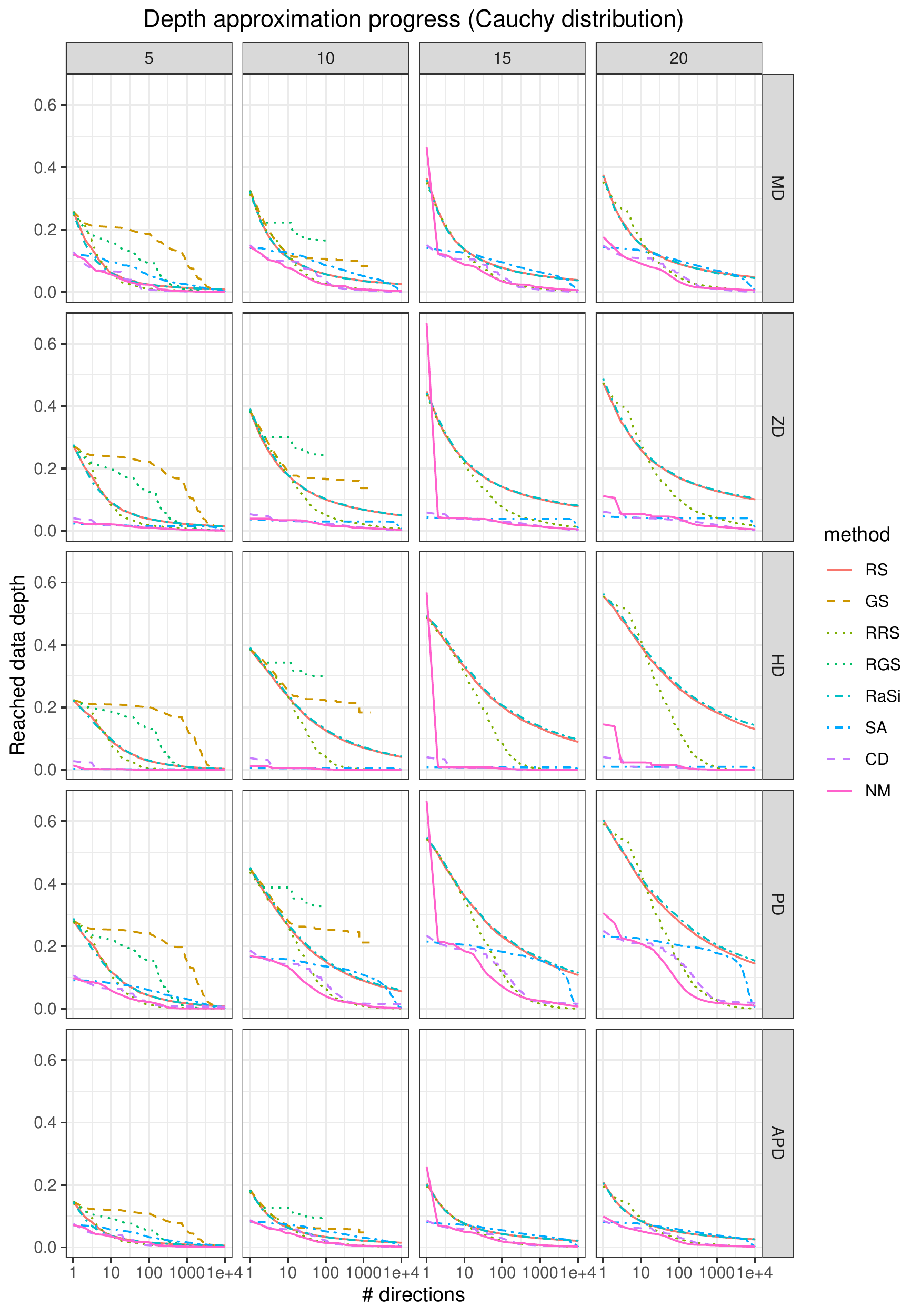}
	\caption{Development of the average minimal depth with the number of directions, using (up to) $N=10000$ random directions for Cauchy distribution.}
\end{figure}

\begin{figure}[h!]\center
	\includegraphics[width=0.9\textwidth,trim=0 0 0 0.75cm,clip=true]{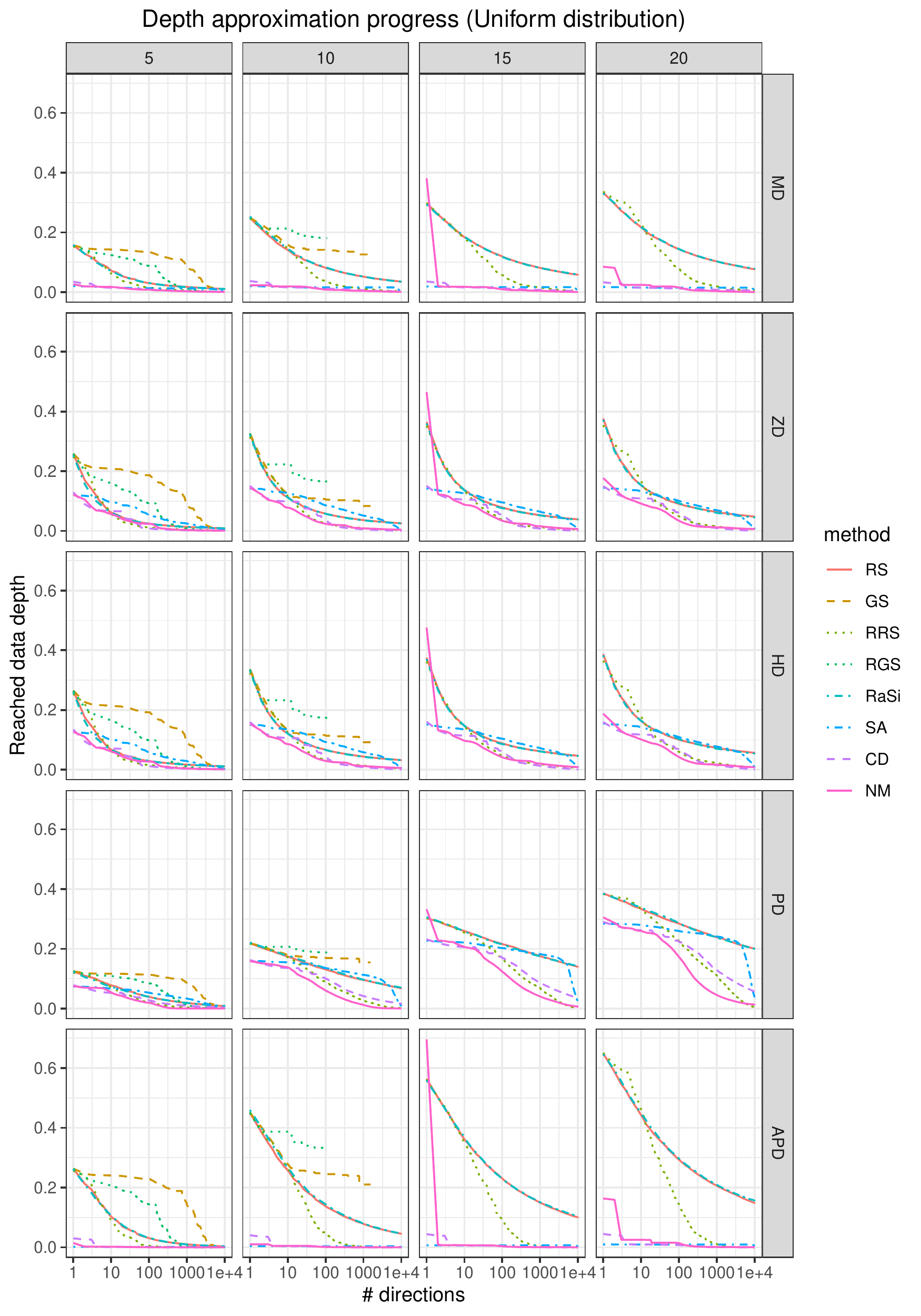}
	\caption{Development of the average minimal depth with the number of directions, using (up to) $N=10000$ random directions for uniform distribution.}
\end{figure}

\begin{figure}[h!]\center
	\includegraphics[width=0.9\textwidth,trim=0 0 0 0.75cm,clip=true]{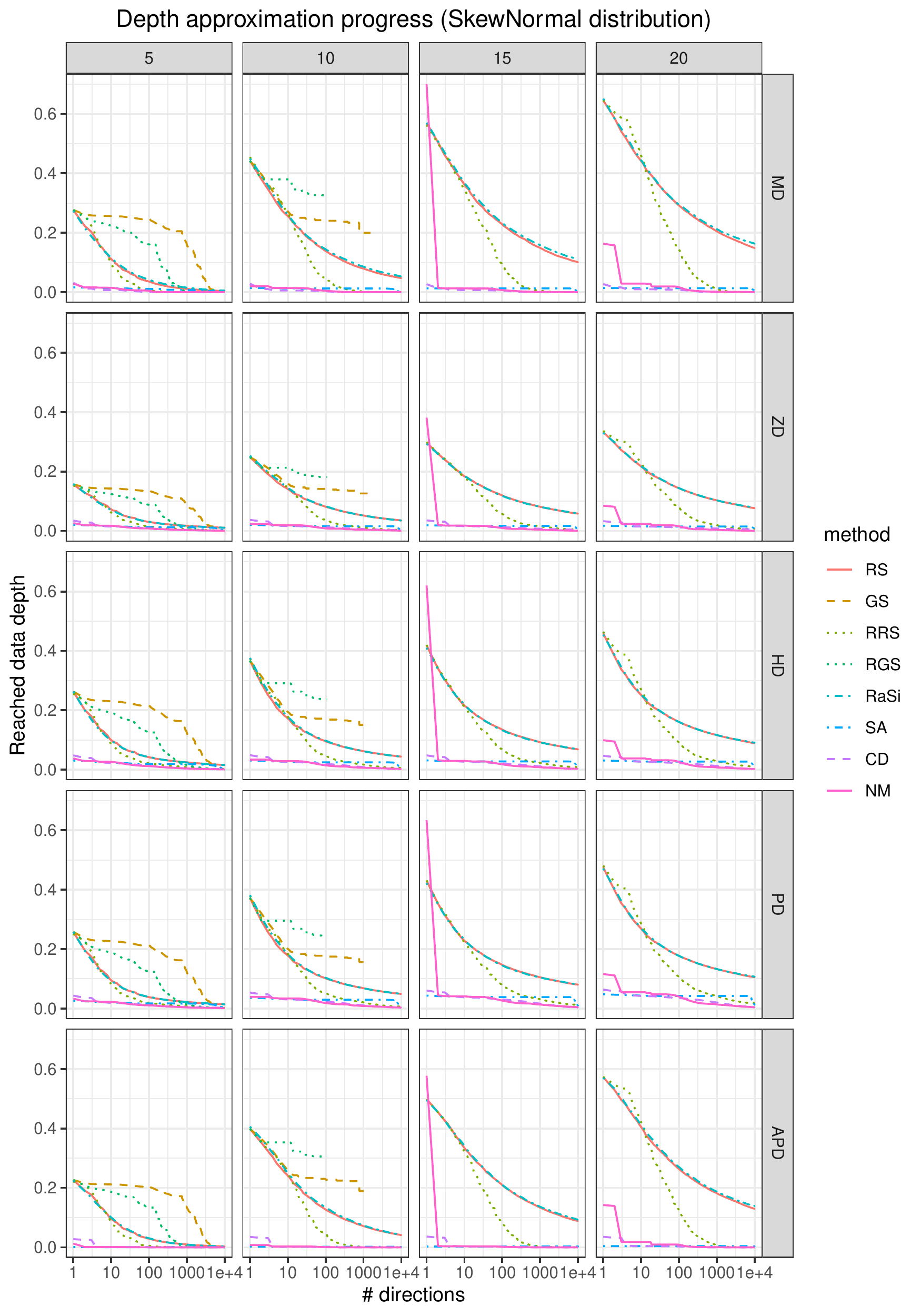}
	\caption{Development of the average minimal depth with the number of directions, using (up to) $N=10000$ random directions for skewed normal distribution.}
\end{figure}

\begin{figure}[h!]\center
	\includegraphics[width=0.9\textwidth,trim=0 0 0 0.75cm,clip=true]{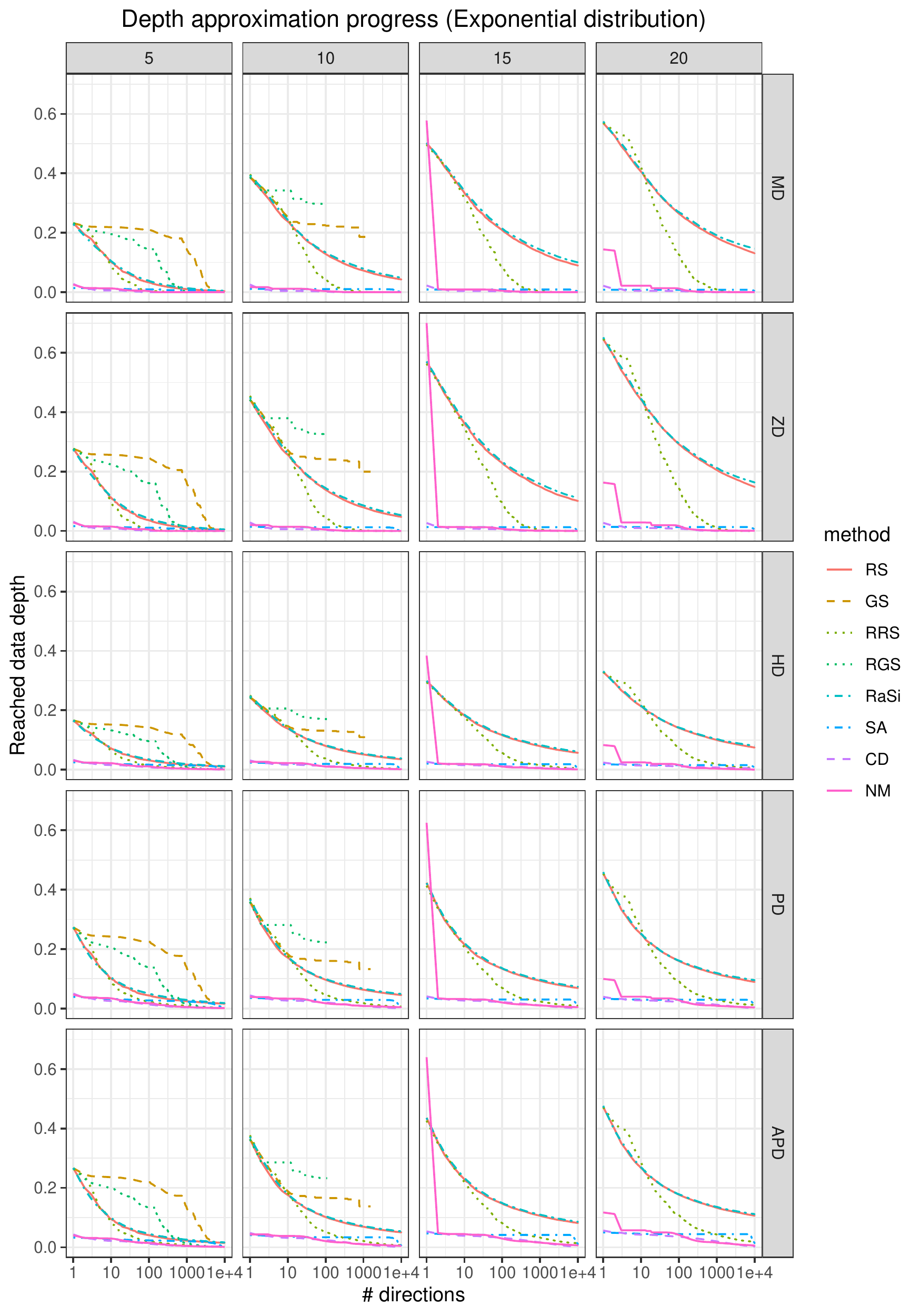}
	\caption{Development of the average minimal depth with the number of directions, using (up to) $N=10000$ random directions for exponential distribution.}
\end{figure}

\clearpage

\section{Tables showing approximated and exact depths and approximation errors}\label{asec:error}


\begin{table}[ht]\footnotesize
\centering
\begin{tabular}{r*{8}{c}}
\toprule
    &    RS &    GS &       RRS &  RaSi &        SA &        CD &        NM & Exact\\
\midrule
  1 & 0.279 & 0.315 & \md 0.273 & 0.288 &     0.275 & \md 0.273 & \md 0.273 & 0.265\\
  2 & 0.149 & 0.166 &     0.148 & 0.160 &     0.138 &     0.138 & \md 0.135 & 0.128\\
  3 & 0.251 & 0.268 &     0.241 & 0.262 &     0.250 &     0.248 & \md 0.240 & 0.236\\
  4 & 0.342 & 0.343 & \md 0.315 & 0.342 &     0.319 &     0.317 &     0.322 & 0.309\\
  5 & 0.242 & 0.289 & \md 0.200 & 0.257 &     0.209 &     0.205 &     0.201 & 0.197\\
  6 & 0.239 & 0.293 & \md 0.224 & 0.255 &     0.228 & \md 0.224 & \md 0.224 & 0.213\\
  7 & 0.258 & 0.261 &     0.237 & 0.258 &     0.240 & \md 0.231 &     0.236 & 0.226\\
  8 & 0.252 & 0.311 &     0.222 & 0.253 &     0.226 &     0.223 & \md 0.218 & 0.215\\
  9 & 0.184 & 0.200 &     0.166 & 0.164 &     0.162 &     0.163 & \md 0.158 & 0.149\\
 10 & 0.245 & 0.260 &     0.239 & 0.256 &     0.232 & \md 0.229 &     0.232 & 0.221\\
 11 & 0.188 & 0.283 &     0.171 & 0.202 &     0.181 & \md 0.170 & \md 0.170 & 0.166\\
 12 & 0.287 & 0.316 &     0.278 & 0.308 &     0.279 & \md 0.273 &     0.278 & 0.268\\
 13 & 0.227 & 0.307 &     0.217 & 0.244 & \md 0.206 &     0.212 &     0.212 & 0.200\\
 14 & 0.192 & 0.260 &     0.163 & 0.181 &     0.163 &     0.162 & \md 0.157 & 0.154\\
 15 & 0.174 & 0.204 &     0.167 & 0.188 &     0.165 & \md 0.163 &     0.164 & 0.157\\
\midrule
MRE & 0.166 & 0.405 &     0.067 & 0.207 &     0.069 &     0.052 & \md 0.046 & 0.000\\
MAE & 0.027 & 0.065 &     0.010 & 0.034 &     0.011 &     0.008 & \md 0.008 & 0.000\\
\bottomrule
\end{tabular}
\caption{Halfspace depth, normal distribution, $n = 1000$, $d = 5$, $N \approx 100$ projections.}
\label{tab.exact_HD_Normal_100}
\end{table}

\begin{table}[ht]\footnotesize
\centering
\begin{tabular}{r*{8}{c}}
\toprule
    &    RS &    GS &       RRS &      RaSi &        SA &        CD &        NM & Exact\\
\midrule
  1 & 0.013 & 0.019 & \md 0.010 &     0.017 &     0.012 & \md 0.010 &     0.011 & 0.010\\
  2 & 0.141 & 0.176 & \md 0.114 &     0.123 &     0.122 &     0.115 &     0.130 & 0.108\\
  3 & 0.302 & 0.315 & \md 0.289 &     0.301 &     0.306 &     0.298 &     0.303 & 0.281\\
  4 & 0.078 & 0.108 &     0.066 &     0.071 &     0.068 & \md 0.065 &     0.068 & 0.058\\
  5 & 0.036 & 0.051 &     0.036 &     0.046 &     0.039 & \md 0.031 &     0.036 & 0.028\\
  6 & 0.140 & 0.141 &     0.129 &     0.145 &     0.216 & \md 0.125 &     0.198 & 0.121\\
  7 & 0.039 & 0.050 &     0.035 &     0.035 &     0.034 &     0.033 & \md 0.030 & 0.028\\
  8 & 0.322 & 0.327 & \md 0.305 &     0.325 &     0.408 &     0.308 &     0.308 & 0.296\\
  9 & 0.075 & 0.081 &     0.069 &     0.070 &     0.089 & \md 0.064 &     0.078 & 0.060\\
 10 & 0.015 & 0.025 & \md 0.014 &     0.017 & \md 0.014 &     0.018 &     0.018 & 0.012\\
 11 & 0.115 & 0.154 &     0.110 &     0.118 & \md 0.108 & \md 0.108 & \md 0.108 & 0.097\\
 12 & 0.126 & 0.172 &     0.114 &     0.131 &     0.140 & \md 0.110 &     0.117 & 0.106\\
 13 & 0.035 & 0.036 & \md 0.031 &     0.034 &     0.038 & \md 0.031 &     0.032 & 0.026\\
 14 & 0.062 & 0.101 &     0.059 &     0.067 &     0.132 &     0.056 & \md 0.055 & 0.052\\
 15 & 0.005 & 0.007 &     0.005 & \md 0.004 & \md 0.004 & \md 0.004 & \md 0.004 & 0.003\\
\midrule
MRE & 0.283 & 0.678 &     0.165 &     0.300 &     0.415 & \md 0.134 &     0.229 & 0.000\\
MAE & 0.015 & 0.032 &     0.007 &     0.015 &     0.030 & \md 0.006 &     0.014 & 0.000\\
\bottomrule
\end{tabular}
\caption{Halfspace depth, Cauchy distribution, $n = 1000$, $d = 5$, $N \approx 100$ projections.}
\label{tab.exact_HD_Cauchy_100}
\end{table}

\begin{table}[ht]\footnotesize
\centering
\begin{tabular}{r*{8}{c}}
\toprule
    &    RS &    GS &       RRS &  RaSi &    SA &        CD &        NM & Exact\\
\midrule
  1 & 0.205 & 0.224 &     0.199 & 0.211 & 0.199 &     0.200 & \md 0.194 & 0.190\\
  2 & 0.146 & 0.187 &     0.137 & 0.161 & 0.137 & \md 0.134 &     0.137 & 0.130\\
  3 & 0.201 & 0.235 & \md 0.190 & 0.204 & 0.195 &     0.191 & \md 0.190 & 0.179\\
  4 & 0.075 & 0.121 &     0.061 & 0.071 & 0.058 &     0.056 & \md 0.054 & 0.050\\
  5 & 0.320 & 0.383 &     0.318 & 0.342 & 0.314 &     0.314 & \md 0.310 & 0.303\\
  6 & 0.323 & 0.347 &     0.296 & 0.311 & 0.299 &     0.298 & \md 0.294 & 0.285\\
  7 & 0.155 & 0.166 & \md 0.135 & 0.155 & 0.138 &     0.136 & \md 0.135 & 0.131\\
  8 & 0.337 & 0.332 &     0.322 & 0.333 & 0.321 & \md 0.312 &     0.314 & 0.305\\
  9 & 0.324 & 0.370 &     0.305 & 0.316 & 0.316 &     0.310 & \md 0.304 & 0.299\\
 10 & 0.267 & 0.313 &     0.255 & 0.308 & 0.259 & \md 0.253 & \md 0.253 & 0.247\\
 11 & 0.135 & 0.151 & \md 0.118 & 0.131 & 0.121 & \md 0.118 &     0.120 & 0.112\\
 12 & 0.192 & 0.224 & \md 0.144 & 0.181 & 0.156 & \md 0.144 &     0.152 & 0.142\\
 13 & 0.208 & 0.219 & \md 0.166 & 0.184 & 0.171 &     0.167 &     0.172 & 0.161\\
 14 & 0.087 & 0.098 &     0.077 & 0.087 & 0.078 & \md 0.076 &     0.077 & 0.072\\
 15 & 0.268 & 0.295 & \md 0.227 & 0.249 & 0.232 &     0.232 &     0.229 & 0.224\\
\midrule
MRE & 0.209 & 0.433 &     0.061 & 0.202 & 0.078 &     0.052 & \md 0.052 & 0.000\\
MAE & 0.028 & 0.056 &     0.008 & 0.028 & 0.011 &     0.007 & \md 0.007 & 0.000\\
\bottomrule
\end{tabular}
\caption{Halfspace depth, t-distribution, $n = 1000$, $d = 5$, $N \approx 100$ projections.}
\label{tab.exact_HD_t-Dist_100}
\end{table}

\begin{table}[ht]\footnotesize
\centering
\begin{tabular}{r*{8}{c}}
\toprule
    &    RS &    GS &       RRS &  RaSi &        SA &        CD &        NM & Exact\\
\midrule
  1 & 0.301 & 0.320 &     0.281 & 0.317 &     0.280 &     0.274 & \md 0.269 & 0.264\\
  2 & 0.291 & 0.377 & \md 0.278 & 0.299 &     0.290 &     0.290 & \md 0.278 & 0.266\\
  3 & 0.253 & 0.302 &     0.251 & 0.283 & \md 0.245 &     0.248 & \md 0.245 & 0.239\\
  4 & 0.370 & 0.392 &     0.358 & 0.367 &     0.369 & \md 0.357 &     0.358 & 0.346\\
  5 & 0.254 & 0.268 &     0.229 & 0.247 &     0.229 &     0.224 & \md 0.222 & 0.215\\
  6 & 0.269 & 0.285 &     0.235 & 0.257 &     0.242 &     0.234 & \md 0.229 & 0.224\\
  7 & 0.195 & 0.234 &     0.184 & 0.193 &     0.189 &     0.183 & \md 0.179 & 0.172\\
  8 & 0.262 & 0.291 &     0.239 & 0.261 &     0.241 & \md 0.237 &     0.242 & 0.231\\
  9 & 0.303 & 0.326 & \md 0.271 & 0.307 &     0.279 &     0.275 &     0.274 & 0.266\\
 10 & 0.253 & 0.295 &     0.236 & 0.249 &     0.240 &     0.226 & \md 0.224 & 0.219\\
 11 & 0.311 & 0.346 &     0.303 & 0.317 &     0.315 &     0.299 & \md 0.298 & 0.291\\
 12 & 0.287 & 0.274 &     0.237 & 0.281 &     0.241 &     0.239 & \md 0.235 & 0.230\\
 13 & 0.269 & 0.305 &     0.261 & 0.278 &     0.262 &     0.260 & \md 0.259 & 0.253\\
 14 & 0.379 & 0.412 &     0.362 & 0.358 &     0.363 &     0.365 & \md 0.355 & 0.342\\
 15 & 0.298 & 0.379 &     0.293 & 0.307 &     0.294 & \md 0.291 &     0.292 & 0.284\\
\midrule
MRE & 0.153 & 0.322 &     0.059 & 0.162 &     0.078 &     0.052 & \md 0.038 & 0.000\\
MAE & 0.030 & 0.064 &     0.012 & 0.032 &     0.016 &     0.011 & \md 0.008 & 0.000\\
\bottomrule
\end{tabular}
\caption{Halfspace depth, uniform distribution, $n = 1000$, $d = 5$, $N \approx 100$ projections.}
\label{tab.exact_HD_Uniform_100}
\end{table}

\begin{table}[ht]\footnotesize
\centering
\begin{tabular}{r*{8}{c}}
\toprule
    &    RS &    GS &       RRS &  RaSi &        SA &        CD &        NM & Exact\\
\midrule
  1 & 0.310 & 0.290 &     0.280 & 0.286 &     0.280 &     0.274 & \md 0.271 & 0.264\\
  2 & 0.336 & 0.338 &     0.328 & 0.338 &     0.322 &     0.323 & \md 0.321 & 0.312\\
  3 & 0.276 & 0.296 &     0.245 & 0.257 &     0.237 & \md 0.217 &     0.234 & 0.202\\
  4 & 0.283 & 0.274 &     0.254 & 0.279 &     0.250 & \md 0.246 &     0.247 & 0.242\\
  5 & 0.244 & 0.263 &     0.251 & 0.273 & \md 0.230 &     0.231 &     0.233 & 0.222\\
  6 & 0.223 & 0.278 &     0.208 & 0.230 &     0.195 & \md 0.191 &     0.193 & 0.185\\
  7 & 0.213 & 0.262 &     0.177 & 0.181 &     0.187 & \md 0.172 &     0.175 & 0.163\\
  8 & 0.299 & 0.345 &     0.291 & 0.297 & \md 0.285 &     0.288 &     0.293 & 0.276\\
  9 & 0.176 & 0.200 &     0.169 & 0.185 &     0.166 & \md 0.160 &     0.164 & 0.157\\
 10 & 0.280 & 0.280 & \md 0.261 & 0.275 &     0.262 &     0.269 & \md 0.261 & 0.249\\
 11 & 0.236 & 0.203 & \md 0.193 & 0.222 &     0.206 &     0.195 &     0.196 & 0.185\\
 12 & 0.350 & 0.337 &     0.329 & 0.349 &     0.351 & \md 0.323 &     0.331 & 0.308\\
 13 & 0.324 & 0.320 & \md 0.304 & 0.318 &     0.312 &     0.309 &     0.305 & 0.295\\
 14 & 0.234 & 0.305 & \md 0.209 & 0.266 &     0.228 &     0.218 &     0.215 & 0.204\\
 15 & 0.147 & 0.214 &     0.105 & 0.200 &     0.116 & \md 0.103 &     0.111 & 0.097\\
\midrule
MRE & 0.231 & 0.370 &     0.092 & 0.260 &     0.105 & \md 0.058 &     0.074 & 0.000\\
MAE & 0.038 & 0.056 &     0.016 & 0.040 &     0.018 & \md 0.011 &     0.013 & 0.000\\
\bottomrule
\end{tabular}
\caption{Halfspace depth, skew normal distribution, $n = 1000$, $d = 5$, $N \approx 100$ projections.}
\label{tab.exact_HD_SkewNormal_100}
\end{table}

\begin{table}[ht]\footnotesize
\centering
\begin{tabular}{r*{8}{c}}
\toprule
    &    RS &    GS &       RRS &      RaSi &        SA &        CD &        NM & Exact\\
\midrule
  1 & 0.243 & 0.266 &     0.233 &     0.243 &     0.233 & \md 0.229 &     0.233 & 0.224\\
  2 & 0.250 & 0.319 &     0.245 &     0.251 &     0.216 & \md 0.205 &     0.208 & 0.199\\
  3 & 0.190 & 0.247 &     0.168 &     0.190 & \md 0.160 &     0.165 &     0.164 & 0.157\\
  4 & 0.376 & 0.399 &     0.366 &     0.372 &     0.460 & \md 0.358 &     0.441 & 0.334\\
  5 & 0.192 & 0.170 & \md 0.162 &     0.183 &     0.167 &     0.166 &     0.167 & 0.157\\
  6 & 0.257 & 0.290 &     0.230 &     0.246 &     0.247 & \md 0.229 &     0.243 & 0.223\\
  7 & 0.161 & 0.198 &     0.140 &     0.149 &     0.140 & \md 0.135 &     0.140 & 0.132\\
  8 & 0.241 & 0.248 &     0.218 &     0.259 &     0.210 &     0.208 & \md 0.207 & 0.199\\
  9 & 0.240 & 0.301 & \md 0.205 &     0.230 &     0.224 &     0.206 &     0.207 & 0.202\\
 10 & 0.202 & 0.230 & \md 0.196 &     0.215 &     0.209 &     0.197 &     0.201 & 0.189\\
 11 & 0.343 & 0.353 &     0.355 & \md 0.312 &     0.441 &     0.332 & \md 0.312 & 0.302\\
 12 & 0.230 & 0.243 &     0.199 &     0.222 & \md 0.186 &     0.191 &     0.191 & 0.183\\
 13 & 0.248 & 0.248 &     0.264 &     0.237 &     0.237 & \md 0.221 &     0.225 & 0.212\\
 14 & 0.312 & 0.368 &     0.304 &     0.341 &     0.326 & \md 0.297 &     0.377 & 0.291\\
 15 & 0.104 & 0.122 &     0.095 &     0.106 &     0.092 &     0.091 & \md 0.090 & 0.088\\
\midrule
MRE & 0.204 & 0.375 &     0.109 &     0.190 &     0.149 & \md 0.051 &     0.104 & 0.000\\
MAE & 0.033 & 0.061 &     0.019 &     0.031 &     0.030 & \md 0.009 &     0.021 & 0.000\\
\bottomrule
\end{tabular}
\caption{Halfspace depth, exponential distribution, $n = 1000$, $d = 5$, $N \approx 100$ projections.}
\label{tab.exact_HD_Exponential_100}
\end{table}

\begin{table}[ht]\footnotesize
\centering
\begin{tabular}{r*{9}{c}}
\toprule
    &    RS &    GS &       RRS &   RGS &  RaSi &        SA &        CD &        NM & Exact\\
\midrule
  1 & 0.279 & 0.286 & \md 0.267 & 0.273 & 0.279 &     0.269 &     0.268 &     0.269 & 0.265\\
  2 & 0.143 & 0.160 &     0.132 & 0.139 & 0.148 &     0.136 &     0.132 & \md 0.130 & 0.128\\
  3 & 0.244 & 0.259 & \md 0.238 & 0.245 & 0.247 &     0.241 & \md 0.238 & \md 0.238 & 0.236\\
  4 & 0.329 & 0.340 & \md 0.310 & 0.336 & 0.324 &     0.316 &     0.311 &     0.311 & 0.309\\
  5 & 0.220 & 0.235 &     0.200 & 0.208 & 0.225 &     0.204 & \md 0.199 &     0.200 & 0.197\\
  6 & 0.236 & 0.259 & \md 0.214 & 0.229 & 0.227 &     0.219 &     0.216 &     0.217 & 0.213\\
  7 & 0.238 & 0.242 &     0.230 & 0.235 & 0.244 &     0.233 & \md 0.227 &     0.229 & 0.226\\
  8 & 0.228 & 0.229 & \md 0.218 & 0.225 & 0.229 &     0.223 &     0.219 & \md 0.218 & 0.215\\
  9 & 0.171 & 0.169 &     0.152 & 0.153 & 0.164 &     0.154 & \md 0.151 &     0.152 & 0.149\\
 10 & 0.241 & 0.248 &     0.224 & 0.230 & 0.228 &     0.225 & \md 0.223 & \md 0.223 & 0.221\\
 11 & 0.187 & 0.205 &     0.170 & 0.182 & 0.185 & \md 0.169 & \md 0.169 & \md 0.169 & 0.166\\
 12 & 0.284 & 0.280 & \md 0.269 & 0.273 & 0.288 &     0.274 &     0.272 &     0.271 & 0.268\\
 13 & 0.221 & 0.228 & \md 0.201 & 0.218 & 0.212 &     0.206 &     0.206 &     0.202 & 0.200\\
 14 & 0.171 & 0.182 &     0.161 & 0.161 & 0.176 &     0.160 &     0.158 & \md 0.157 & 0.154\\
 15 & 0.168 & 0.188 & \md 0.157 & 0.158 & 0.168 &     0.161 &     0.161 & \md 0.157 & 0.157\\
\midrule
MRE & 0.105 & 0.169 &     0.017 & 0.063 & 0.099 &     0.035 &     0.019 & \md 0.016 & 0.000\\
MAE & 0.017 & 0.027 & \md 0.003 & 0.011 & 0.016 &     0.006 &     0.003 & \md 0.003 & 0.000\\
\bottomrule
\end{tabular}
\caption{Halfspace depth, normal distribution, $n = 1000$, $d = 5$, $N \approx 1000$ projections.}
\label{tab.exact_HD_Normal_1000}
\end{table}

\begin{table}[ht]\footnotesize
\centering
\begin{tabular}{r*{9}{c}}
\toprule
    &    RS &    GS &       RRS &       RGS &  RaSi &    SA &        CD &        NM & Exact\\
\midrule
  1 & 0.012 & 0.012 &     0.011 &     0.011 & 0.013 & 0.011 & \md 0.010 & \md 0.010 & 0.010\\
  2 & 0.119 & 0.127 & \md 0.109 &     0.116 & 0.120 & 0.110 &     0.110 &     0.110 & 0.108\\
  3 & 0.302 & 0.305 & \md 0.284 &     0.289 & 0.296 & 0.289 &     0.286 &     0.290 & 0.281\\
  4 & 0.068 & 0.069 & \md 0.058 &     0.061 & 0.068 & 0.067 &     0.061 & \md 0.058 & 0.058\\
  5 & 0.032 & 0.035 & \md 0.030 &     0.034 & 0.036 & 0.034 & \md 0.030 & \md 0.030 & 0.028\\
  6 & 0.130 & 0.141 & \md 0.123 &     0.126 & 0.137 & 0.128 &     0.124 & \md 0.123 & 0.121\\
  7 & 0.034 & 0.035 & \md 0.028 &     0.032 & 0.032 & 0.033 &     0.029 &     0.030 & 0.028\\
  8 & 0.316 & 0.327 &     0.302 & \md 0.298 & 0.321 & 0.310 &     0.299 &     0.299 & 0.296\\
  9 & 0.068 & 0.065 & \md 0.062 &     0.066 & 0.070 & 0.066 &     0.063 &     0.063 & 0.060\\
 10 & 0.015 & 0.016 &     0.014 &     0.016 & 0.014 & 0.014 & \md 0.013 & \md 0.013 & 0.012\\
 11 & 0.110 & 0.121 & \md 0.098 &     0.110 & 0.114 & 0.104 &     0.099 &     0.099 & 0.097\\
 12 & 0.113 & 0.119 &     0.112 &     0.113 & 0.115 & 0.112 & \md 0.108 &     0.110 & 0.106\\
 13 & 0.033 & 0.036 & \md 0.028 &     0.031 & 0.031 & 0.029 &     0.029 &     0.029 & 0.026\\
 14 & 0.060 & 0.060 & \md 0.052 &     0.064 & 0.057 & 0.055 &     0.055 &     0.055 & 0.052\\
 15 & 0.004 & 0.006 &     0.004 &     0.005 & 0.004 & 0.004 &     0.004 & \md 0.003 & 0.003\\
\midrule
MRE & 0.168 & 0.262 &     0.062 &     0.165 & 0.176 & 0.119 &     0.063 & \md 0.041 & 0.000\\
MAE & 0.009 & 0.013 & \md 0.002 &     0.006 & 0.009 & 0.005 &     0.002 &     0.002 & 0.000\\
\bottomrule
\end{tabular}
\caption{Halfspace depth, Cauchy distribution, $n = 1000$, $d = 5$, $N \approx 1000$ projections.}
\label{tab.exact_HD_Cauchy_1000}
\end{table}

\begin{table}[ht]\footnotesize
\centering
\begin{tabular}{r*{9}{c}}
\toprule
    &    RS &    GS &       RRS &       RGS &  RaSi &        SA &        CD &        NM & Exact\\
\midrule
  1 & 0.204 & 0.224 & \md 0.192 &     0.194 & 0.209 &     0.199 &     0.194 &     0.194 & 0.190\\
  2 & 0.144 & 0.139 &     0.135 &     0.138 & 0.143 &     0.135 & \md 0.133 &     0.134 & 0.130\\
  3 & 0.195 & 0.206 &     0.188 &     0.191 & 0.198 &     0.193 &     0.188 & \md 0.187 & 0.179\\
  4 & 0.065 & 0.074 &     0.051 &     0.052 & 0.061 &     0.054 & \md 0.050 &     0.051 & 0.050\\
  5 & 0.320 & 0.331 & \md 0.308 &     0.316 & 0.314 &     0.313 &     0.311 &     0.309 & 0.303\\
  6 & 0.299 & 0.304 &     0.291 &     0.298 & 0.301 &     0.293 &     0.290 & \md 0.289 & 0.285\\
  7 & 0.144 & 0.164 & \md 0.132 &     0.135 & 0.154 & \md 0.132 &     0.133 &     0.133 & 0.131\\
  8 & 0.322 & 0.331 & \md 0.307 &     0.315 & 0.318 &     0.311 &     0.310 & \md 0.307 & 0.305\\
  9 & 0.316 & 0.327 & \md 0.304 &     0.313 & 0.316 &     0.305 & \md 0.304 & \md 0.304 & 0.299\\
 10 & 0.254 & 0.264 &     0.251 & \md 0.249 & 0.265 &     0.255 & \md 0.249 &     0.251 & 0.247\\
 11 & 0.125 & 0.136 & \md 0.114 & \md 0.114 & 0.126 &     0.116 &     0.116 & \md 0.114 & 0.112\\
 12 & 0.163 & 0.162 & \md 0.143 &     0.153 & 0.156 &     0.146 & \md 0.143 & \md 0.143 & 0.142\\
 13 & 0.186 & 0.173 & \md 0.164 &     0.170 & 0.172 &     0.166 & \md 0.164 &     0.165 & 0.161\\
 14 & 0.086 & 0.086 &     0.075 & \md 0.074 & 0.084 &     0.077 & \md 0.074 &     0.075 & 0.072\\
 15 & 0.236 & 0.258 &     0.226 &     0.231 & 0.234 &     0.229 &     0.228 & \md 0.225 & 0.224\\
\midrule
MRE & 0.121 & 0.178 & \md 0.023 &     0.048 & 0.113 &     0.044 &     0.024 &     0.023 & 0.000\\
MAE & 0.015 & 0.023 & \md 0.003 &     0.008 & 0.015 &     0.006 &     0.004 & \md 0.003 & 0.000\\
\bottomrule
\end{tabular}
\caption{Halfspace depth, t-distribution, $n = 1000$, $d = 5$, $N \approx 1000$ projections.}
\label{tab.exact_HD_t-Dist_1000}
\end{table}

\begin{table}[ht]\footnotesize
\centering
\begin{tabular}{r*{9}{c}}
\toprule
    &    RS &    GS &       RRS &   RGS &  RaSi &        SA &        CD &        NM & Exact\\
\midrule
  1 & 0.280 & 0.295 &     0.272 & 0.276 & 0.289 &     0.274 &     0.274 & \md 0.267 & 0.264\\
  2 & 0.290 & 0.294 &     0.280 & 0.283 & 0.289 &     0.280 &     0.275 & \md 0.273 & 0.266\\
  3 & 0.252 & 0.252 &     0.244 & 0.245 & 0.257 &     0.245 &     0.245 & \md 0.241 & 0.239\\
  4 & 0.358 & 0.363 & \md 0.350 & 0.369 & 0.367 &     0.352 &     0.351 &     0.351 & 0.346\\
  5 & 0.234 & 0.230 &     0.220 & 0.223 & 0.241 &     0.228 &     0.219 & \md 0.217 & 0.215\\
  6 & 0.239 & 0.246 &     0.229 & 0.232 & 0.236 &     0.230 &     0.227 & \md 0.226 & 0.224\\
  7 & 0.189 & 0.196 & \md 0.176 & 0.183 & 0.187 &     0.177 & \md 0.176 &     0.177 & 0.172\\
  8 & 0.250 & 0.251 &     0.235 & 0.237 & 0.245 &     0.237 & \md 0.234 &     0.235 & 0.231\\
  9 & 0.288 & 0.281 &     0.273 & 0.273 & 0.277 &     0.274 &     0.272 & \md 0.270 & 0.266\\
 10 & 0.239 & 0.258 & \md 0.224 & 0.236 & 0.236 & \md 0.224 & \md 0.224 & \md 0.224 & 0.219\\
 11 & 0.304 & 0.320 &     0.297 & 0.311 & 0.312 &     0.303 & \md 0.296 &     0.298 & 0.291\\
 12 & 0.252 & 0.251 & \md 0.233 & 0.246 & 0.243 &     0.239 &     0.234 &     0.234 & 0.230\\
 13 & 0.269 & 0.269 &     0.260 & 0.267 & 0.274 &     0.258 &     0.259 & \md 0.256 & 0.253\\
 14 & 0.364 & 0.363 & \md 0.347 & 0.354 & 0.358 &     0.351 & \md 0.347 &     0.350 & 0.342\\
 15 & 0.298 & 0.307 & \md 0.285 & 0.288 & 0.301 &     0.289 &     0.286 &     0.286 & 0.284\\
\midrule
MRE & 0.089 & 0.112 &     0.028 & 0.059 & 0.090 &     0.039 &     0.025 & \md 0.021 & 0.000\\
MAE & 0.018 & 0.022 &     0.006 & 0.012 & 0.018 &     0.008 &     0.005 & \md 0.004 & 0.000\\
\bottomrule
\end{tabular}
\caption{Halfspace depth, uniform distribution, $n = 1000$, $d = 5$, $N \approx 1000$ projections.}
\label{tab.exact_HD_Uniform_1000}
\end{table}

\begin{table}[ht]\footnotesize
\centering
\begin{tabular}{r*{9}{c}}
\toprule
    &    RS &    GS &       RRS &       RGS &  RaSi &        SA &        CD &        NM & Exact\\
\midrule
  1 & 0.277 & 0.282 &     0.271 &     0.276 & 0.282 &     0.277 &     0.271 & \md 0.268 & 0.264\\
  2 & 0.327 & 0.332 & \md 0.313 &     0.324 & 0.323 &     0.321 &     0.315 &     0.314 & 0.312\\
  3 & 0.227 & 0.233 & \md 0.205 &     0.211 & 0.239 &     0.211 & \md 0.205 &     0.207 & 0.202\\
  4 & 0.257 & 0.274 & \md 0.244 &     0.246 & 0.259 &     0.246 &     0.246 & \md 0.244 & 0.242\\
  5 & 0.240 & 0.241 & \md 0.225 & \md 0.225 & 0.238 &     0.227 & \md 0.225 & \md 0.225 & 0.222\\
  6 & 0.200 & 0.215 &     0.190 &     0.188 & 0.205 &     0.190 & \md 0.186 &     0.189 & 0.185\\
  7 & 0.179 & 0.183 & \md 0.163 &     0.166 & 0.178 &     0.169 &     0.165 &     0.165 & 0.163\\
  8 & 0.294 & 0.299 &     0.283 &     0.291 & 0.294 &     0.285 & \md 0.278 &     0.280 & 0.276\\
  9 & 0.172 & 0.177 &     0.162 &     0.168 & 0.174 &     0.161 & \md 0.160 & \md 0.160 & 0.157\\
 10 & 0.267 & 0.280 &     0.253 &     0.257 & 0.275 & \md 0.250 &     0.253 &     0.255 & 0.249\\
 11 & 0.206 & 0.201 & \md 0.187 &     0.192 & 0.200 &     0.189 &     0.191 &     0.189 & 0.185\\
 12 & 0.340 & 0.337 &     0.315 &     0.323 & 0.346 &     0.316 & \md 0.312 &     0.321 & 0.308\\
 13 & 0.312 & 0.320 &     0.306 &     0.307 & 0.309 &     0.306 & \md 0.297 &     0.299 & 0.295\\
 14 & 0.228 & 0.225 &     0.210 &     0.223 & 0.215 &     0.212 & \md 0.208 &     0.211 & 0.204\\
 15 & 0.123 & 0.127 & \md 0.099 &     0.124 & 0.133 &     0.102 & \md 0.099 &     0.101 & 0.097\\
\midrule
MRE & 0.115 & 0.144 &     0.024 &     0.067 & 0.126 &     0.038 & \md 0.019 &     0.025 & 0.000\\
MAE & 0.019 & 0.024 &     0.004 &     0.011 & 0.021 &     0.007 & \md 0.003 &     0.004 & 0.000\\
\bottomrule
\end{tabular}
\caption{Halfspace depth, skew normal distribution, $n = 1000$, $d = 5$, $N \approx 1000$ projections.}
\label{tab.exact_HD_SkewNormal_1000}
\end{table}

\begin{table}[ht]\footnotesize
\centering
\begin{tabular}{r*{9}{c}}
\toprule
    &    RS &    GS &       RRS &       RGS &  RaSi &    SA &        CD &        NM & Exact\\
\midrule
  1 & 0.235 & 0.254 &     0.228 &     0.230 & 0.239 & 0.228 &     0.227 & \md 0.226 & 0.224\\
  2 & 0.222 & 0.238 &     0.202 &     0.204 & 0.228 & 0.203 &     0.202 & \md 0.201 & 0.199\\
  3 & 0.174 & 0.183 &     0.163 &     0.160 & 0.164 & 0.160 &     0.159 & \md 0.158 & 0.157\\
  4 & 0.356 & 0.349 &     0.366 &     0.343 & 0.347 & 0.338 & \md 0.336 &     0.339 & 0.334\\
  5 & 0.171 & 0.170 &     0.163 &     0.164 & 0.171 & 0.162 & \md 0.160 &     0.161 & 0.157\\
  6 & 0.241 & 0.240 & \md 0.223 &     0.235 & 0.246 & 0.225 &     0.225 &     0.225 & 0.223\\
  7 & 0.143 & 0.145 & \md 0.134 &     0.135 & 0.144 & 0.135 &     0.135 & \md 0.134 & 0.132\\
  8 & 0.223 & 0.221 &     0.203 & \md 0.200 & 0.215 & 0.202 &     0.201 & \md 0.200 & 0.199\\
  9 & 0.219 & 0.212 &     0.204 &     0.204 & 0.216 & 0.205 & \md 0.203 &     0.204 & 0.202\\
 10 & 0.202 & 0.208 &     0.191 &     0.197 & 0.208 & 0.192 &     0.197 & \md 0.190 & 0.189\\
 11 & 0.315 & 0.336 &     0.350 &     0.307 & 0.312 & 0.338 &     0.329 & \md 0.305 & 0.302\\
 12 & 0.195 & 0.219 & \md 0.184 &     0.185 & 0.199 & 0.186 & \md 0.184 &     0.185 & 0.183\\
 13 & 0.224 & 0.235 & \md 0.218 &     0.219 & 0.225 & 0.221 & \md 0.218 &     0.219 & 0.212\\
 14 & 0.307 & 0.305 & \md 0.294 &     0.354 & 0.307 & 0.296 &     0.295 &     0.295 & 0.291\\
 15 & 0.094 & 0.099 &     0.091 &     0.093 & 0.099 & 0.091 &     0.091 & \md 0.090 & 0.088\\
\midrule
MRE & 0.092 & 0.131 &     0.041 &     0.050 & 0.095 & 0.033 &     0.026 & \md 0.016 & 0.000\\
MAE & 0.015 & 0.021 &     0.008 &     0.009 & 0.015 & 0.006 &     0.005 & \md 0.003 & 0.000\\
\bottomrule
\end{tabular}
\caption{Halfspace depth, exponential distribution, $n = 1000$, $d = 5$, $N \approx 1000$ projections.}
\label{tab.exact_HD_Exponential_1000}
\end{table}

\begin{table}[ht]\footnotesize
\centering
\begin{tabular}{r*{9}{c}}
\toprule
    &    RS &    GS &       RRS &       RGS &  RaSi &    SA &        CD &        NM & Exact\\
\midrule
  1 & 0.276 & 0.281 &     0.268 &     0.268 & 0.277 & 0.268 &     0.268 & \md 0.267 & 0.265\\
  2 & 0.141 & 0.140 & \md 0.129 & \md 0.129 & 0.133 & 0.130 &     0.131 & \md 0.129 & 0.128\\
  3 & 0.241 & 0.251 &     0.238 & \md 0.237 & 0.244 & 0.238 & \md 0.237 & \md 0.237 & 0.236\\
  4 & 0.321 & 0.334 &     0.311 &     0.313 & 0.324 & 0.313 & \md 0.310 & \md 0.310 & 0.309\\
  5 & 0.210 & 0.224 &     0.200 & \md 0.198 & 0.209 & 0.199 & \md 0.198 & \md 0.198 & 0.197\\
  6 & 0.224 & 0.223 &     0.215 &     0.221 & 0.227 & 0.215 & \md 0.214 & \md 0.214 & 0.213\\
  7 & 0.236 & 0.234 &     0.228 &     0.228 & 0.239 & 0.228 & \md 0.227 &     0.228 & 0.226\\
  8 & 0.224 & 0.235 & \md 0.215 &     0.218 & 0.224 & 0.218 &     0.217 & \md 0.215 & 0.215\\
  9 & 0.158 & 0.159 & \md 0.150 &     0.151 & 0.157 & 0.151 &     0.151 & \md 0.150 & 0.149\\
 10 & 0.231 & 0.238 & \md 0.222 & \md 0.222 & 0.228 & 0.224 & \md 0.222 & \md 0.222 & 0.221\\
 11 & 0.175 & 0.184 & \md 0.166 &     0.168 & 0.175 & 0.169 &     0.169 &     0.168 & 0.166\\
 12 & 0.276 & 0.278 & \md 0.269 &     0.270 & 0.277 & 0.270 &     0.270 &     0.270 & 0.268\\
 13 & 0.219 & 0.217 & \md 0.200 &     0.202 & 0.212 & 0.202 & \md 0.200 & \md 0.200 & 0.200\\
 14 & 0.165 & 0.167 &     0.160 &     0.160 & 0.160 & 0.159 & \md 0.155 &     0.156 & 0.154\\
 15 & 0.164 & 0.172 & \md 0.157 & \md 0.157 & 0.167 & 0.158 &     0.161 & \md 0.157 & 0.157\\
\midrule
MRE & 0.064 & 0.093 &     0.010 &     0.015 & 0.059 & 0.015 &     0.011 & \md 0.007 & 0.000\\
MAE & 0.010 & 0.016 &     0.002 &     0.003 & 0.010 & 0.003 &     0.002 & \md 0.001 & 0.000\\
\bottomrule
\end{tabular}
\caption{Halfspace depth, normal distribution, $n = 1000$, $d = 5$, $N \approx 10000$ projections.}
\label{tab.exact_HD_Normal_10000}
\end{table}

\begin{table}[ht]\footnotesize
\centering
\begin{tabular}{r*{9}{c}}
\toprule
    &        RS &    GS &       RRS &       RGS &  RaSi &        SA &        CD &        NM & Exact\\
\midrule
  1 &     0.012 & 0.012 & \md 0.010 & \md 0.010 & 0.012 & \md 0.010 & \md 0.010 & \md 0.010 & 0.010\\
  2 &     0.115 & 0.120 & \md 0.108 &     0.109 & 0.119 &     0.110 &     0.109 & \md 0.108 & 0.108\\
  3 &     0.290 & 0.296 &     0.284 &     0.288 & 0.291 &     0.286 & \md 0.283 &     0.286 & 0.281\\
  4 &     0.066 & 0.065 & \md 0.058 & \md 0.058 & 0.063 &     0.059 & \md 0.058 & \md 0.058 & 0.058\\
  5 &     0.032 & 0.033 & \md 0.029 &     0.031 & 0.034 & \md 0.029 & \md 0.029 & \md 0.029 & 0.028\\
  6 &     0.129 & 0.131 & \md 0.122 &     0.124 & 0.131 &     0.124 &     0.124 &     0.123 & 0.121\\
  7 &     0.030 & 0.031 &     0.029 & \md 0.028 & 0.031 &     0.030 & \md 0.028 & \md 0.028 & 0.028\\
  8 &     0.314 & 0.310 &     0.302 & \md 0.298 & 0.312 &     0.299 & \md 0.298 &     0.299 & 0.296\\
  9 &     0.064 & 0.067 & \md 0.062 & \md 0.062 & 0.065 & \md 0.062 & \md 0.062 & \md 0.062 & 0.060\\
 10 &     0.014 & 0.014 & \md 0.012 &     0.013 & 0.014 &     0.013 &     0.013 &     0.013 & 0.012\\
 11 &     0.103 & 0.111 & \md 0.097 &     0.098 & 0.104 &     0.098 &     0.098 & \md 0.097 & 0.097\\
 12 &     0.113 & 0.114 & \md 0.107 &     0.110 & 0.113 &     0.108 &     0.108 & \md 0.107 & 0.106\\
 13 &     0.028 & 0.031 &     0.029 &     0.031 & 0.030 &     0.028 & \md 0.026 &     0.027 & 0.026\\
 14 &     0.056 & 0.056 & \md 0.052 & \md 0.052 & 0.056 &     0.053 &     0.055 &     0.053 & 0.052\\
 15 & \md 0.003 & 0.004 & \md 0.003 & \md 0.003 & 0.004 & \md 0.003 &     0.004 & \md 0.003 & 0.003\\
\midrule
MRE &     0.092 & 0.141 &     0.019 &     0.037 & 0.128 &     0.031 &     0.043 & \md 0.019 & 0.000\\
MAE &     0.006 & 0.007 &     0.001 &     0.002 & 0.006 &     0.002 &     0.001 & \md 0.001 & 0.000\\
\bottomrule
\end{tabular}
\caption{Halfspace depth, Cauchy distribution, $n = 1000$, $d = 5$, $N \approx 10000$ projections.}
\label{tab.exact_HD_Cauchy_10000}
\end{table}

\begin{table}[ht]\footnotesize
\centering
\begin{tabular}{r*{9}{c}}
\toprule
    &    RS &    GS &       RRS &       RGS &  RaSi &        SA &        CD &        NM & Exact\\
\midrule
  1 & 0.201 & 0.211 & \md 0.193 &     0.194 & 0.201 & \md 0.193 & \md 0.193 & \md 0.193 & 0.190\\
  2 & 0.135 & 0.138 & \md 0.131 &     0.132 & 0.137 & \md 0.131 & \md 0.131 & \md 0.131 & 0.130\\
  3 & 0.195 & 0.199 &     0.186 &     0.187 & 0.193 &     0.184 & \md 0.180 &     0.185 & 0.179\\
  4 & 0.057 & 0.064 &     0.052 &     0.052 & 0.055 &     0.051 & \md 0.050 & \md 0.050 & 0.050\\
  5 & 0.312 & 0.323 & \md 0.305 & \md 0.305 & 0.314 &     0.309 &     0.307 &     0.307 & 0.303\\
  6 & 0.296 & 0.297 &     0.288 &     0.289 & 0.295 &     0.287 & \md 0.286 & \md 0.286 & 0.285\\
  7 & 0.138 & 0.140 & \md 0.132 & \md 0.132 & 0.142 & \md 0.132 &     0.133 & \md 0.132 & 0.131\\
  8 & 0.319 & 0.322 & \md 0.305 &     0.306 & 0.314 &     0.309 &     0.306 &     0.306 & 0.305\\
  9 & 0.312 & 0.318 &     0.302 &     0.302 & 0.313 &     0.303 &     0.301 & \md 0.300 & 0.299\\
 10 & 0.254 & 0.259 &     0.249 &     0.250 & 0.257 &     0.249 &     0.249 & \md 0.248 & 0.247\\
 11 & 0.122 & 0.118 & \md 0.114 & \md 0.114 & 0.122 &     0.115 & \md 0.114 & \md 0.114 & 0.112\\
 12 & 0.158 & 0.159 & \md 0.143 & \md 0.143 & 0.155 &     0.145 & \md 0.143 & \md 0.143 & 0.142\\
 13 & 0.171 & 0.175 &     0.163 &     0.168 & 0.172 &     0.163 &     0.164 & \md 0.162 & 0.161\\
 14 & 0.078 & 0.085 & \md 0.072 &     0.078 & 0.081 &     0.074 &     0.074 &     0.074 & 0.072\\
 15 & 0.233 & 0.235 & \md 0.224 &     0.225 & 0.231 &     0.225 & \md 0.224 &     0.225 & 0.224\\
\midrule
MRE & 0.074 & 0.107 &     0.014 &     0.025 & 0.075 &     0.018 & \md 0.012 &     0.012 & 0.000\\
MAE & 0.010 & 0.014 &     0.002 &     0.003 & 0.010 &     0.003 & \md 0.002 &     0.002 & 0.000\\
\bottomrule
\end{tabular}
\caption{Halfspace depth, t-distribution, $n = 1000$, $d = 5$, $N \approx 10000$ projections.}
\label{tab.exact_HD_t-Dist_10000}
\end{table}

\begin{table}[ht]\footnotesize
\centering
\begin{tabular}{r*{9}{c}}
\toprule
    &    RS &    GS &       RRS &       RGS &  RaSi &        SA &        CD &        NM & Exact\\
\midrule
  1 & 0.277 & 0.273 &     0.268 &     0.268 & 0.277 & \md 0.267 &     0.268 & \md 0.267 & 0.264\\
  2 & 0.281 & 0.292 & \md 0.267 &     0.275 & 0.283 &     0.268 & \md 0.267 & \md 0.267 & 0.266\\
  3 & 0.247 & 0.252 &     0.242 &     0.242 & 0.248 &     0.242 &     0.242 & \md 0.241 & 0.239\\
  4 & 0.358 & 0.363 & \md 0.347 &     0.348 & 0.358 &     0.350 &     0.348 & \md 0.347 & 0.346\\
  5 & 0.227 & 0.223 &     0.218 & \md 0.217 & 0.229 &     0.218 & \md 0.217 & \md 0.217 & 0.215\\
  6 & 0.237 & 0.235 & \md 0.225 & \md 0.225 & 0.235 &     0.227 &     0.227 & \md 0.225 & 0.224\\
  7 & 0.183 & 0.187 & \md 0.173 &     0.174 & 0.183 &     0.175 &     0.175 &     0.174 & 0.172\\
  8 & 0.242 & 0.244 & \md 0.232 &     0.234 & 0.243 &     0.234 &     0.233 &     0.233 & 0.231\\
  9 & 0.278 & 0.278 &     0.269 & \md 0.267 & 0.277 &     0.271 &     0.270 &     0.268 & 0.266\\
 10 & 0.233 & 0.237 & \md 0.219 & \md 0.219 & 0.234 &     0.221 &     0.221 & \md 0.219 & 0.219\\
 11 & 0.304 & 0.307 &     0.295 & \md 0.292 & 0.307 &     0.295 &     0.295 &     0.294 & 0.291\\
 12 & 0.246 & 0.246 & \md 0.232 & \md 0.232 & 0.243 &     0.233 & \md 0.232 &     0.233 & 0.230\\
 13 & 0.265 & 0.266 & \md 0.255 &     0.258 & 0.263 &     0.256 &     0.256 & \md 0.255 & 0.253\\
 14 & 0.356 & 0.357 & \md 0.343 & \md 0.343 & 0.356 &     0.344 &     0.344 & \md 0.343 & 0.342\\
 15 & 0.297 & 0.304 & \md 0.284 & \md 0.284 & 0.295 &     0.286 &     0.286 &     0.285 & 0.284\\
\midrule
MRE & 0.063 & 0.074 &     0.009 &     0.012 & 0.063 &     0.015 &     0.013 & \md 0.009 & 0.000\\
MAE & 0.013 & 0.015 &     0.002 &     0.002 & 0.013 &     0.003 &     0.003 & \md 0.002 & 0.000\\
\bottomrule
\end{tabular}
\caption{Halfspace depth, uniform distribution, $n = 1000$, $d = 5$, $N \approx 10000$ projections.}
\label{tab.exact_HD_Uniform_10000}
\end{table}

\begin{table}[ht]\footnotesize
\centering
\begin{tabular}{r*{9}{c}}
\toprule
    &    RS &    GS &       RRS &       RGS &  RaSi &        SA &        CD &        NM & Exact\\
\midrule
  1 & 0.274 & 0.273 &     0.271 & \md 0.267 & 0.281 & \md 0.267 & \md 0.267 & \md 0.267 & 0.264\\
  2 & 0.325 & 0.335 &     0.315 &     0.319 & 0.323 &     0.315 & \md 0.314 & \md 0.314 & 0.312\\
  3 & 0.217 & 0.224 &     0.208 &     0.209 & 0.220 &     0.206 & \md 0.205 &     0.206 & 0.202\\
  4 & 0.254 & 0.252 & \md 0.243 & \md 0.243 & 0.255 &     0.245 & \md 0.243 & \md 0.243 & 0.242\\
  5 & 0.233 & 0.236 &     0.225 &     0.225 & 0.227 &     0.225 &     0.225 & \md 0.224 & 0.222\\
  6 & 0.194 & 0.201 &     0.188 &     0.187 & 0.194 &     0.189 & \md 0.186 &     0.187 & 0.185\\
  7 & 0.176 & 0.178 &     0.166 & \md 0.163 & 0.175 &     0.165 & \md 0.163 &     0.164 & 0.163\\
  8 & 0.290 & 0.288 &     0.284 &     0.279 & 0.287 &     0.279 & \md 0.277 &     0.280 & 0.276\\
  9 & 0.166 & 0.173 & \md 0.157 &     0.165 & 0.170 &     0.159 & \md 0.157 &     0.158 & 0.157\\
 10 & 0.261 & 0.266 & \md 0.250 & \md 0.250 & 0.255 &     0.251 &     0.253 &     0.251 & 0.249\\
 11 & 0.200 & 0.205 &     0.188 &     0.190 & 0.200 & \md 0.187 & \md 0.187 & \md 0.187 & 0.185\\
 12 & 0.327 & 0.327 &     0.315 &     0.315 & 0.325 &     0.312 & \md 0.308 &     0.309 & 0.308\\
 13 & 0.308 & 0.312 &     0.303 & \md 0.296 & 0.306 &     0.301 &     0.297 &     0.297 & 0.295\\
 14 & 0.221 & 0.221 &     0.209 &     0.206 & 0.215 &     0.210 & \md 0.205 &     0.208 & 0.204\\
 15 & 0.117 & 0.111 &     0.100 & \md 0.099 & 0.112 &     0.101 & \md 0.099 & \md 0.099 & 0.097\\
\midrule
MRE & 0.081 & 0.094 &     0.022 &     0.020 & 0.073 &     0.020 & \md 0.010 &     0.013 & 0.000\\
MAE & 0.013 & 0.016 &     0.004 &     0.003 & 0.012 &     0.003 & \md 0.002 &     0.002 & 0.000\\
\bottomrule
\end{tabular}
\caption{Halfspace depth, skew normal distribution, $n = 1000$, $d = 5$, $N \approx 10000$ projections.}
\label{tab.exact_HD_SkewNormal_10000}
\end{table}

\begin{table}[ht]\footnotesize
\centering
\begin{tabular}{r*{9}{c}}
\toprule
    &    RS &    GS &       RRS &       RGS &  RaSi &    SA &        CD &        NM & Exact\\
\midrule
  1 & 0.234 & 0.238 & \md 0.225 &     0.227 & 0.234 & 0.228 &     0.227 &     0.226 & 0.224\\
  2 & 0.212 & 0.221 &     0.201 &     0.202 & 0.217 & 0.203 &     0.201 & \md 0.200 & 0.199\\
  3 & 0.169 & 0.169 &     0.158 & \md 0.157 & 0.164 & 0.160 & \md 0.157 &     0.158 & 0.157\\
  4 & 0.342 & 0.348 & \md 0.334 &     0.337 & 0.345 & 0.338 &     0.336 &     0.335 & 0.334\\
  5 & 0.168 & 0.169 & \md 0.158 &     0.160 & 0.165 & 0.160 &     0.160 & \md 0.158 & 0.157\\
  6 & 0.236 & 0.233 & \md 0.223 &     0.225 & 0.232 & 0.225 &     0.224 &     0.224 & 0.223\\
  7 & 0.140 & 0.141 &     0.135 &     0.135 & 0.138 & 0.134 &     0.134 & \md 0.133 & 0.132\\
  8 & 0.211 & 0.214 &     0.201 &     0.200 & 0.209 & 0.201 &     0.200 & \md 0.199 & 0.199\\
  9 & 0.214 & 0.216 & \md 0.203 & \md 0.203 & 0.212 & 0.204 & \md 0.203 & \md 0.203 & 0.202\\
 10 & 0.198 & 0.207 & \md 0.190 &     0.200 & 0.198 & 0.193 & \md 0.190 & \md 0.190 & 0.189\\
 11 & 0.315 & 0.328 & \md 0.305 &     0.326 & 0.312 & 0.323 &     0.308 & \md 0.305 & 0.302\\
 12 & 0.193 & 0.201 &     0.184 &     0.184 & 0.190 & 0.185 & \md 0.183 & \md 0.183 & 0.183\\
 13 & 0.224 & 0.228 &     0.217 & \md 0.213 & 0.223 & 0.216 &     0.216 &     0.215 & 0.212\\
 14 & 0.301 & 0.305 &     0.293 & \md 0.292 & 0.299 & 0.293 & \md 0.292 & \md 0.292 & 0.291\\
 15 & 0.093 & 0.101 & \md 0.089 & \md 0.089 & 0.093 & 0.090 & \md 0.089 & \md 0.089 & 0.088\\
\midrule
MRE & 0.065 & 0.094 &     0.010 &     0.021 & 0.056 & 0.023 &     0.011 & \md 0.007 & 0.000\\
MAE & 0.011 & 0.015 &     0.002 &     0.004 & 0.009 & 0.004 &     0.002 & \md 0.001 & 0.000\\
\bottomrule
\end{tabular}
\caption{Halfspace depth, exponential distribution, $n = 1000$, $d = 5$, $N \approx 10000$ projections.}
\label{tab.exact_HD_Exponential_10000}
\end{table}

\clearpage


\begin{table}[ht]\footnotesize
\centering
\resizebox{0.99\linewidth}{!}{%
\begin{tabular}{ll*{7}{c}}
\toprule
&&\multicolumn{7}{c}{Approximation algorithm}\\
\cmidrule{3-9}
$d$ & Distribution &           RS &           GS &          RRS &         RaSi &           SA &           CD &           NM\\
\midrule
 5 &        Normal &     0.031370 &     0.085732 &     0.000252 &     0.031346 &     0.001046 & \md 0.000044 &     0.000278\\
   &        Cauchy &     0.041369 &     0.094434 & \md 0.006588 &     0.041998 &     0.044832 &     0.017824 &     0.020638\\
   &       Uniform &     0.032574 &     0.089988 &     0.000284 &     0.033469 &     0.001285 & \md 0.000056 &     0.000329\\
   &        t-Dist &     0.030983 &     0.085391 &     0.000307 &     0.031952 &     0.001257 & \md 0.000050 &     0.000308\\
   &    SkewNormal &     0.035904 &     0.092107 &     0.001406 &     0.041867 &     0.009238 & \md 0.000329 &     0.004313\\
   &   Exponential &     0.033265 &     0.088961 &     0.000580 &     0.033646 &     0.005773 & \md 0.000256 &     0.002109\\
\bottomrule
\end{tabular}}
\caption{Mean absolute error (MAE) for the approximation of the zonoid depth, $n = 1000$ data points, $N \approx 100$ projections.}
\label{tab.MAE_ZD_100}
\end{table}

\begin{table}[ht]\footnotesize
\centering
\resizebox{0.99\linewidth}{!}{%
\begin{tabular}{ll*{8}{c}}
\toprule
&&\multicolumn{8}{c}{Approximation algorithm}\\
\cmidrule{3-10}
$d$ & Distribution &           RS &           GS &          RRS &          RGS &         RaSi &           SA &           CD &           NM\\
\midrule
 5 &        Normal &     0.009714 &     0.021233 &     0.000001 &     0.001457 &     0.010003 &     0.000291 &     0.000002 & \md 0.000001\\
   &        Cauchy &     0.016155 &     0.029608 &     0.000170 &     0.006634 &     0.016788 &     0.005390 &     0.006615 & \md 0.000020\\
   &       Uniform &     0.010280 &     0.022672 &     0.000001 &     0.001605 &     0.010222 &     0.000302 &     0.000004 & \md 0.000001\\
   &        t-Dist &     0.010006 &     0.021637 &     0.000001 &     0.001720 &     0.009553 &     0.000295 &     0.000002 & \md 0.000001\\
   &    SkewNormal &     0.011078 &     0.023798 &     0.000001 &     0.003045 &     0.015119 &     0.000340 &     0.000003 & \md 0.000001\\
   &   Exponential &     0.010899 &     0.023762 &     0.000001 &     0.001487 &     0.010426 &     0.000337 &     0.000004 & \md 0.000001\\
\bottomrule
\end{tabular}}
\caption{Mean absolute error (MAE) for the approximation of the zonoid depth, $n = 1000$ data points, $N \approx 1000$ projections.}
\label{tab.MAE_ZD_1000}
\end{table}

\begin{table}[ht]\footnotesize
\centering
\resizebox{0.99\linewidth}{!}{%
\begin{tabular}{ll*{8}{c}}
\toprule
&&\multicolumn{8}{c}{Approximation algorithm}\\
\cmidrule{3-10}
$d$ & Distribution &           RS &           GS &          RRS &          RGS &         RaSi &           SA &           CD &           NM\\
\midrule
 5 &        Normal &     0.003030 &     0.007351 &     0.000001 &     0.000001 &     0.003137 &     0.000088 &     0.000001 & \md 0.000001\\
   &        Cauchy &     0.006423 &     0.012417 &     0.000028 &     0.000231 &     0.006782 &     0.000417 &     0.005745 & \md 0.000001\\
   &       Uniform &     0.003164 &     0.007488 &     0.000001 &     0.000001 &     0.003349 &     0.000094 &     0.000001 & \md 0.000001\\
   &        t-Dist &     0.003121 &     0.007218 &     0.000001 &     0.000001 &     0.003166 &     0.000087 &     0.000001 & \md 0.000001\\
   &    SkewNormal &     0.003771 &     0.008543 &     0.000001 &     0.000001 &     0.005117 &     0.000097 &     0.000001 & \md 0.000001\\
   &   Exponential &     0.003501 &     0.008011 &     0.000001 &     0.000001 &     0.003436 &     0.000092 &     0.000001 & \md 0.000001\\
\bottomrule
\end{tabular}}
\caption{Mean absolute error (MAE) for the approximation of the zonoid depth, $n = 1000$ data points, $N \approx 10000$ projections.}
\label{tab.MAE_ZD_10000}
\end{table}

\begin{table}[ht]\footnotesize
\centering
\resizebox{0.99\linewidth}{!}{%
\begin{tabular}{ll*{7}{c}}
\toprule
&&\multicolumn{7}{c}{Approximation algorithm}\\
\cmidrule{3-9}
$d$ & Distribution &           RS &           GS &          RRS &         RaSi &           SA &           CD &           NM\\
\midrule
 5 &        Normal &     0.060247 &     0.166683 &     0.000481 &     0.060159 &     0.001987 & \md 0.000084 &     0.000525\\
   &        Cauchy &     0.115329 &     0.292067 & \md 0.017188 &     0.117511 &     0.119828 &     0.044392 &     0.049441\\
   &       Uniform &     0.066873 &     0.188808 &     0.000574 &     0.069567 &     0.002667 & \md 0.000113 &     0.000658\\
   &        t-Dist &     0.059165 &     0.164331 &     0.000550 &     0.060612 &     0.002379 & \md 0.000094 &     0.000578\\
   &    SkewNormal &     0.070097 &     0.181868 &     0.002772 &     0.082358 &     0.017901 & \md 0.000617 &     0.008739\\
   &   Exponential &     0.066097 &     0.179209 &     0.001146 &     0.067098 &     0.010217 & \md 0.000514 &     0.003898\\
\bottomrule
\end{tabular}}
\caption{Mean relative error (MRE) for the approximation of the zonoid depth, $n = 1000$ data points, $N \approx 100$ projections.}
\label{tab.MRE_ZD_100}
\end{table}

\begin{table}[ht]\footnotesize
\centering
\resizebox{0.99\linewidth}{!}{%
\begin{tabular}{ll*{8}{c}}
\toprule
&&\multicolumn{8}{c}{Approximation algorithm}\\
\cmidrule{3-10}
$d$ & Distribution &           RS &           GS &          RRS &          RGS &         RaSi &           SA &           CD &           NM\\
\midrule
 5 &        Normal &     0.018782 &     0.040681 &     0.000002 &     0.002805 &     0.019107 &     0.000535 &     0.000004 & \md 0.000001\\
   &        Cauchy &     0.043137 &     0.080894 &     0.000465 &     0.016957 &     0.045022 &     0.010639 &     0.017534 & \md 0.000056\\
   &       Uniform &     0.021093 &     0.047018 &     0.000002 &     0.003281 &     0.021418 &     0.000597 &     0.000008 & \md 0.000001\\
   &        t-Dist &     0.018799 &     0.041453 &     0.000002 &     0.003367 &     0.018021 &     0.000537 &     0.000004 & \md 0.000001\\
   &    SkewNormal &     0.021533 &     0.046762 &     0.000002 &     0.005977 &     0.029959 &     0.000643 &     0.000005 & \md 0.000001\\
   &   Exponential &     0.021569 &     0.047682 &     0.000002 &     0.003068 &     0.020915 &     0.000653 &     0.000007 & \md 0.000001\\
\bottomrule
\end{tabular}}
\caption{Mean relative error (MRE) for the approximation of the zonoid depth, $n = 1000$ data points, $N \approx 1000$ projections.}
\label{tab.MRE_ZD_1000}
\end{table}

\begin{table}[ht]\footnotesize
\centering
\resizebox{0.99\linewidth}{!}{%
\begin{tabular}{ll*{8}{c}}
\toprule
&&\multicolumn{8}{c}{Approximation algorithm}\\
\cmidrule{3-10}
$d$ & Distribution &           RS &           GS &          RRS &          RGS &         RaSi &           SA &           CD &           NM\\
\midrule
 5 &        Normal &     0.005834 &     0.014136 &     0.000001 &     0.000002 &     0.005985 &     0.000161 &     0.000002 & \md 0.000001\\
   &        Cauchy &     0.017120 &     0.032837 &     0.000076 &     0.000619 &     0.017988 &     0.001450 &     0.015270 & \md 0.000002\\
   &       Uniform &     0.006517 &     0.015546 &     0.000001 &     0.000002 &     0.006914 &     0.000183 &     0.000002 & \md 0.000001\\
   &        t-Dist &     0.005958 &     0.013804 &     0.000001 &     0.000002 &     0.005981 &     0.000157 &     0.000002 & \md 0.000001\\
   &    SkewNormal &     0.007359 &     0.016835 &     0.000001 &     0.000002 &     0.009915 &     0.000181 &     0.000002 & \md 0.000001\\
   &   Exponential &     0.007000 &     0.015922 &     0.000001 &     0.000002 &     0.006924 &     0.000179 &     0.000002 & \md 0.000001\\
\bottomrule
\end{tabular}}
\caption{Mean relative error (MRE) for the approximation of the zonoid depth, $n = 1000$ data points, $N \approx 10000$ projections.}
\label{tab.MRE_ZD_10000}
\end{table}

\end{document}